\documentclass[useAMS,usenatbib]{mn2e}
\newcommand \kms{km s$^{-1}$}
\newcommand \zabs{z$_{\rm abs}$}

\newcommand \url{}

\newcommand{\ion}[1]{~\textsc{#1}}

\usepackage{longtable}
\usepackage{graphicx,subcaption}
\usepackage{lscape}
\usepackage{rotating}
\usepackage{amssymb}
\usepackage{amsmath}
\usepackage{hyperref}
\usepackage[T1]{fontenc} 
\usepackage{lmodern} 
\rmfamily 
\DeclareFontShape{T1}{lmr}{b}{sc}{<->ssub*cmr/bx/sc}{}
\DeclareFontShape{T1}{lmr}{bx}{sc}{<->ssub*cmr/bx/sc}{}

\def \afe{[$\alpha$/Fe]}
\def \HI{H\textsc{i}}

\def \omegaM{log($\Omega_{\rm M}$)}

\def \IC{IC($X_{i}$)}
\def \ZIC{[M/H]${\rm _{IC}}$}
\def \meanZ{$\langle Z/Z_{\odot} \rangle$}
\hypersetup{draft}

\begin{document}

\title[Chemical evolution of XQ-100 subDLAs]{Sub-damped Lyman $\alpha$ systems in the XQ-100 survey II -- Chemical evolution at $2.4 \leq {\rm z} \leq 4.3$}

\author[Berg et al.] {
\parbox[t]{\textwidth}{
Trystyn A. M. Berg$^{1}$, Michele Fumagalli$^{2,3,4}$, Valentina D'Odorico$^{5,6}$, Sara L. Ellison$^7$, Sebasti\'an L\'opez$^{8}$, George D. Becker$^{9}$, Lise Christensen$^{10}$, Guido Cupani$^{5}$, Kelly D. Denney$^{11}$, Rub\'en S\'anchez-Ram\'irez$^{12,13}$, G\'abor Worseck$^{14,15}$\\\\}
\\
$^{1}$European Southern Observatory, Alonso de Cordova 3107, Casilla 19001, Santiago, Chile.\\
$^{2}$Dipartimento di Fisica G. Occhialini, Universit\`a degli Studi di Milano Bicocca, Piazza della Scienza 3, 20126 Milano, Italy \\
$^{3}$Institute for Computational Cosmology, Durham University, South Road, Durham, DH1 3LE, UK \\
$^{4}$Centre for Extragalactic Astronomy, Durham University, South Road, Durham, DH1 3LE, UK \\
$^{5}$INAF-Osservatorio Astronomico di Trieste, Via Tiepolo 11, I-34143 Trieste, Italy.\\
$^{6}$Scuola Normale Superiore Piazza dei Cavalieri, 7 I-56126 Pisa, Italy.\\
$^{7}$Department of Physics and Astronomy, University of Victoria, Victoria, British Columbia, V8P 1A1, Canada.\\
$^{8}$Departamento de Astronom\'{\i}a, Universidad de Chile, Casilla 36-D, Santiago, Chile.\\
$^{9}$Department of Physics \& Astronomy, University of California, Riverside, California, 92521, USA.\\
$^{10}$DARK, Niels Bohr Institute, University of Copenhagen, Lyngbyvej 2, 2100 Copenhagen, Denmark\\
$^{11}$Department of Astronomy, The Ohio State University, 140 West 18th Avenue, Columbus, OH 43210, USA.\\
$^{12}$INAF, Istituto di Astrofisica e Planetologia Spaziali, Via Fosso del Cavaliere 100, I-00133 Roma, Italy.\\
$^{13}$Instituto de Astrof\'{\i}sica de Andaluc\'{\i}a (IAA-CSIC), Glorieta de la Astronom\'{\i}a s/n, E-18008, Granada, Spain.\\
$^{14}$Max-Planck-Institut f\"{u}r Astronomie, K\"{o}nigstuhl 17, D-69117 Heidelberg, Germany.\\
$^{15}$Institut f\"{u}r Physik und Astronomie, Universit\"{a}t Potsdam, Karl-Liebknecht-Str. 24/25, D-14476 Potsdam, Germany.\\
}

\maketitle
\begin{abstract}
We present the measured gas-phase metal column densities in 155 sub-damped Ly$\alpha$ systems (subDLAs) with the aim to investigate the contribution of subDLAs to the chemical evolution of the Universe.  The sample was identified within the absorber-blind XQ-100 quasar spectroscopic survey over the redshift range $2.4\leq{\rm z_{abs}}\leq4.3$. Using all available column densities of the ionic species investigated (mainly C\ion{iv}, Si\ion{ii}, Mg\ion{ii}, Si\ion{iv}, Al\ion{ii}, Fe\ion{ii}, C\ion{ii}, and O\ion{i}; in order of decreasing detection frequency), we estimate the ionization-corrected gas-phase metallicity of each system using Markov Chain Monte Carlo techniques to explore a large grid of \textsc{Cloudy} ionization models. Without accounting for ionization and dust depletion effects, we find that the \HI{}-weighted gas-phase metallicity evolution of subDLAs are consistent with damped Ly$\alpha$ systems  (DLAs). When ionization corrections are included, subDLAs are systematically more metal-poor than DLAs (between $\approx 0.5\sigma$ and $\approx 3\sigma$ significance) by up to $\approx1.0$ dex over the redshift range $3\leq{\rm z_{abs}}\leq4.3$. The correlation of gas-phase [Si/Fe] with metallicity in subDLAs appears to be consistent with that of DLAs, suggesting that the two classes of absorbers have a similar relative dust depletion pattern. As previously seen for Lyman limit systems, the gas-phase [C/O] in subDLAs remains constantly solar for all metallicities indicating that both subDLAs and Lyman limit systems could trace carbon-rich ejecta, potentially in circumgalactic environments.
\end{abstract}

\begin{keywords}
galaxies: high redshift -- galaxies: ISM -- quasars: absorption lines
\end{keywords}

\section{Introduction}

The production and dispersal of metals is a key product of the astrophysical processes that grow and regulate the evolution of galaxies. With the bulk of metals being produced in stars, studying the metallicity evolution of galactic systems within the Universe provides a tracer of the systems' integrated star formation history \citep{McWilliam97,Tolstoy09}. Galactic-scale winds and outflows are thought to redistribute the stellar material both within and outside galaxies, depositing material from the interstellar medium (ISM) into both the circumgalactic (CGM) and intergalactic (IGM) media \citep[e.g.][]{Steidel10,Lehner13,Peroux13,Rubin14,Kacprzak15}. Thus characterizing the distribution and evolution of metals in these gaseous reservoirs places fundamental constraints on models of galaxy evolution \citep{Pontzen08,Dave11,Peeples14}.

Gaseous absorbers seen along quasar (QSO) sightlines provide a unique method of probing the evolution of the neutral gas content of the Universe across cosmic time. Both damped and sub-damped Lyman $\alpha$ systems (DLAs and subDLAs respectively; \HI{} column densities of logN(\HI{})~$\geq20.3$ and $19.0\leq$~logN(\HI{})~$<20.3$) dominate the cosmic \HI{} mass density \citep{Peroux03,Prochaska09,Noterdaeme12,Zafar13,Berg19}. Studies of subDLAs and DLAs with high resolution spectrographs provide detailed chemical abundances and kinematic structure of the absorbing gas, enabling studies of the chemical enrichment of the Universe \citep{Pettini97,Rafelski12,Jorgenson13,Som15,Fumagalli16, Quiret16} and kinematic properties of absorbing systems \citep{Prochaska97,Ledoux06,Neeleman13,Christensen14}.

Despite the wealth of spectroscopic information obtained from individual quasar sightlines, identifying the physical properties of QSO absorbers has remained a challenge. Several studies have focussed on distinguishing the masses of the host galaxies associated with subDLAs and DLAs. By assuming a mass-metallicity relation and comparing the metallicity and velocity dispersion \citep[e.g.~as traced by the velocity width of the metal line, $v_{90}$;][]{Prochaska97} of subDLAs and DLAs, it has been proposed that subDLAs are associated with more massive galaxies than DLAs, as subDLAs are typically found to be more metal rich than DLAs, particularly at redshifts $z\lesssim3$ \citep{Khare07,Kulkarni10}.

An alternative explanation for the elevated metallicities in subDLAs is a bias in their identification that stems from selection effects. At $z\lesssim1.5$, when the Ly$\alpha$ is observed in the UV from space-based observatories, the bulk of subDLAs in the literature have been selected based on the presence of strong Mg~\textsc{ii} absorption that is easily observed from the ground.  While the Mg\ion{ii} pre-selection provides an efficient use of space-based resources \citep{Rao17}, it potentially leads to the selection of metal-rich systems \citep{Ellison06,Murphy07,DZavadsky09,Berg17}. Using simulations, \cite{Rhodin19} demonstrated that the mass-difference of DLAs and subDLA host is an effect of the gas cross section selections, primarily at $z<1$. Despite several large literature compilations of blind \citep[][]{Rafelski12} or uniformly-selected DLA studies \citep[][]{Jorgenson13}, there are only a few blind literature compilations or homogeneous samples of subDLA metal abundances available \citep[][with respective sample sizes of 12, 99, and 92 subDLAs]{DZavadsky03, Prochaska15,Quiret16}. These homogeneous samples of subDLAs have typically shown stronger metallicity evolution in subDLAs with respect to DLAs at \zabs{}~$\approx2-3$ \citep{Peroux07,Fumagalli16,Quiret16}.

In concert with the interest in subDLAs for their impact on global metallicity evolution, detailed chemical abundance patterns and their nucleosynthetic origins have been used to understand the physical properties of the host system. The evolution of [$\alpha$/Fe] (where $\alpha$ represents the elements from from $\alpha$-capture, such as O, Si, and Mg) and [C/O] in the local Universe, as measured using stellar and nebular abundances, can be used as a diagnostic to differentiate dwarf-like and Milky Way-like chemical enrichment patterns \citep{McWilliam97, Venn04, Tolstoy09, DBerg19}. While similar arguments have been made for subDLAs and DLAs by comparing [$\alpha$/Fe], [C/O], [Zn/Fe] and [Mn/Fe] to stellar abundances measured in the Local Group \citep{Meiring09, Rafelski12,Som13, Berg15II, Cooke17,Skuladottir17}, separating nucleosynthetic patterns from the  effects of dust depletion \citep{Pettini97,Vladilo01,Berg15II, deCia18}, ionization corrections \citep[][]{Vladilo01,DZavadsky09,Milutinovic10} and absorber selection effects remains a challenge, particularly for subDLAs.

Here, we present the chemical abundances of a blind sample of 155 \HI{}-selected subDLAs across redshifts $2.4<{\rm z_{abs}}<4.3$, as identified in the XQ-100 survey \citep{Lopez16}. In a previous paper \citep{Berg19} we have assessed the contribution of subDLAs to the cosmic \HI{} budget, finding that they contribute 10--20 per cent at $2.4<{\rm z_{abs}}<4.3$. In the current paper, we present metal column density measurements of these absorbers in order to quantify the contribution of subDLAs to the chemical enrichment of the Universe. Through-out this paper we assume a flat $\Lambda$CDM Universe with ${\rm H_{0}=70.0~km~s^{-1}~Mpc^{-1}}$ and ${\rm \Omega_{M,0}=0.3}$.

\section{Data}
\subsection{XQ-100 survey design and data}
\label{sec:Data}
The XQ-100 Legacy Survey \citep{Lopez16} observed 100 QSO sightlines between redshifts $3.5\leq z{\rm_{em}}\leq 4.5$  with the X-Shooter Spectrograph \citep{Vernet11} on the Very Large Telescope (VLT). Each QSO was observed for $\approx0.5$ or $\approx1$ hr based on the QSO's brightness; providing a median signal-to-noise ratio of $\approx30$ per pixel. The survey was designed for multiple science cases; from studying the associated absorbers near QSOs \citep{Perrotta16, Perrotta18} to characterizing the Ly$\alpha$ forest \citep{Irsic17a, Irsic17b}. As the QSOs were purposefully chosen to be blind to any intervening absorption line systems, the XQ-100 dataset provides a randomly-selected sample of subDLAs and DLAs to assess their cosmological contribution to the evolution of neutral gas at redshifts modestly probed in past surveys \citep{SanchezRamirez16,Berg16,Berg17,Christensen17,Berg19}. Details about the XQ-100 survey design and data reduction are summarized in \cite{Lopez16}.

The XQ-100 absorber catalogue consists of 155 subDLAs and 43 DLAs \citep{SanchezRamirez16,Berg19}. In brief, the absorbers were visually identified in the XQ-100 spectra by searching for simultaneous Ly$\alpha$, Ly$\beta$ and Ly$\gamma$ absorption consistent with simulated Voigt profiles of column density of logN(\HI{})~$\geq19.0$. The search for subDLA Ly$\alpha$ absorption was truncated bluewards of the observed-frame wavelength corresponding to the Lyman limit of the QSO due to the lack of continuum to properly fit any Ly$\alpha$ absorption. \HI{} column densities and absorption redshifts (\zabs{}) for the identified systems were measured by simultaneously fitting the continuum and Voigt profiles to the corresponding absorption for several Lyman series lines (typically Ly$\alpha$, Ly$\beta$, Ly$\gamma$, and Ly$\delta$). For more details about the identification and \HI{} properties of the XQ-100 subDLAs, see \cite{Berg19}. 

With the increasing density of Lyman series absorption towards bluer wavelengths, there is an increasing chance of false-positive identification of logN(\HI{})~$\lesssim 19.4$, \zabs{}~$\lesssim 3.2$ subDLAs due to blending of lower column density systems. Although using the presence of metal lines to remove false positives is common practice in the literature, the method introduces potential metallicity biases by preferentially removing low metallicity systems. Given the resolution and signal-to-noise ratio of the XQ-100 data, requiring the presence of metal lines would potentially remove systems with [M/H]~$\lesssim -2$ from the sample. Furthermore, $\approx40$ per cent of the subDLAs identified in \cite{Berg19} have the commonly identified C\ion{iv} and Si\ion{iv} in blended or noisy regions, which can also produce selection effects at various redshifts when these key metal lines shift into the Ly$\alpha$ forest.

To provide an upper and lower limit on the true distribution of subDLAs within the XQ-100 survey, we follow the method outlined in \cite{Berg19} and split the XQ-100 subDLA absorbers into two samples; the full catalogue of subDLAs (abbreviated FS; 155 subDLAs) and those with at least one confirmed detection of a metal species from the method outlined in Section \ref{sec:Mets} (referred to as the metal-selected sample (MS), see Table \ref{tab:mets}; 80 systems). Throughout the paper we remove all proximate absorbers to the host QSO (i.e.~absorbers who's \zabs{} is within 5000~\kms{} of the QSO's redshift), as proximate systems have been shown to have different metal abundance patterns and ionization conditions \citep[][]{Ellison02, Ellison10,Ellison11, Berg16, Perrotta16}.

\subsection{Metal column density measurements}
\label{sec:Mets}
For each subDLA in the XQ-100, we searched for absorption corresponding to the 20 species presented in Table \ref{tab:ionSum} within $\pm250$ km/s of the \HI{} absorption redshift \zabs{} where there is sufficient signal-to-noise ratio ($snr\gtrsim5$ pixel$^{-1}$) and little contamination from saturated or broad absorption features (such as the Ly$\alpha$ forest or telluric lines). gas-phase metal column densities were measured using the continuum-normalized spectrum presented in \cite{Lopez16}. For detected absorption lines, column densities were estimated using the apparent optical depth method \citep[AODM;][]{Savage91}. In brief, the AODM uses the integrated optical depth ($\tau$) across an unsaturated absorption line to estimate the column density N, i.e.
\begin{equation}
\label{eq:AODM}
{\rm N} = \frac{m_{e}c}{\pi e^{2} f \lambda} \int_{v_{min}}^{v_{max}} \tau dv
\end{equation}
where $m_{e}$ and $e$ are the electron mass and charge, $c$ is the speed of light, and $f$ and $\lambda$ are the oscillator strength and rest-frame wavelength of the atomic transition being measured \citep[taken from][]{Morton03}. For each subDLA, the velocity bounds $v_{min}$ and $v_{max}$ used for the AODM column density derivation were selected to encompass the full range of absorption of the strongest metal line detected in the system. However, if any absorption not associated with the system was present in a region associated with clean continuum or lengthy regions of clean continuum and minimal absorption, the velocity bounds were reduced to encompass only the observed, associated absorption. On rare occasions, we also shrunk the velocity bounds to avoid substantial blending to provide a more accurate lower limit on the column density.

Due to the modest resolution of X-Shooter (R=5100, 8800, and 5300 for the UVB, VIS, and NIR arms; respectively), it is possible that there is unseen contamination or saturation which would lead to incorrect AODM measurements. Following previous studies using X-Shooter and similar resolution spectrographs \citep{Prochaska15,Krogager16,Berg16}, we conservatively flagged the line as a lower limit if the strongest absorption component dropped below a continuum-normalized flux of 0.5  unless a weaker absorption line of the same species could confirm the measured column density is accurate. Additionally, any absorption containing some blending was flagged as an upper limit.  

If no absorption was detected for the metal species, we derive a 3$\sigma$ upper limit assuming the metal absorption line is an unresolved single component with a FHWM equal to the instrument resolution \citep{Pettini94}. The resulting column density limit is given by:

\begin{equation}
\label{eq:limit}
N=\frac{3 m_{e} c^2 ~ FWHM}{\pi e^{2} f \lambda^{2} (1+z_{abs}) snr},
\end{equation}

where the FWHM corresponds to the resolution (measured in the same units as $\lambda$) of the corresponding X-Shooter arm and $snr$ is the average signal-to-noise ratio (per pixel) of the continuum within the velocity bounds for the AODM measurement. Note that the measured $snr$ and thus the column density upper limits are sensitive to continuum level \citep{Berg16}. Errors on the AODM are derived using the variance obtained from the corresponding error spectrum. For marginally-detected absorption lines, we demand a 5$\sigma$ detection to be considered as a detection. Anything below this threshold, we adopt the $3\sigma$ upper limit instead, unless a stronger line from the same species could confirm the detected AODM column density is accurate. Through visually inspecting the data, we confirmed that requiring a 5$\sigma$ detection for our sample only removes uncertain cases where small dips in the continuum close to the expected absorption can be consistent with noise.

The column density for each absorption line is tabulated in Table \ref{tab:Naodm} (full version of Table \ref{tab:Naodm} is available online-only), along with the corresponding AODM velocity bounds, and flag indicating whether the absorption is adopted ($+1$), saturated or a lower limit ($+2$), undetected ($+4$), or blended ($+8$). Note that these flags can be combined, such that lines that are both blended and saturated would have a flag of $+10$. For reference, Table \ref{tab:ionSum} summarizes the number of subDLAs where: the species is detected (including both unsaturated and saturated absorption; ${\rm n_{det}}$), and there is sufficient data quality to obtain a reliable column density or upper limit measurement (${\rm n_{covered}}$). The frequency of metal line detections for species ($f_{\rm det} = \frac{{\rm n_{det}}}{{\rm n_{covered}}}$) is also provided in Table \ref{tab:ionSum}. Figure \ref{fig:velprof} provides the absorption line profiles in each subDLA (only metal species with $f_{\rm det}>0$).

\begin{table}
\begin{center}
\caption{XQ-100 subDLA ion detection statistics}
\label{tab:ionSum}
\begin{tabular}{lccc}
\hline
Ion & ${\rm n_{det}}$& ${\rm n_{covered}}$ & $f_{\rm det}$\\
\hline
Al\ion{ii} & 37 & 123 & 0.30\\
Al\ion{iii} & 4 & 109 & 0.04\\
C\ion{i} & 0 & 120 & 0.00\\
C\ion{ii} & 26 & 35 & 0.74\\
C\ion{iv} & 61 & 101 & 0.60\\
Ca\ion{i} & 0 & 87 & 0.00\\
Ca\ion{ii} & 0 & 72 & 0.00\\
Cr\ion{ii} & 0 & 124 & 0.00\\
Fe\ion{ii} & 29 & 152 & 0.19\\
Mg\ion{i} & 0 & 43 & 0.00\\
Mg\ion{ii} & 39 & 89 & 0.44\\
Mn\ion{ii} & 0 & 104 & 0.00\\
Ni\ion{ii} & 0 & 143 & 0.00\\
O\ion{i} & 21 & 31 & 0.68\\
S\ion{ii} & 0 & 11 & 0.00\\
Si\ion{ii} & 45 & 136 & 0.33\\
Si\ion{iii} & 0 & 107 & 0.00\\
Si\ion{iv} & 38 & 56 & 0.68\\
Ti\ion{ii} & 0 & 102 & 0.00\\
Zn\ion{ii} & 0 & 124 & 0.00\\
\hline
\end{tabular}
\end{center}
\end{table}

\begin{table*}
\begin{center}
\caption{XQ-100 subDLA metal column densities (shown for one absorber; full version online-only)}
\label{tab:Naodm}
\begin{tabular}{lccccccccc}
\hline
QSO& \zabs{}& Ion& $\lambda$& $v_{min}$& $v_{max}$& log(N)& Flag& log(N$_{\rm adopt}$)& log(N$_{\rm adopt}+$~IC)\\
& & & \AA{}& km s$^{-1}$& km s$^{-1}$& log(atoms cm$^{-2}$)& & log(atoms cm$^{-2}$)& log(atoms cm$^{-2}$)\\
\hline
J0003-2603& 3.05490& AlII& 1670& -100& 100& $<11.53$& $+5$& --& --\\
J0003-2603& 3.05490& AlII& --& --& --& --& \textbf{$+5$}& \textbf{$<11.53$}& \textbf{$<12.26^{+0.05}_{-0.05}$}\\
J0003-2603& 3.05490& AlIII& 1854& -100& 100& $<11.65$& $+5$& --& --\\
J0003-2603& 3.05490& AlIII& 1862& -100& 100& $<12.09$& $+4$& --& --\\
J0003-2603& 3.05490& AlIII& --& --& --& --& \textbf{$+5$}& \textbf{$<11.65$}& \textbf{$<12.34^{+0.05}_{-0.05}$}\\
J0003-2603& 3.05490& CIV& 1548& -275& -15& $13.02 \pm 0.03$& $+1$& --& --\\
J0003-2603& 3.05490& CIV& 1550& -275& -25& $13.14 \pm 0.04$& $+1$& --& --\\
J0003-2603& 3.05490& CIV& --& --& --& --& \textbf{$+1$}& \textbf{$13.08 \pm 0.05$}& \textbf{$14.22^{+0.06}_{-0.06}$}\\
J0003-2603& 3.05490& CaI& 2276& -100& 100& $<12.62$& $+4$& --& --\\
J0003-2603& 3.05490& CaI& 4227& -100& 100& $<10.93$& $+5$& --& --\\
J0003-2603& 3.05490& CaI& --& --& --& --& \textbf{$+5$}& \textbf{$<10.93$}& \textbf{$<12.19^{+0.06}_{-0.06}$}\\
J0003-2603& 3.05490& CaII& 3934& -100& 100& $<11.76$& $+5$& --& --\\
J0003-2603& 3.05490& CaII& 3969& -100& 100& $<12.26$& $+4$& --& --\\
J0003-2603& 3.05490& CaII& --& --& --& --& \textbf{$+5$}& \textbf{$<11.76$}& \textbf{$<12.31^{+0.05}_{-0.05}$}\\
J0003-2603& 3.05490& CrII& 2056& -100& 100& $<12.25$& $+5$& --& --\\
J0003-2603& 3.05490& CrII& 2062& -100& 100& $<12.57$& $+4$& --& --\\
J0003-2603& 3.05490& CrII& 2066& -100& 100& $<12.65$& $+4$& --& --\\
J0003-2603& 3.05490& CrII& --& --& --& --& \textbf{$+5$}& \textbf{$<12.25$}& \textbf{$<12.31^{+0.01}_{-0.01}$}\\
J0003-2603& 3.05490& FeII& 1608& -100& 100& $<12.77$& $+4$& --& --\\
J0003-2603& 3.05490& FeII& 2586& -70& 70& $12.64 \pm 0.17$& $+1$& --& --\\
J0003-2603& 3.05490& FeII& 2600& -70& 70& $12.51 \pm 0.06$& $+1$& --& --\\
J0003-2603& 3.05490& FeII& --& --& --& --& \textbf{$+1$}& \textbf{$12.57 \pm 0.06$}& \textbf{$13.34^{+0.06}_{-0.06}$}\\
J0003-2603& 3.05490& MnII& 2576& -100& 100& $<11.89$& $+5$& --& --\\
J0003-2603& 3.05490& MnII& 2594& -100& 100& $<12.08$& $+4$& --& --\\
J0003-2603& 3.05490& MnII& 2606& -100& 100& $<12.30$& $+4$& --& --\\
J0003-2603& 3.05490& MnII& --& --& --& --& \textbf{$+5$}& \textbf{$<11.89$}& \textbf{$<11.97^{+0.01}_{-0.01}$}\\
J0003-2603& 3.05490& NiII& 1709& -100& 100& $<13.25$& $+4$& --& --\\
J0003-2603& 3.05490& NiII& 1751& -150& 50& $<13.07$& $+5$& --& --\\
J0003-2603& 3.05490& NiII& --& --& --& --& \textbf{$+5$}& \textbf{$<13.07$}& \textbf{$<13.10^{+0.01}_{-0.00}$}\\
J0003-2603& 3.05490& TiII& 3073& -100& 100& $<12.25$& $+5$& --& --\\
J0003-2603& 3.05490& TiII& 3242& -100& 100& $<12.30$& $+4$& --& --\\
J0003-2603& 3.05490& TiII& --& --& --& --& \textbf{$+5$}& \textbf{$<12.25$}& \textbf{$<12.26^{+0.00}_{-0.00}$}\\
J0003-2603& 3.05490& ZnII& 2026& -100& 100& $<11.67$& $+5$& --& --\\
J0003-2603& 3.05490& ZnII& 2062& -100& 100& $<12.00$& $+4$& --& --\\
J0003-2603& 3.05490& ZnII& --& --& --& --& \textbf{$+5$}& \textbf{$<11.67$}& \textbf{$<11.69^{+0.00}_{-0.00}$}\\
\hline
\end{tabular}
\end{center}
\end{table*}

Following our previous metal column density measurements for the XQ-100 DLAs \citep{Berg16}, the final adopted column density ($N_{adopt}$) for each ion was computed using the following criteria:
\begin{itemize}
\item[(i)] If only one transition line has a flag of $+1$, the corresponding column density is adopted. 
\item[(ii)] If multiple transition lines of a given ionization species have a flag of $+1$, the error-weighted average column density using all cleanly detected lines is taken. The error on the adopted column density is computed using the standard deviation of the column densities. Note that the errors on individual column density measurements are typically much smaller than the difference in column densities measured.
\item[(iii)] If there are no transition lines with a flag of $+1$, but the difference between the most constraining upper and lower limits on the column density is $<0.4$ dex, the average of the two constraining column densities is taken. The error in the adopted column density is replaced by half the difference between the corresponding limits.
\item[(iv)] If there are no transition lines with a flag of $+1$, the most constraining limit is adopted. If there is both an upper and lower limit, the upper limit is preferentially adopted (unless the conditions of Step (iii) are met).
\end{itemize}
We adopt a minimum error on $N_{adopt}$ of 0.05 dex to account for errors in the continuum placement as was measured using multiple continuum fits to the XQ-100 quasars \citep{Berg16}.

\subsection{Metallicities and ionization modelling}
\label{sec:MetIon}

To enable the comparison of the gas-phase metallicity evolution between subDLAs and DLAs, we adopt the same method as previous studies \citep{Rafelski12,Rafelski14,Quiret16} that choose a single low-ionization species to represent the total gas-phase metallicity of the system ([M/H]), with preference for volatile elements to avoid effects of dust depletion. Following \cite{Quiret16}, we adopt the following ion abundance in decreasing order of preference (for both DLAs and subDLAs): Zn\ion{ii}, O\ion{i}, S\ion{ii}, Si\ion{ii}, Mg\ion{ii}, Fe\ion{ii}, and Ni\ion{ii}. Note that Zn\ion{ii} and S\ion{ii} lines are not detected in the XQ-100 subDLA sample, but upper limits for these two ions are used. The gas-phase metallicity for all the XQ-100 subDLAs following this method are tabulated in Table \ref{tab:mets}, and assume the \cite{Asplund09} solar abundances.

Although the described method works well for DLAs due the presence of self-shielded gas, it is unclear for a given subDLA how much of the material is ionized to a higher state from the UV background radiation, as corrections typically range between $\approx0-0.7$ dex \citep{DZavadsky03,Milutinovic10}. To account for these ionization effects in this work, we used the same method as \cite{Crighton15,Fumagalli16} to run Markov chain Monte Carlo (MCMC) modelling on each subDLA to estimate an ionization-corrected gas-phase metallicity (\ZIC{}). In brief, the MCMC modelling method uses a grid of \textsc{Cloudy} models \citep[version 17;][]{Cloudy17} that maps each set of \textsc{Cloudy} input parameters (gas density, \HI{} column density, redshift, and \ZIC{}) to the corresponding column densities of all metal ions that we considered in this work. The MCMC sampler searches this input parameter space for models with the column density pattern that best matches the \HI{} and metal ion column densities measured (exluding C\ion{i} as it probes cold gas), including lower and upper limits. The output of the MCMC modelling is a probability distribution of the independent sets of input parameters that can best match the observed column densities.  In this paper, we adopt the minimal model parameters of \cite{Fumagalli16} which assume a single phase slab of gas at constant gas density ($n_{H}$) illuminated on one side by both the \cite{HM12} UV and the cosmic microwave backgrounds. All metals are assumed to be in the gas phase with a solar abundance pattern of \cite{Asplund09}. We refer the reader to \cite{Fumagalli16} for a thorough assessment of the model and its implications for CGM gas.

For each subDLA, the input priors include the measured N(\HI{}) and error, \zabs{} (assuming an error of 0.001), and all adopted metal column densities with their corresponding errors. The precision of the estimated output parameters depends on how many column density measurements are provided to constrain the sampled models. In Table \ref{tab:mets} (full version online-only) we report the values of the ionization-corrected gas-phase metallicity (\ZIC{}) and gas density corresponding to the median of the probability distribution function and the $1\sigma$ percentiles errors derived from the MCMC analysis. We note that for the candidate subDLAs where no metals are detected (Section \ref{sec:Data}), the MCMC modelling of the \ZIC{} is only constrained by upper limits on a couple of key metal lines. In these cases, the MCMC modelling is poorly constrained and it is thus possible that the true metallicity of the subDLA is below the minimum metallicity of the \textsc{cloudy} grid (\ZIC{}$<-4$). Although we include the output from the MCMC modelling in Table \ref{tab:mets} for these poorly-constrained systems, we caution the user that the corresponding \ZIC{} are potentially upper limits to the true metallicity.

\begin{figure}
\begin{center}
\includegraphics[width=0.5\textwidth]{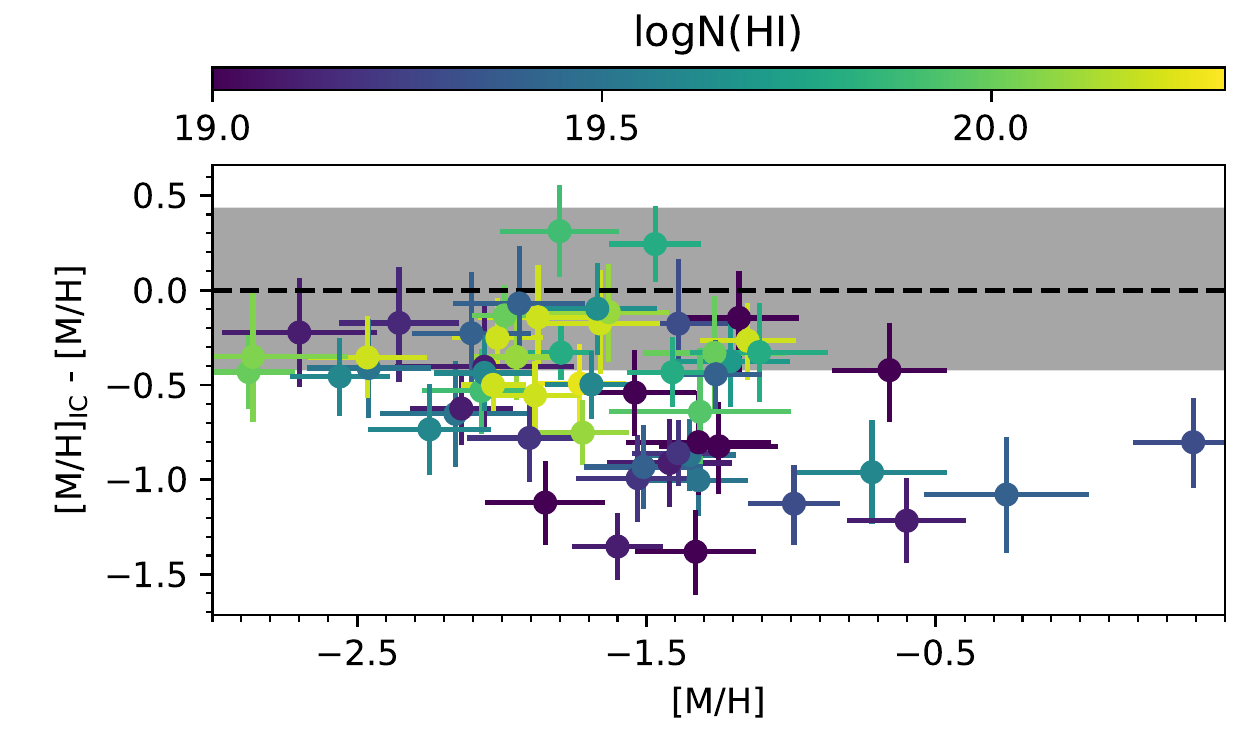}
\caption{The difference between the metallicities measured using the MCMC modelling (\ZIC) and using a single representative ion ([M/H]) as a function of [M/H]. Points are colour coded by their \HI{} column density. The grey band spans the median error in the metallicity difference, such that any points outside the grey band imply a large ionization correction is likely required.}
\label{fig:Zcompare}
\end{center}
\end{figure}

The resulting difference between [M/H] to \ZIC{} for all the metal-selected sample is shown in Figure \ref{fig:Zcompare}, as a function of both [M/H] and logN(\HI{}) (as shown by the colour of the points). As expected, the largest differences between [M/H] and \ZIC{} occur more frequently at logN(\HI{})$\lesssim19.6$, where the effects of self-shielding are lost \citep{Zheng02}.

We also estimate ionization corrections (IC) on the column densities for each ion by comparing the measured column density to the total column density of all ionization states within a \textsc{cloudy} model ($N_{c}(X)$). For ion $X_{i}$, the corresponding ionization correction to the column density [\IC{}] is defined as:
\begin{equation}
\rm{IC}(X_{i}) =  N_{c}(X) - N(X_{i}).
\end{equation}
Using the marginalized probability distribution functions (PDFs) of \textsc{cloudy} models from the MCMC modelling of each system to obtain a distribution of $N_{c}(X)$ values, we adopt the median and $1\sigma$ percentiles of the resulting \IC{} distribution for each system as the ionization correction and error for every species. The ionization-corrected column densities [log(N+IC)] and associated errors are provided in Table \ref{tab:Naodm} only for systems with detected metals (i.e.~within the MS). Note that the ionization correction PDFs in this work are consistent with the range of ionization corrections derived for other subDLAs in previous works that use \textsc{cloudy} \citep[e.g.][]{DZavadsky03, Milutinovic10, Battisti12, Som15}. We emphasize that differences in the ionization correction PDFs will arise when a particular ion (or set of ions) is used, especially if the selected ion(s) traces a particular ionization state of the gas, as our statistical approach combines all available ionization species. Below we discuss the effects of combining multiple ionization states on the derived ionization corrections (Section \ref{sec:ionAssump}).

\begin{table*}
\begin{center}
\caption{XQ-100 subDLA metallicities (full version online-only)}
\label{tab:mets}
\begin{tabular}{lccccccc}
\hline
QSO& \zabs{}& logN(\HI{}) [atoms cm$^{-2}$]& [M/H]& [M/H] ion& \ZIC{}& log($n_{H}$) [cm$^{-3}$]& Sample\\
\hline
J0003$-$2603& 2.78140& $19.00\pm0.20$& $<-2.68$& Mg\ion{ii}& $-3.83_{-0.45}^{+0.57}$& $-1.17_{-2.04}^{+1.45}$& FS\\
J0003$-$2603& 3.05490& $20.00\pm0.15$& $-2.88\pm0.16$& Fe\ion{ii}& $-3.31_{-0.12}^{+0.12}$& $-2.80_{-0.08}^{+0.08}$& MS\\
J0006$-$6208& 3.45890& $19.30\pm0.20$& $<-2.54$& Mg\ion{ii}& $-3.68_{-0.55}^{+0.56}$& $-1.39_{-2.00}^{+1.62}$& FS\\
J0034+1639& 2.99540& $19.00\pm0.15$& $<0.57$& Zn\ion{ii}& $-3.57_{-0.62}^{+0.83}$& $-1.04_{-2.12}^{+1.38}$& FS\\
J0034+1639& 3.02170& $19.30\pm0.15$& $<0.60$& Zn\ion{ii}& $-3.59_{-0.61}^{+0.75}$& $-1.15_{-2.01}^{+1.45}$& FS\\
J0034+1639& 3.22750& $19.80\pm0.15$& $<-0.33$& Zn\ion{ii}& $-3.83_{-0.45}^{+0.48}$& $-0.97_{-1.25}^{+1.31}$& FS\\
J0034+1639& 3.75350& $20.25\pm0.15$& $-1.73\pm0.16$& Si\ion{ii}& $-2.22_{-0.12}^{+0.13}$& $-2.70_{-0.08}^{+0.13}$& MS\\
J0034+1639& 3.82470& $19.00\pm0.20$& $<0.73$& Zn\ion{ii}& $-3.46_{-0.70}^{+0.87}$& $-0.70_{-1.36}^{+1.14}$& FS\\
J0034+1639& 4.22715& $19.60\pm0.20$& $<-3.28$& Si\ion{ii}& $-3.68_{-0.10}^{+0.13}$& $-4.40_{-0.07}^{+0.29}$& MS\\
J0042$-$1020& 2.75482& $20.10\pm0.15$& $>-1.35$& Mg\ion{ii}& $-1.09_{-0.23}^{+0.17}$& $-4.38_{-0.09}^{+0.22}$& MS\\
\hline

\end{tabular}
\end{center}
\end{table*}

\subsubsection{Effects of dust depletion}
\label{sec:dustAssump}

The depletion of gas onto dust grains plagues metallicity measurements in QSO absorber analyses, as a component of the total metal content is missed. Such an effect is suspected to have an impact on the total metallicity evolution of subDLAs and DLAs \citep{deCia18,Krogager19, Poudel20}. Corrections for dust depletion can be computed, but require the detection of multiple chemical elements in the absorber to either model the physical and chemical properties \citep{Jenkins09,deCia16} or measure abundance ratios of elements with similar nucleosynthetic origins \citep{Berg15}. For the XQ-100 subDLAs, there are insufficient detections of multiple elements to appropriately model the dust depletion in these system.

\begin{figure}
\begin{center}
\includegraphics[width=0.5\textwidth]{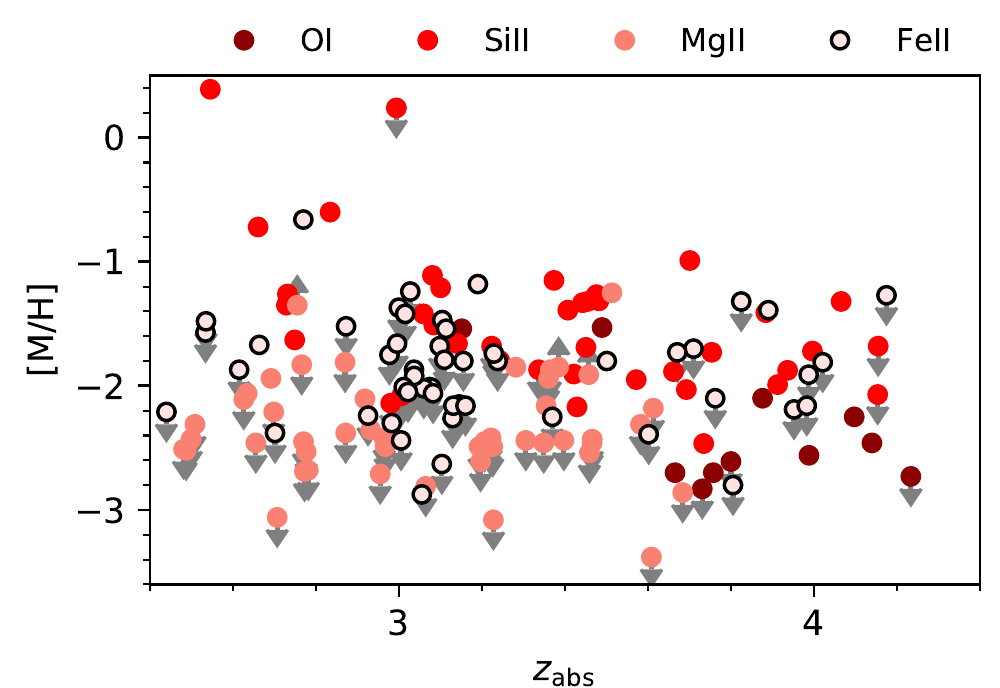}
\caption{The [M/H] of the XQ-100 subDLAs as a function of redshift. Each point is coloured by what ion was used to determine [M/H], (O\ion{i} as dark red, Si\ion{ii} as red, Mg\ion{ii} as light red, and Fe\ion{ii} as outlined circles) such that the darker the point the more volatile the ion that was used to measure [M/H].}
\label{fig:zZion}
\end{center}
\end{figure}

In such cases, the most appropriate course of action is to use available volatile metal tracers to measure the metallicity \citep{Rafelski12}. As described above, we implemented such a strategy to measure [M/H].  Figure \ref{fig:zZion} shows the XQ-100 subDLAs  [M/H] as a function of redshift, colour-coded by the ion used. The colour coding was chosen such that volatile elements are darker than refractory elements. From Figure \ref{fig:zZion} it is evident that there are clear selection effects as a function of redshift in what ion is chosen to measure [M/H] in the XQ-100 subDLAs. For example, systems that have [M/H] estimated by O\ion{i} are preferentially found at higher redshifts. This is likely due to a combination of absorbers being more metal-poor on average at high redshifts (and thus the strong O\ion{i} 1302\AA{} is not saturated) and the line is often found outside of the Ly$\alpha$ forest given the XQ-100 quasar redshifts. On the other hand, the more commonly-observed refractory ions Si\ion{ii} and Fe\ion{ii} have a variety of wavelengths and oscillator strengths, making them easier to identify across all redshifts covered by the XQ-100 absorbers. Coupled with underestimates in [M/H] due to dust depletion, we note that using a single ion as a [M/H] tracer can lead to potential selection effects that complicate the interpretation of metallicity evolution.

In their MCMC ionization modelling, \cite{Fumagalli16} experimented with changing the underlying solar abundance pattern by the \cite{Jenkins09} dust depletion pattern. From their experiment, \cite{Fumagalli16} found that both the inferred ionization-corrected metallicity and gas density between their nominal and dust-depleted abundance patterns models are highly correlated, with strong metallicity differences at low redshifts and high metallicity (\ZIC{}$\gtrsim-1.0$). For the 85 subDLAs in the \cite{Fumagalli16} sample that match the redshift and metallicity range of the XQ-100 subDLAs, the median \ZIC{} offset (with $\pm 1\sigma$ percentile confidence intervals) after including the effects of dust depletion is $0.00^{+0.10}_{-0.06}$. Note that for the majority of the absorbers in \cite{Fumagalli16}, the systematic offsets in \ZIC{} after including dust depletion are much smaller than the error in the ionization modelling. We have not included any dust corrections to our derived \ZIC{}. As we are using the same MCMC-based \ZIC{} modelling as \cite{Fumagalli16}, we expect the effects of including dust depletion and/or nucleosynthesis in the ionization modelling to be equally similar for the XQ-100 sample and expect the same statistical systematic offsets of $0.00^{+0.10}_{-0.06}$ to apply for the XQ-100 subDLAs.

\subsubsection{Effects of ionization assumptions}
\label{sec:ionAssump}

\begin{figure*}
\begin{center}
\includegraphics[width=\textwidth]{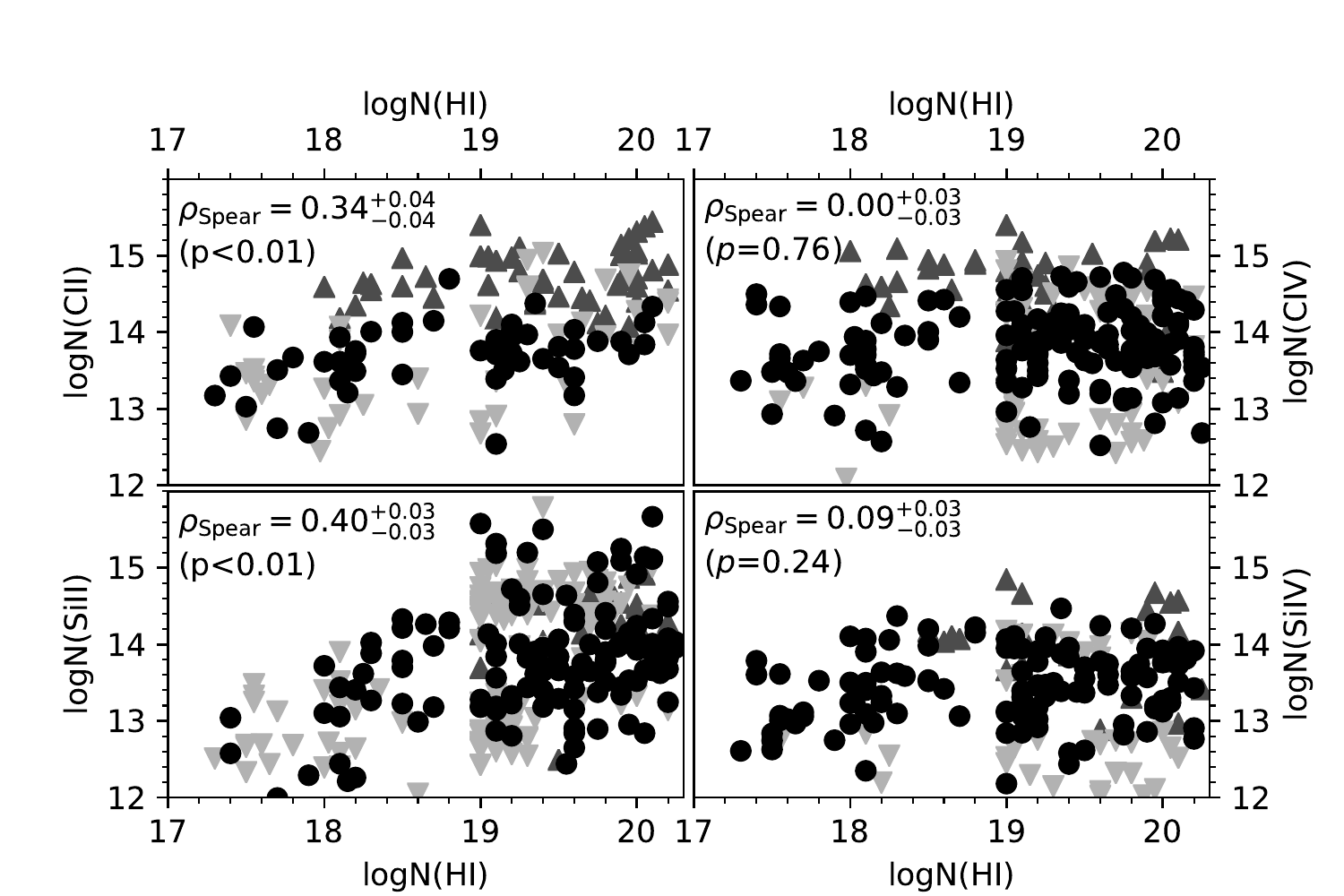}
\caption{The metal column density of four commonly observed species (C\ion{ii}, Si\ion{ii}, C\ion{iv}, Si\ion{iv}; top left, bottom left, top right, and bottom right, respectively) as a function of logN(\HI{}) for LLSs and subDLAs from XQ-100 and \citet{Prochaska15}. Upper and lower limits are denoted by the light and dark grey triangles (respectively), while clean detections are shown as black circles. The Spearman-rank correlation parameters ($\rho_{\rm Spear}$) and its $1\sigma$ confidence intervals, and associated p-values are shown in the top left corner of each panel. It is clear from the Spearman-rank correlation tests that there is no evidence of correlation of logN(C\ion{iv}) nor logN(Si\ion{iv}) with logN(\HI{}), but a modest correlation exists for C\ion{ii} and Si\ion{ii}.}
\label{fig:highion}
\end{center}
\end{figure*}

While we assume a typical quasar absorption line model of a single phase medium in our \textsc{Cloudy} modelling, it is possible that the subDLAs probe multiphase gas or gas exposed to a local stellar radiation source. For example, the lack of correlation in column density of high ionization species (e.g.~C\ion{iv} an Si\ion{iv}) across more than three orders of magnitude in N(\HI{}) can potentially be explained as a kinematically-distinct outer shell of CGM-like material surrounding the increasing amount of neutral, ISM-like gas within a galaxy \citep[][]{Fox07,Milutinovic10,Prochaska15}. For reference, Figure \ref{fig:highion} shows logN(C\ion{iv})  and logN(Si\ion{iv}) (top right and bottom right panels, respectively) as a function of logN(\HI{}) for Lyman Limit systems (LLS; QSO absorbers with $17.0\leq$~logN(\HI{})~$<19.0$) and subDLAs from XQ-100 and \cite{Prochaska15}. A Spearman-rank correlation test\footnote{The Spearman-rank correlation parameter ($\rho_{\rm Spear}$) was computed using a Monte Carlo simulation of 10 000 iterations by varying each datapoint within its Gaussian errorbars. Limits on column densities are varied uniformly between their measured value and $\pm1$~dex from the limiting value (depending on the limit type). The results are consistent irrespective of treating the limits as assumed detections.} demonstrates a lack of correlation of high ion column density with increasing N(\HI{}); $\rho_{\rm Spear}$ and its associated p-value are shown in the respective panels of Figure \ref{fig:highion}.   For comparison, the low-ion counterparts (logN(C\ion{ii}) and logN(Si\ion{ii}); top left and bottom left panels, respectively) show an increase in metal column with increasing logN(\HI{}) \citep[with the null hypothesis that there is no correlation rejected at $>3\sigma$ significance using a Spearman-rank correlation test, as previously seen by][]{Jorgenson13,Prochaska15}. Therefore if there is indeed a multiphase structure to subDLA gas, it may not be appropriate to mix both high and low ions' states as constraints in the modelling.

\begin{figure}
\begin{center}
\includegraphics[width=0.5\textwidth]{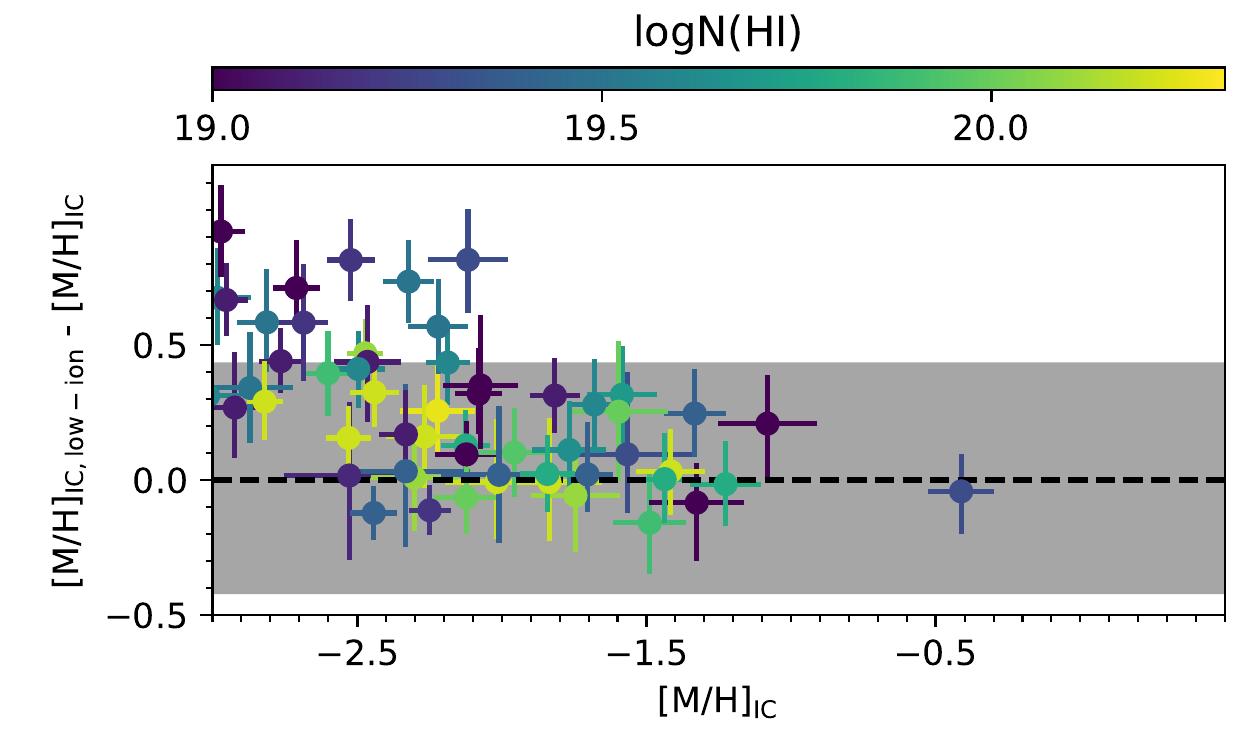}
\caption{The relative difference in the gas-phase metallicity measured upon treating logN(C\ion{iv})  and logN(Si\ion{iv}) as upper limits ([M/H]$_{\rm IC, low-ion}$) relative to the original run (\ZIC{}). The difference in metallicity is shown as a function of \ZIC{}, with the points colour-coded by their \HI{} column density. The shaded grey band shows the region in which the measured difference is consistent within the typical 17th and 83rd percentiles of the \ZIC{} distribution, while the error bars on the individual points represent the same percentile range for the individual system. SubDLAs with logN(\HI{})~$\lesssim19.5$ and \ZIC{}$\lesssim-2$ are the only systems that are significantly sensitive to a multi-phase prescription.}
\label{fig:lowvsall}
\end{center}
\end{figure}

To test how much influence the high-ion component has on the MCMC modelling, we reran our code assuming all logN(C\ion{iv})  and logN(Si\ion{iv}) measurements as upper limits, assuming only a portion of the high ion material is co-existent with the low ionized species. Figure \ref{fig:lowvsall} shows the difference in the gas-phase metallicity measured when treating logN(C\ion{iv})  and logN(Si\ion{iv}) as upper limits ([M/H]$_{\rm IC, low-ion}$) relative to the original run (\ZIC{}), as a function of \ZIC{} and logN(\HI{}) (represented by the colour of the points). Note that the absorbers shown in these plots represent a biased subset of the subDLAs, in that they must also have a detection (including lower limits) of at least one low-ionization species to be able to properly assess [M/H]$_{\rm IC, low-ion}$. Comparing to the typical error in \ZIC{} (grey band in both panels of Figure \ref{fig:lowvsall}),  subDLAs with logN(\HI{})~$\lesssim19.5$ and \ZIC{}~$\lesssim-2$ tend to have underestimated metallicities when assuming a single phase, and potentially need to be modelled with a multi-phase prescription. However, many of these systems with large offsets in [M/H]$_{\rm IC, low-ion} -$~\ZIC{} have [M/H]$_{\rm IC, low-ion}$ constrained only by either a single ion's column density or rely solely on lower limits on the metal column densities, leading to highly uncertain ionization corrections (see discussion above). Despite this systematic increase in the gas-phase metallicity from assuming a multiphase medium, we emphasize that when including this affect in the results presented in Section \ref{sec:Results}, our conclusions remain consistent within the $1\sigma$ scatter of the measurements from assuming a single phase, and thus do not change our interpretations.

\begin{figure}
\begin{center}
\includegraphics[width=0.5\textwidth]{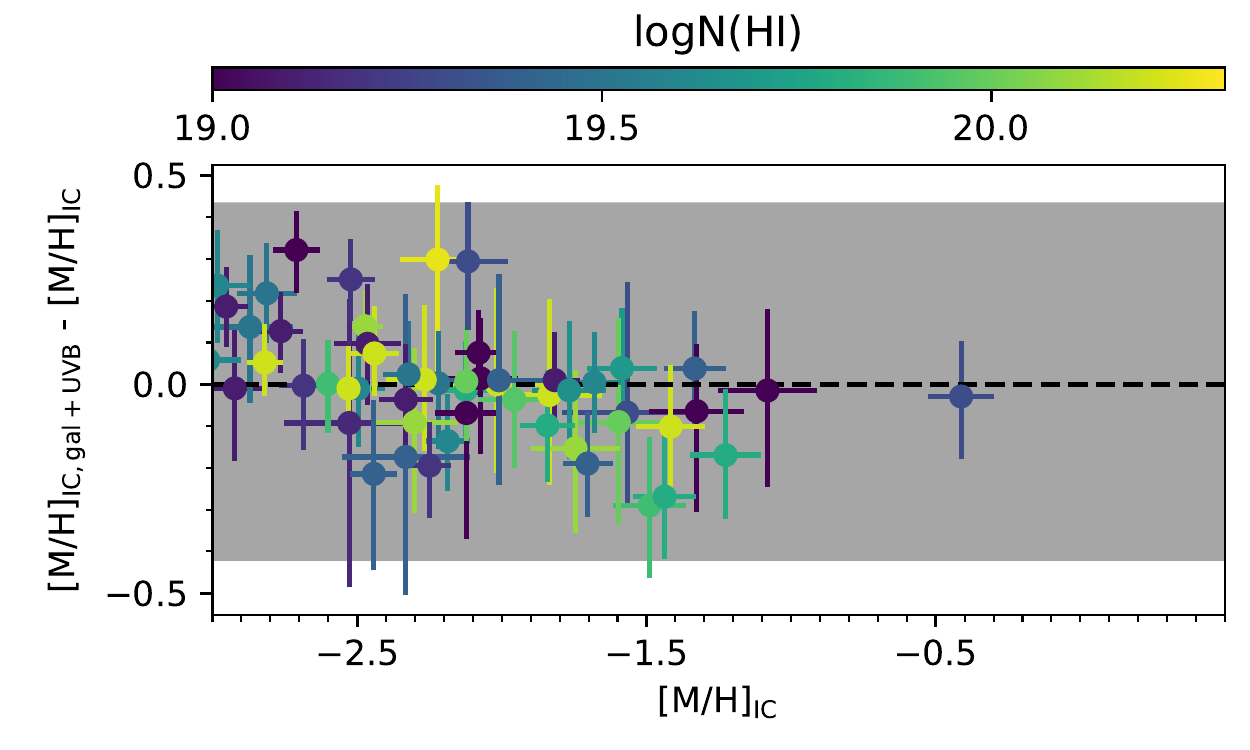}
\caption{The relative difference in the gas-phase metallicity measured upon including a stellar radiation component ([M/H]$_{\rm IC, gal+UVB}$) relative to the original run with only the extragalactic UV background (\ZIC{}). The notation of this figure is the same as Figure \ref{fig:lowvsall}. There appears to be no significant effect of including a stellar component in the \textsc{Cloudy} modelling.}
\label{fig:galvsuvb}
\end{center}
\end{figure}

We also reran our MCMC modelling with the inclusion of a galaxy radiation component on top of the UV background. We used the same local galaxy source models as described in \cite{Fumagalli16}, using an updated model grid with the same resolution as the minimal model. In brief, the galaxy source term in our models was derived using a \textsc{Starburst99} \citep{Starburst99} galaxy model undergoing a continuous star formation rate of 1~M$_{\odot}$~yr$^{-1}$, at a fixed distance from the slab of gas. The distance of the source to the gas slab is accounted for by varying the flux of the stellar radiation term within the MCMC ionization modelling. Similar to Figure \ref{fig:lowvsall}, Figure \ref{fig:galvsuvb} shows the relative difference in \ZIC{} when including the stellar radiation component ([M/H]$_{\rm IC, gal+UVB}$) to the fiducial \ZIC{} for the FS subDLAs. We remind the reader that, as with the low-ionization modelling above, several systems are poorly constrained due to a lack of detection of metals, particularly those at low metallicity. Differences in metallicity for these systems are highly uncertain. As \ZIC{} is predominantly constrained by the ionization parameter $U \propto \phi\, n_{H}^{-1}$ (where $\phi$ is the ionizing photon flux), the $\lesssim1\sigma$ difference  between  [M/H]$_{\rm IC, gal+UVB}$ and [M/H] for the bulk of the subDLAs is suggestive that the metallicity and ionization corrections are less sensitive to the inclusion of the stellar ionizing flux for the range of $n_{H}$ probed \citep[see also][]{Fumagalli16}.

\section{Results}
\label{sec:Results}

\subsection{Mean metallicity}
Within a given redshift bin, we define the \HI{}-weighted mean gas-phase metallicity (\meanZ) as:

\begin{equation}
\langle Z/Z_{\odot} \rangle = {\rm log} \left( \frac{\sum_{i} {\rm N(HI)}_{i} 10^{{\rm [M/H]}_{i}} }{\sum_{i} {\rm N(HI)}_{i}} \right)
\end{equation}

where N(\HI{})$_{i}$ and [M/H]$_{i}$ are the respective N(\HI) and metallicity of the $i$th subDLA in the redshift bin. To properly include both sampling and measurement errors in our computation of \meanZ{}, we use a bootstrap-Monte Carlo (B-MC) resampling technique, as was implemented in \cite{SanchezRamirez16,Berg19}. In summary, for each iteration of the B-MC method, the measured [M/H] and logN(\HI{}) for each subDLA is varied within their errors (assuming a Gaussian distribution). Within each iteration, a mock XQ-100 subDLA sample (i.e.~150 systems) is then randomly selected from these varied measurements (with replacement). This process is repeated 10 000 times to create 10 000 simulated catalogues of absorbers.

We experimented with including and excluding [M/H] limits into the B-MC simulation by adopting a random metallicity from a uniform distribution of values between an arbitrary minimum or maximum metallicity (for upper or lower limits, respectively) and the limiting value. Given that there is no true metallicity floor or ceiling, the choice of these arbitrary bounds can affect the calculation of \meanZ{}. We experimented with different metallicity floors ([M/H]~$=-3$, $-5$, and $-7$) and ceilings ([M/H]~$=+0.25$, $+0.5$, $+0.75$, and $+1.0$). Once the metallicity upper limits are included (rather than being ignored), there is a significant drop in the computed \meanZ{} of XQ-100 by up to $\approx-0.4$ dex. The choice of metallicity floor does not significantly affect \meanZ{} ($\lesssim 0.1$ dex differences comparing the metallicity floors of [M/H]~$=-3$ and $-\infty$). We therefore include upper limits in our B-MC simulations following this methodology, using a metallicity floor of [M/H]~$=-7$ \citep[in line with the most metal-poor stars known; e.g.~{[Fe/H]}~$\approx-7$ from][]{Keller14}. For lower limits, the choice of the maximum metallicity adopted has a significant impact on the \meanZ{} by up to $+0.25$ dex (assuming a ceiling of [M/H]~$\geq+1.0$; see thick and thin purple lines in the top panel of Figure \ref{fig:zZ}). Given that subDLAs and DLAs at such high metallicities are rarely observed \citep{Rafelski12,Berg15II,Quiret16,Fumagalli16} and that the typical $1\sigma$ variation in \meanZ{} is $\approx0.3$ dex from measurement errors and sampling effects, we chose to exclude metallicity lower limits in our calculations as they should have little effect on the computation \meanZ{}.

\subsection{Mean metallicity evolution of subDLAs}
\label{sec:meanZ}
In order to compare the gas-phase metallicity evolution of subDLAs with DLAs, we applied the B-MC methodology to the XQ-100 subDLAs and a statistical DLA sample composed of DLAs from \cite{Rafelski12,Rafelski14} and the XQ-100 survey \citep{Berg16}. Given the uncertainty in the number of metal-poor subDLAs without metal line detections (i.e.~absorbers only in the full sample catalogue), we also computed the mean gas-phase metallicity evolution of both the full and metal-selected subDLAs. To compute the evolution of \meanZ{} with redshift, we used a sliding bin technique, where \meanZ{} is computed within redshift bin width of $\Delta$\zabs{}~$=0.5$. The bin is then moved in redshift increments of $\delta$\zabs{}~$=0.05$ in order to finely sample the effects of absorbers being added and removed from the bin and remove effect from the choice of bins. The results using the sliding bin technique are broadly independent of the choice of $\Delta$\zabs{} and $\delta$\zabs{}, assuming reasonable values for these two parameters. 

\begin{figure*}
\begin{center}
\includegraphics[scale=1.0]{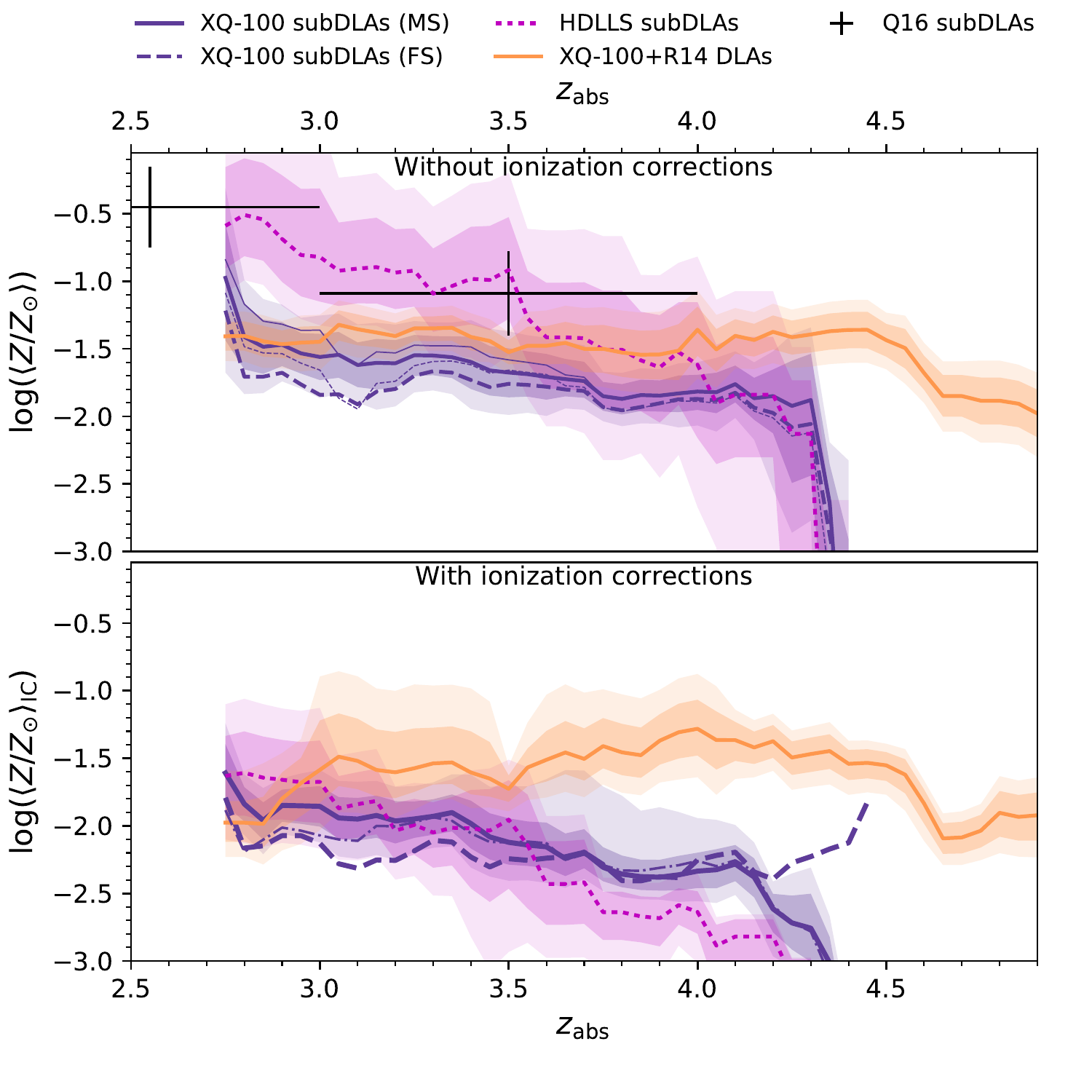}
\caption{gas-phase Metallicity evolution of the subDLAs (purple) and DLAs (orange) without (top panel) and with (bottom panel) ionization corrections applied. SubDLAs are split into full (purple dashed line) and metal-selected (purple solid line) samples. Curves are generated from sliding bin approach used in previous XQ-100 DLA/subDLA papers. The DLA sample consists of the XQ-100 DLA measurements \citep{Berg16} plus the literature catalogue of \citet[R14]{Rafelski14}. The mean gas-phase metallicity for the two highest redshift bins of \citet[Q16]{Quiret16} are shown as black crosses in the top panel. The B-MC method is also applied to the subDLAs from the HD-LLS sample \citep{Fumagalli16}, and is shown by the dashed magenta line. The thin purple lines in the upper panel denote the effects of including lower limits in the XQ-100 subDLA dataset, assuming a metallicity ceiling of $+1.0$ dex. Note that for the full-sample of subDLAs with no metal lines detected, the best-fitting ionization models are poorly constrained, typically reaching the lowest metallicity limit of the \textsc{cloudy} ionization grid ([M/H]~$=-4$), but could still be lower if such low metallicity systems exist. The thin dashed-dotted line in the lower panel shows the \meanZ{} evolution of the FS subDLAs upon the inclusion of an additional local galaxy radiation source.}
\label{fig:zZ}
\end{center}

\end{figure*}

The top panel of Figure \ref{fig:zZ} shows the median \meanZ{} evolution for the subDLA (purple lines) and DLA (orange line) samples. The subDLA is split into the full sample (FS; solid purple line) and metal-selected sample (MS; dashed purple line). The darker and lighter shaded regions represent the 67 and 90 per cent confidence intervals (respectively) for the DLA and FS subDLAs.

To compare to previous subDLA studies over a similar redshift range, we also included the sample from the High Dispersion LLS survey \citep[HD-LLS;][magenta dotted line, computed using the same B-MC method without ionization or dust corrections]{Prochaska15, Fumagalli16}. In brief, the HD-LLS survey selected all absorbers (including subDLAs) from QSO sightlines targetted for \HI{} absorption studies, with the subDLAs being identified solely based on the presence of \HI{}.  We also show the two high redshift bins of the mean gas-phase metallicity for subDLAs from the ESO UVES Advance Data Products Quasar Sample \citep[EUADP;][]{Zafar13,Quiret16} as the black crosses. As the EUADP sample typically probes lower redshifts than XQ-100 or the HD-LLS surveys and the EUADP subDLAs were selected based on metal absorption lines, it is not possible to make a direct comparison to the EUADP sample.  We point out that the sudden drop in gas-phase metallicity for subDLAs at \zabs{}$\gtrsim4.2$ is a consequence of a decreasing number of systems at these redshifts, and is likely not real.

There are three striking results in the top panel of Figure \ref{fig:zZ}. First, there appears to be very little difference in the gas-phase metallicity evolution between the XQ-100 subDLAs (either MS and FS) and DLAs, with both populations exhibiting an approximately  constant \HI{}-weighted gas-phase metallicity between $2.7 \lesssim$~\zabs{}~$\lesssim 4.3$. However, the subDLAs are systematically more metal-poor on average than the DLAs by 0.2--0.3 dex (albeit at $<1\sigma$ significance) across the entire redshift range. The low metallicities that we find in \zabs{}~$>3$ subDLAs contrast with \zabs{}~$\lesssim3$ subDLAs, which have been previously found to have metallicities that are higher than DLAs \citep[][]{Peroux08,Kulkarni10,Som15,Quiret16}. Our results are also in contrast with the HD-LLS subDLAs, which show a much steeper evolution, matching the high metallicities seen in the literature at \zabs{}$\lesssim3$, but quickly dropping to the low metallicities traced by the XQ-100 subDLAs. The differences between the various samples could potentially be explained by at least one of the following issues: selection biases between the three subDLA samples, dust depletion, or ionization corrections \citep[][]{DZavadsky09}. We discuss each of these in turn below.

\subsubsection*{Sample selection effects}
As pointed out in Section \ref{sec:Data} and discussed in more detail in \cite{Berg19}, metal-selection techniques of \HI{} absorbers are typically employed to discern between genuine subDLAs and false positives (such as two blended Ly$\alpha$ forest clouds), and may potentially introduce biases into the selection of metal abundances. For example, these metallicity-selection methods can miss metal-poor subDLAs if the observations are not sufficiently deep, and thus artificially increases the \meanZ{} measurement. Additionally, \zabs{}$\lesssim1.7$ subDLAs are selected using strong Mg\ion{ii} absorption, which can potentially lead to biases in the overall metallicity of these systems \citep[][conversely see discussions in \citet{Kulkarni07,Rao17} for counterarguments]{Murphy07,DZavadsky09,Berg17} and thus lead to higher metallicities and/or stronger metallicity evolution relative to DLAs. While these metal line selection methods are effective at finding systems, particularly at low redshifts when the metallicity is higher, we must proceed with caution at higher redshift where low column density, metal-poor systems are more frequent.

We tested the significance of metal-selection techniques using the XQ-100 data by splitting the subDLAs into the MS and FS (solid and dashed purple lines in the top panel of Figure \ref{fig:zZ}), effectively providing bounds on the \meanZ{} evolution from using metal selection techniques or not. It is clear that, although the measured \meanZ{} for the MS is indeed higher than the FS subDLAs, it is not sufficient to explain the higher gas-phase metallicity of the \cite{Quiret16} best-fit evolution at lower redshifts (i.e. black line in Figure \ref{fig:zZ}). As discussed in \cite{Berg19}, Ly$\alpha$ absorption from low column density subDLAs seen at \zabs{}~$\lesssim3.2$ become increasingly difficult to separate from higher Lyman series lines. In addition, strong metal lines (e.g. C\ion{iv} and Si\ion{iv} doublets) common in subDLAs shift into the Ly$\alpha$ forest. While the FS subDLAs may have a significant number of false-positives, the FS provides a lower bound on \meanZ{} from metal-blind selection effects.

\begin{figure}
\begin{center}
\includegraphics[scale=0.7]{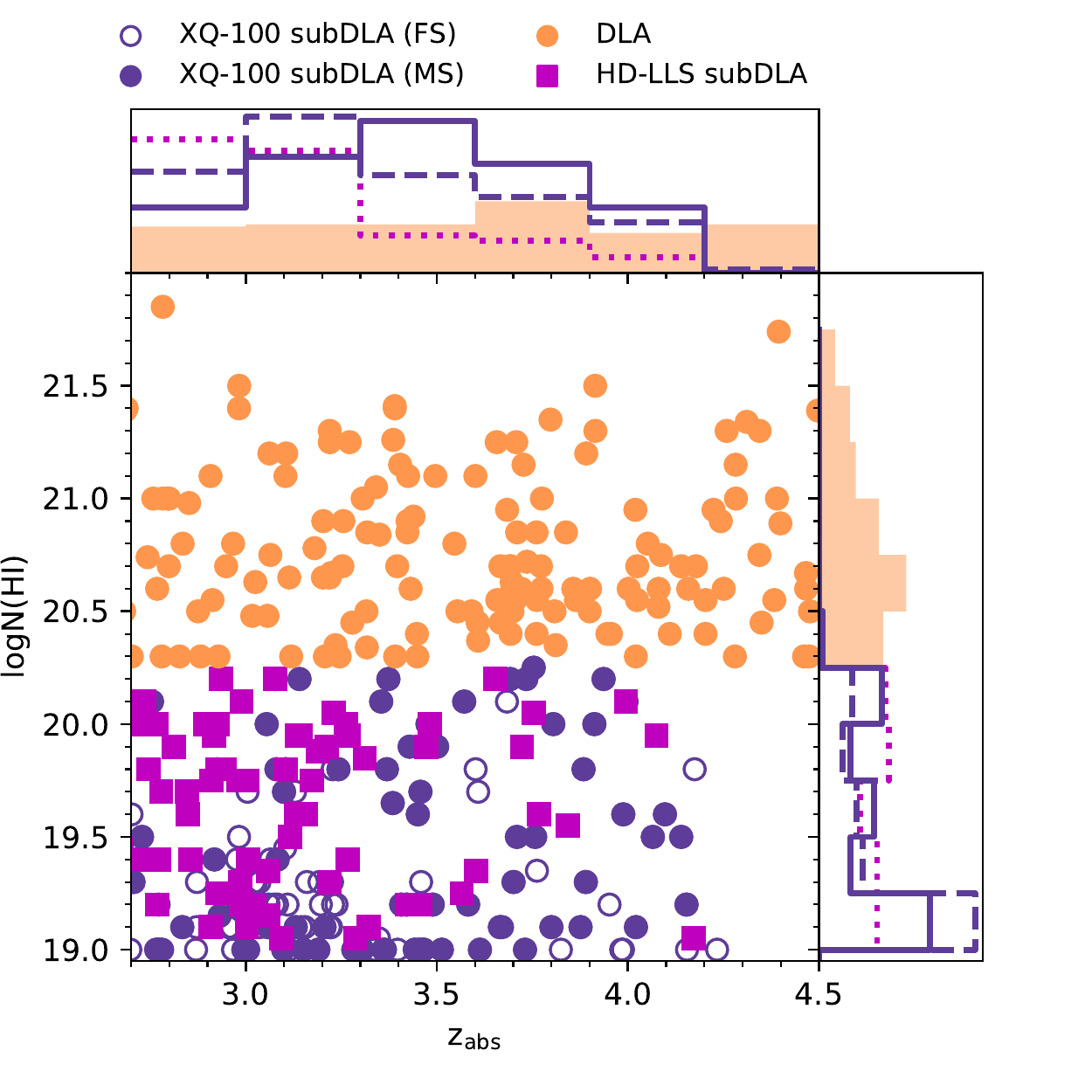}
\caption{The distribution of logN(\HI{}) as a function \zabs{} for the XQ-100  (purple circles) and HD-LLS (magenta squares) subDLA and DLA (orange circles) samples. The open and filled purple circles denote the full (FS) and metal-selected (MS) subDLAs. The normalized distributions of logN(\HI{}) and \zabs{} for the respective samples are also shown in the right and top panel. The notation of the lines follows that of Figure \ref{fig:zZ} for the subDLAs (solid purple is for the FS subDLAs, dashed purple for the MS subdlas, dotted magenta for the HD-LLS subDLAS), while the DLAs distributions are denoted by the filled orange histograms.}
\label{fig:zH}
\end{center}
\end{figure}

To highlight the properties of the various samples, Figure \ref{fig:zH} shows the distribution of logN(\HI{}) as a function \zabs{} for the three absorber samples (XQ-100 subDLA, HD-LLS subDLA, and DLA
 samples; purple circles, magenta squares, and orange circles, respectively). We highlight two differences in the sampling of the two subDLA samples: the bulk of the HD-LLS subDLAs are at \zabs{}~$\lesssim3.3$ with only 14 subDLAs at higher redshift, and the XQ-100 sample contains only 5 logN(\HI)~$\gtrsim19.5$ absorbers at redshift \zabs{}~$\leq3$ (compared the 18 HD-LLS systems in the same redshift and \HI{} column density range). The poor sampling of subDLAs in the respective regimes can potentially lead to the significant difference between the XQ-100 and HD-LLS subDLA samples, particularly as the XQ-100 subDLAs are mostly at [M/H]~$\lesssim-1.5$ at \zabs{}$\leq3$. The differences in the column density distributions of both XQ-100 and HD-LLS subDLA samples likely implies the two samples are probing different physical conditions (such as different gas densities), resulting in different ionization corrections. As discussed below, the inclusion of ionization modelling can correct for these differences in physical conditions between samples.

Another possibility for the large discrepancies could come from variations in data quality. While the XQ-100 survey provides a consistently modest signal-to-noise ratio across all 100 QSO spectra, both the HD-LLS and EUADP samples use data from various observing programs of different data quality, instrumental resolutions (HD-LLS only), and target selection (such as pre-selecting for absorbers). For example, the signal-to-noise ratio of the HD-LLS measurements vary between $\approx 10-40$~pixel$^{-1}$, using resolving powers R~$\approx5000-10000$ \citep{Prochaska15}, introducing variations in metallicity sensitivity by up to $\lesssim 1$ dex across the HD-LLS data. Thus some HD-LLS sightlines are more sensitive to metal-poor objects than XQ-100. Although the higher resolution of Keck/HIRES (HD-LLS sample) and VLT/UVES (EUADP) should both aid in the removal of equivalent metal-poor false positive absorbers found in XQ-100 at \zabs{}$\lesssim3.2$ and potentially explain the discrepancy in the steeper \meanZ{} evolution implied by the HD-LLS and EUADP samples, we emphasize that most of the absorbers exclusively in the FS have logN(\HI{})~$\leq19.4$ and are at \zabs{}~$\leq3.2$ \citep[i.e. where the prominent metal lines are within the Ly$\alpha$ forest;][]{Berg19}; the FS systems should not contribute significantly to the \HI{}-weighted \meanZ{} due to their low \HI{} column densities.

\subsubsection*{Dust depletion}
\label{sec:dust}
The differences in sample properties (Section \ref{sec:dustAssump}), data quality and resolution between the XQ-100 and HD-LLS surveys can greatly influence the detection of different elements, as some of the typical volatile elements used as total metallicity tracers are also weak lines (i.e.~ S\ion{ii} and Zn\ion{ii}) or are often blended with the Ly$\alpha$ forest (O\ion{i}). For redshifts \zabs{}~$\leq3.2$ (where the discrepancy between the XQ-100 and HD-LLS \meanZ{} evolution is biggest), only one of 24 of the [M/H] detections for the XQ-100 subDLAs makes use of a volatile element compared to $\approx25$ per cent of the HD-LLS subDLAs. When only refractory elements are used (Si, Fe, Mg) in our computation of \meanZ{} for the XQ-100 and HD-LLS samples at \zabs{}~$\geq3.2$,  we found \meanZ{} changed by at most $0.2$ dex in a given redshift bin. Assuming there is little redshift evolution in the dust content in absorbers \citep{deCia18}, it is unlikely that the treatment of dust can explain the discrepancy in \meanZ{} between the three subDLA samples \citep{Peroux07}.

We note that the bulk of QSOs observed in the XQ-100, EUADP, and HD-LLS surveys were selected based on optical photometry. Various studies have demonstrated that optically-selected QSOs can have a significant effect on the measured \HI{} column density distribution function and \meanZ{} by missing some of the most metal and \HI{} rich systems \citep{Jorgenson13,Krogager19}. Given that the two surveys being compared in this work all use optically-selected QSOs and the systems' metallicities are measured with the same methodology, their relative evolution of \meanZ{} should be consistent with each other; but may not completely encompass the total metallicity evolution of subDLAs.

\subsubsection*{Ionization effects}
As discussed in Section \ref{sec:MetIon}, subDLAs are not completely self-shielded and can be affected by external ionizing radiation sources. To study the effects of ionizing radiation, we re-computed the \meanZ{} for all the subDLAs and DLA samples using the MCMC-derived metallicities (\ZIC{}) presented in Section \ref{sec:MetIon}. We elected to draw \ZIC{} values from the resulting MCMC posterior probability density distribution for each subDLA in the B-MC simulation to account for the errors in the metallicity. The bottom panel of Figure \ref{fig:zZ} shows the resulting \meanZ{} curves using \ZIC{} in place of [M/H] for the two subDLA samples. We remind the reader that the MCMC simulations poorly constrain \ZIC{} for systems without metal detections such that the solid purple line in the bottom panel of Figure \ref{fig:zZ} is only used as a reference to study the relative effects on \meanZ{} from including these metal-poor systems.

With the inclusion of the estimated ionization corrections for subDLAs, it is clear from the bottom panel of Figure \ref{fig:zZ} that the difference at low redshifts in \meanZ{} between the HD-LLS and XQ-100 subDLA samples disappears.  The inclusion of ionization corrections demonstrates, with respect to DLAs, that the XQ-100 subDLAs exhibit a steeper gas-phase metallicity evolution for redshifts $2.7\leq$~\zabs{}~$\leq4.4$. Additionally, subDLAs still appear to be more metal poor than DLAs  at $\gtrsim 1\sigma$ significance ($-2.5\leq$~\ZIC{}~$\leq-1.7$; compared to [M/H]$\approx-1.5$ in DLAs over the same redshift range), consistent with what is seen in \cite{Fumagalli16}. 

The thin dashed-dotted line in the bottom panel of Figure \ref{fig:zZ} shows the FS subDLA sample using ionization corrections derived from the inclusion of a galaxy radiation source (Section \ref{sec:ionAssump}). There is a $\lesssim0.3$ dex offset in \meanZ{} once a stellar component to the radiation field is added. Nevertheless, the general trend of \meanZ{} from the inclusion of a galaxy radiation source remains consistent with the trends seen for the MS and FS samples. With the dash-dotted metallicity curve bounded by the FS and MS samples, the effects of sample selection have a much more significant effect than the choice of radiation source.

Independent of potential ionization and sample selection effects, it is evident from the top and bottom panels of Figure \ref{fig:zZ} that subDLAs at $3.0\leq$~\zabs{}~$\leq4.3$ are on average more metal-poor than DLAs across the same redshift range, which is in contrast to what is seen at redshifts \zabs{}~$\lesssim2$ \citep{Peroux07,DZavadsky09,Som13,Quiret16}. The inclusion of ionization corrections (i.e.~bottom panel of \ref{fig:zZ}), which are required to remove significant systematic errors, can also reconcile the differences between the observed \meanZ{} evolution from the HD-LLS and XQ-100 subDLAs.  At face value, the inclusion of ionization corrections extends the redshift range (from \zabs{}~$\gtrsim3.5$) at which the \meanZ{} of subDLAs evolves much more rapidly than DLAs \citep{Peroux07,Meiring09,Som13,Quiret16}. However, as discussed in Section  \ref{sec:MetIon}, the interpretation of the nature of the redshift evolution of \meanZ{} hinges on correct assumptions in the ionization modelling \citep[as previously noted, e.g.][]{DZavadsky09,Fumagalli16} along with a proper characterization of the false positive rate of \HI{} selected absorbers (i.e.~the difference in the solid and dashed purple curves in the bottom panel of Figure \ref{fig:zZ}), and consistent dust corrections.  The evidence of multiphase gas \citep[][and demonstrated in Figure \ref{fig:highion}]{Fox07,Milutinovic10,Prochaska15}  requires more complex modelling \citep[see \zabs{}~$\approx0$ example in][]{Buie20}. As noted in Figure \ref{fig:lowvsall}, assuming a multi-phase model may increase the metallicities of the subDLAs by $\approx0.2-0.3$ dex. Including such a shift in the subDLA gas-phase metallicity would be insufficient to raise the subDLA curves in the bottom panel of Figure \ref{fig:zZ} to completely match the higher metallicity DLA curve for redshifts \zabs{}~$\gtrsim3.5$.

Studies comparing the metallicity of the CGM to the host galaxies ISM at $z\leq0.2$ have typically found lower metallicities in the CGM \citep{Prochaska17,Kacprzak19}, which is also found for the host galaxies of $z<1$ subDLAs \citep{Rahmani16,Rhodin18}. Combining with the results of subDLAs being on average more metal-poor than DLAs, this could potentially imply that subDLAs tend to trace CGM-like rather than ISM-like environments more frequently than DLAs \citep{Fumagalli11}. \cite{Rhodin19} analysed zoom in simulations of a Milky Way-like progenitor and found sub-DLAs and DLAs arising in the galaxy CGM, but with higher metallicities than measured from the XQ-100 sub-DLAs. Future cosmological simulations with improved resolution in CGM environments \citep[e.g.][]{Peeples19, Vandevoort19} could provide more detailed predictions for the distributions of metallicities as a function of logN(\HI{}) in both the ISM and outer regions of the  CGM, bringing more context to interpreting these differences in metallicity of subDLA and DLA absorbers.

\subsection{The mass-density of metals in subDLAs}
\label{sec:omegaM}

In order to compare the relative contribution of subDLAs and DLAs to the overall metal budget of the Universe, we computed the mass-density of metals (\omegaM{}). For absorbers with an N(\HI{}) between N$_{1}$ and N$_{2}$, $\Omega_{M}$ is given by:

\begin{equation}
\label{eq:omegaM}
\Omega_{M} = \frac{1.3 m_{p} H_{0}}{c \rho_{\rm crit}} \int_{N_{1}}^{N_{2}} \, \frac{ N \, f(N)}{\rm X_{HI}}  \, 10^{\rm log(\langle Z/Z_{\odot} \rangle_{IC})} Z_{\odot} \, dN,
\end{equation}

where $m_{p}$ is the mass of the proton, $\rho_{\rm crit}$ is the critical density of the Universe at redshift 0, ${\rm X_{HI}}$ is the neutral fraction of atomic hydrogen gas, $f(N)$ is the \HI{} column density distribution function of the absorber, and $Z_{\odot}$ is the metallicity of the Sun in mass units  \citep[$Z_{\odot}=0.0142$;][]{Asplund09}.  The factor of 1.3 accounts for the mass of both atomic H and He. For each logN(\HI{}) bin in equation \ref{eq:omegaM}, we compute the \HI{}-weighted mean of ${\rm X_{HI}}\langle Z/Z_{\odot} \rangle_{IC}$.  To estimate the errors on \omegaM{}, we follow the same B-MC method by sampling the posterior distributions of the gas-phase metallicity and ${\rm X_{HI}}$ of each system's ionization modelling to reconstruct 10 000 samples to compute \omegaM{}. For each system, we randomly draw pairs of neutral fraction and metallicity from the MCMC modelling output to include any possible degeneracies between the two parameters in our computation of \omegaM{}. We also sample the respective $f(N)$ distributions from \cite{Berg19} for the FS subDLAs, MS subDLAs, and DLAs. Note that as the HD-LLS subDLA sample is selected in a similar fashion to the XQ-100 FS subDLAs, we adopt the FS $f(N)$ distribution for the computation of \omegaM{} for the HD-LLS subDLAs. \omegaM{} is computed within the same two redshift bins of $f(N)$ from \citet[$2.3 \leq$~\zabs{}~$<3.2$ and $3.2 \leq$~\zabs{}~$<4.5$]{Berg19}. 

Table \ref{tab:omegaM} presents the median and $1\sigma$ percentile confidence intervals of \omegaM{} for subDLAs and DLAs for the respective subDLA and DLA samples. We find DLAs and subDLAs contribute approximately the same amount of metals to the cosmic metal budget, with DLAs contributing between $\approx0.8$--1.1$\times$ as many metals compared to subDLAs  (depending on the subDLA sample and type of ionization modelling). We note that the errors on \omegaM{} for the XQ-100 sample are larger than for the other subDLA and DLA samples as a result of a significant fraction of XQ-100 absorbers having poorly-constrained ionization corrections due to the few number of line detections. Our result are inconsistent with similar measurements at $2 \lesssim$~\zabs{}~$\lesssim3$ in the literature \citep{Prochaska06,Peroux07,Fumagalli16}, where subDLAs contribute $\gtrsim3\times$ the amount of metals than DLAs \citep[as measured by][]{Rafelski14}.

\begin{figure}
\begin{center}
\includegraphics[width=0.49\textwidth]{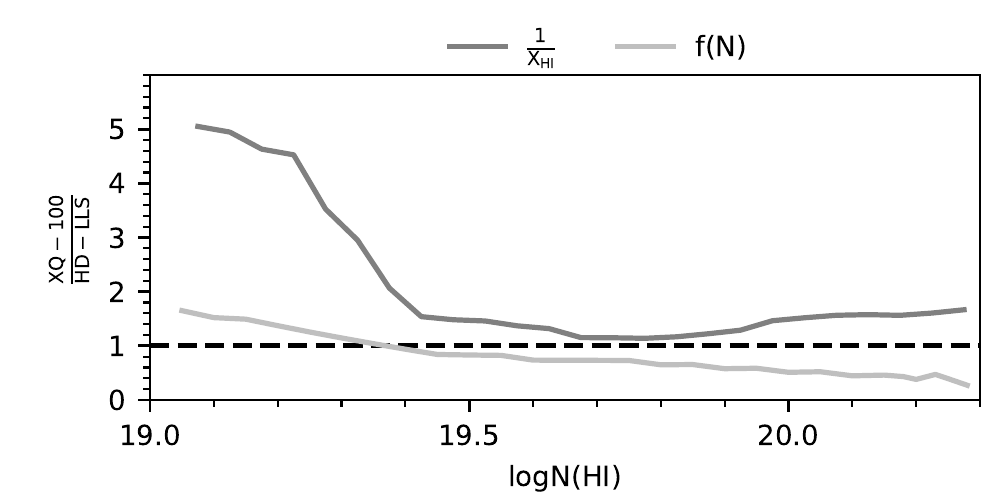}
\caption{The fractional difference of the assumed neutral fraction ($\frac{1}{{\rm X_{HI}}}$, dark grey line) and f(N) (light grey line) of this work relative to \citet[${\rm \frac{XQ-100}{HD-LLS}}$]{Fumagalli16} as a function of logN(\HI{}). Values of ${\rm \frac{XQ-100}{HD-LLS}}>1$ (i.e. above the dashed black line) imply the measured quantity from the method used in this paper is larger than that from \citet{Fumagalli16}. While f(N) has a small difference between the two methods, the neutral fraction can vary by up to a factor of 5 for low column density subDLAs (logN(\HI{})~$\lesssim19.4$).}
\label{fig:HDomegaM}
\end{center}
\end{figure}

Although our method recovers the same \omegaM{} value as \cite{Rafelski14} for the DLAs, we point out that we underestimate the \omegaM{} derived for the HD-LLS sample by almost an order of magnitude compared to the value of \citet[\omegaM{}~$\approx-5.8$]{Fumagalli16} using the same metallicity data. Using the functional forms of f(N) and neutral fraction from \cite{Fumagalli16}, we are able to reproduce the \omegaM{} for the HD-LLS sample, highlighting that the effects of the underlying functional forms of f(N) and neutral fractions of the sample have a significant effect on computing \omegaM{}. As stated above, we elected to randomly drawn pairs of neutral fraction and metallicity from the MCMC ionzation modelling done by \cite{Fumagalli16} over using a functional form fitted to the data, as this accounts for any possible degeneracies between metallicity and neutral fraction in the posterior distributions of the ionization modelling. Furthermore, the functional form of the neutral fraction used by \cite{Fumagalli16} is a linear regression spanning 3 orders of magnitude of \HI{} to cover both LLS and subDLAs in the HD-LLS sample; and thus may not be the best description of the neutral fraction in the subDLAs. To investigate the discrepancy in the derived \omegaM{} between \cite{Fumagalli16} and this paper, Figure \ref{fig:HDomegaM} shows the fractional difference in both the adopted neutral fraction (shown as $\frac{1}{{\rm X_{HI}}}$ following Equation \ref{eq:omegaM}; dark grey line) and f(N) (light grey line) as a function of logN(\HI{}) using \emph{the same HD-LLS data}. It is clear from Figure \ref{fig:HDomegaM} that the neutral fraction is at most $\approx5\times$ larger when using our method compared to the linear regression of \cite{Fumagalli16}, while the choice of f(N) has a much smaller effect on the calculation of \omegaM{}. We emphasize that, while  our \ZIC{} analysis is robust to the choice of ionization modelling (Sections \ref{sec:MetIon} and \ref{sec:meanZ}), the calculation of \omegaM{} is very sensitive to a combination of both the method used to describe the neutral fraction of gas in subDLAs and LLS samples \citep[][]{Fumagalli16}, and the small sample sizes which do not adequately sample the diverse range in physical environments probed by subDLAs (see discussion above about differences in the HD-LLS and XQ-100 subDLA samples).

\begin{table*}
\begin{center}
\caption{\omegaM{} estimates}
\label{tab:omegaM}
\begin{tabular}{l|ccc|ccc}
\hline
Sample & \multicolumn{3}{c}{$2.3 \leq$~\zabs{}~$<3.2$} & \multicolumn{3}{c}{$3.2 \leq$~\zabs{}~$<4.5$} \\
 & \omegaM$_{17}$ & \omegaM$_{50}$ & \omegaM$_{83}$ & \omegaM$_{17}$ & \omegaM$_{50}$ & \omegaM$_{83}$ \\
 \hline

XQ-100 subDLAs (FS) & $-6.8$ & $-5.8$ & $-4.7$ & $-7.0$ & $-6.3$ & $-5.5$\\
XQ-100 subDLAs (MS) & $-7.1$ & $-5.9$ & $-4.8$ & $-7.0$ & $-6.4$ & $-5.5$\\
XQ-100 subDLAs (MS; low-ions only) & $-7.7$ & $-6.9$ & $-6.3$ & $-7.8$ & $-7.4$ & $-6.9$\\
HD-LLS subDLAs & $-6.9$ & $-6.7$ & $-6.5$ & $-7.4$ & $-7.1$ & $-6.9$\\
XQ-100+R14 DLAs & $-6.9$ & $-6.5$ & $-6.0$ & $-6.4$ & $-6.0$ & $-5.4$\\

\hline
\end{tabular}
\end{center}
\end{table*}

\begin{figure*}
\begin{center}
\includegraphics[width=0.95\textwidth]{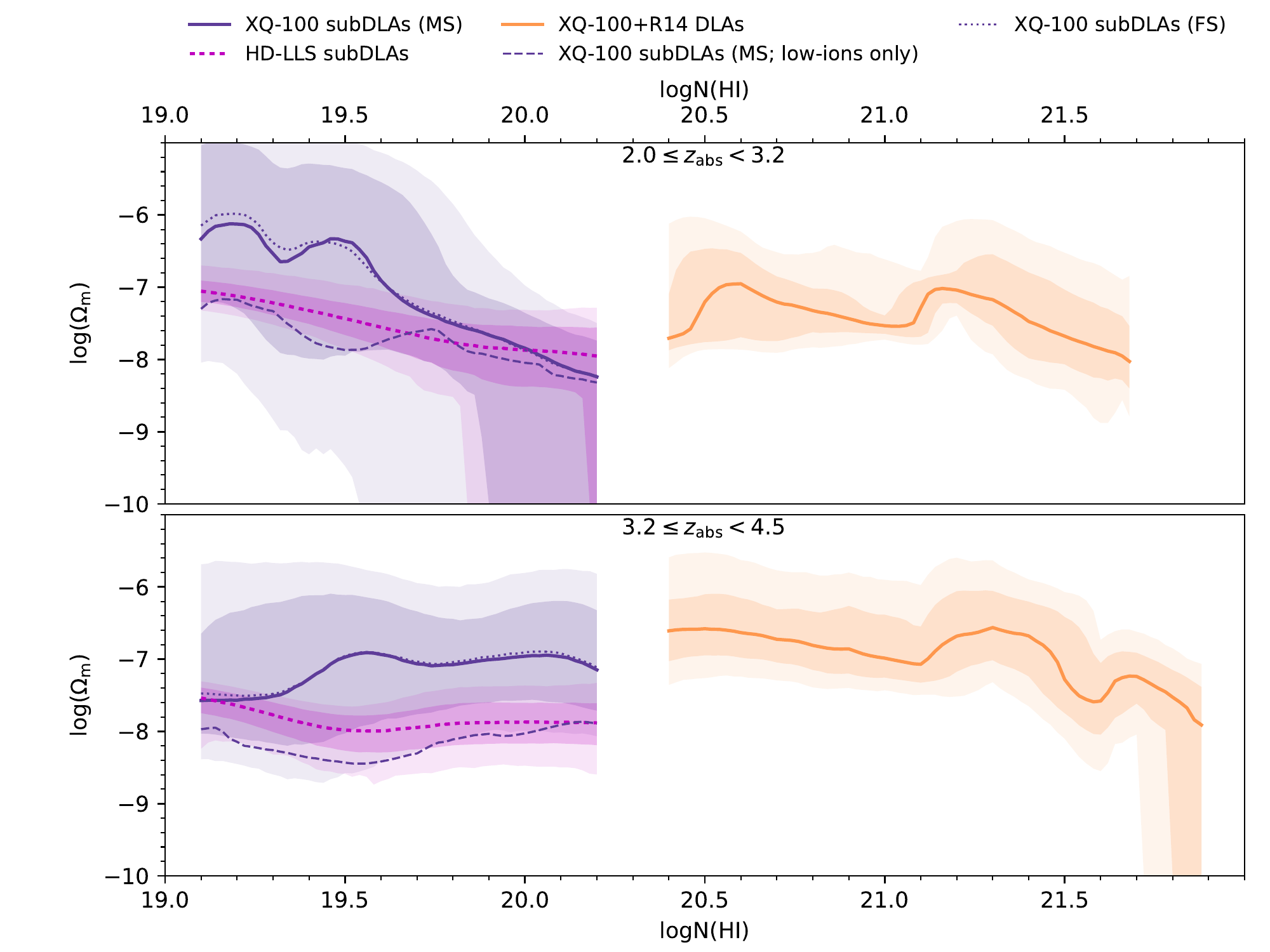}
\caption{THe mass-density of metals (\omegaM{}) as a function of logN(\HI{}) for both subDLAs (magenta and purple curves) and DLAs (orange curves). The purple lines represent different subsamples of the XQ-100 subDLAs, while the curves for the HD-LLS are shown by the dotted magenta lines. The darker and lighter shaded regions represent the $1\sigma$ and $2\sigma$ percentile confidence intervals on the respective curves. The top and bottom panels show the evolution for two different redshift bins ($2.3 \leq$~\zabs{}~$<3.2$ and $3.2 \leq$~\zabs{}~$<4.5$; respectively). All curves in both the top and lower panels are consistent within $1\sigma$. However, the increased contribution to \omegaM{} from logN(\HI{})$\lesssim19.5$ absorbers at lower redshifts (top panel) is suggestive that low density environments are being enriched with metals faster than higher density regions.}
\label{fig:omegaM}
\end{center}
\end{figure*}

To better quantify the contribution of metals from absorbers of different logN(\HI{}) and redshift, Figure \ref{fig:omegaM} shows \omegaM{} as a function of logN(\HI{}) across the two redshift bins ($2.3 \leq$~\zabs{}~$<3.2$, and $3.2 \leq$~\zabs{}~$<4.5$). The \omegaM{} curves are generated using a sliding bin of width $\Delta$logN(\HI{})~$=$~0.2 dex and a sliding step size of 0.02 dex. Although the respective curves between the top and bottom panels are consistent with the $1\sigma$ distributions, we note that all the \omegaM{} curves for the subDLAs (XQ-100 and HD-LLS samples; purple and magenta lines, respectively) show a steeper dependence on logN(\HI{}) in the low redshift bin compared to the higher redshift curves, in contrast to the DLA \omegaM{} curves which remain unchanged.

While the absolute contribution to \omegaM{} is dependent on the implementation of the ionization modelling (such as assuming a single versus multiphase medium; solid and dashed purple lines of Figure \ref{fig:omegaM}, respectively), absorbers with logN(\HI{})$\lesssim19.5$ appear to contribute $\approx0.4-0.8$ dex more metals at lower redshifts while higher column density systems show a small decrease in \omegaM{} (see also the total values in Table \ref{tab:omegaM}). This tentative redshift evolution is likely independent of absorber selection effects, as the \omegaM{} curves for the FS and MS subDLAs (dotted and solid purple lines, respectively) are nearly identical. If such a redshift evolution exists, it would imply feedback processes are efficient at rapidly ejecting metals from higher density environments (such as the ISM) into lower density environments (CGM). However, larger statistical samples of subDLAs in both redshift regimes are needed to truly constrain this result.

\subsection{[$\alpha$/Fe] in subDLAs}
\label{sec:aFe}

\afe{} is a commonly used diagnostic to probe the star formation history within a galaxy \citep{Tinsley79}, tracing the relative nucleosynthetic contribution of massive stars (the primary source of $\alpha$ elements; such as O, Si, S) to Type Ia supernovae (the main source of the Fe-peak elements).  As the metallicity of the system increases, \afe{} remains constant at the Type II supernovae yield (\afe{}~$\approx+0.3$ dex) until the time-delayed contribution of Fe from Type Ia supernovae. The metallicity at which \afe{} begins to include Type Ia contributions (typically referred to as the \afe{} `knee') provides information on how many stars were formed in the past, and can be used as a probe of relative galaxy mass \citep[e.g.~dwarfs vs Milky Way-like galaxies;][]{Tolstoy09}.

\begin{figure*}
\begin{center}
\includegraphics[scale=0.95]{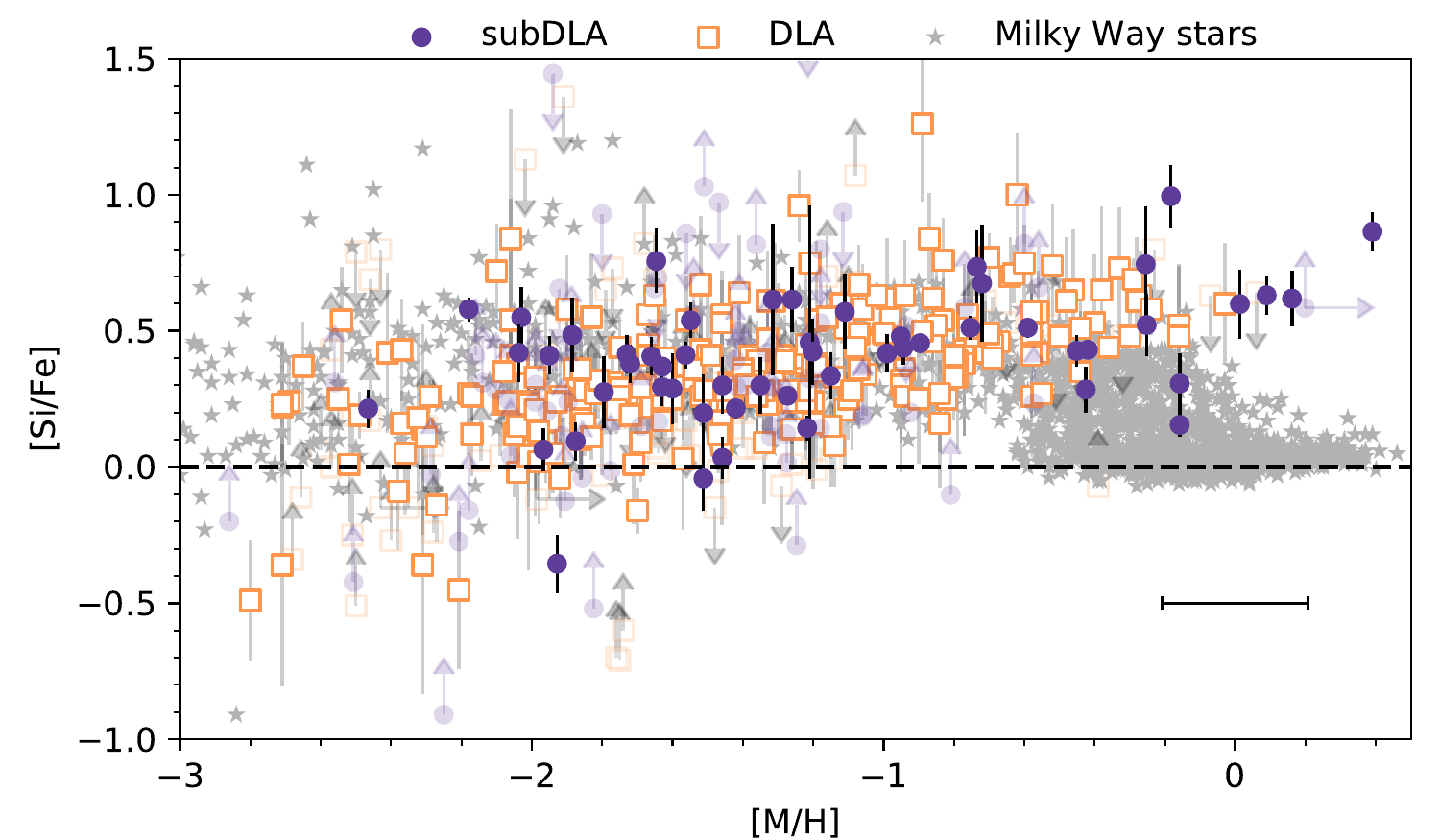}
\caption{The metallicity evolution of gas-phase [Si/Fe] for subDLAs (purple circles, this work), DLAs \citep[orange squares][]{Berg15}, and Milky Way stars \citep[grey stars][]{Berg15}. The increasing trend of [Si/Fe] with metallicity in DLAs is typically explained by the increasing contribution of dust depletion of Fe with metallicity. The ionization corrections for [Si/Fe] are typically small in magnitude ($\approx0.02$ dex) and are not included in the plot. Errors in [Si/Fe] are denoted by the vertical lines, and include the error in the ionization correction. The horizontal error bar in the bottom right shows the median error in metallicity for the subDLAs. Despite the inclusion of ionization corrections, subDLAs appear to follow the same trend as the DLAs and potentially indicating that the two populations have similar dust depletion patterns.}
\label{fig:AFe}
\end{center}
\end{figure*}

Figure \ref{fig:AFe} displays the evolution of the gas-phase [Si/Fe] for subDLAs (both XQ-100 and HD-LLS absorbers; purple circles) compared to a literature compilation of DLAs and Milky Way stars \citep[][orange squares and grey stars, respectively]{Berg15}. Ionization corrections are not included on the subDLAs and DLAs as the observed \afe{} is typically only overestimated by at most $\approx0.02$ dex based on our modelled ionization corrections (see Table \ref{tab:Naodm}). It is well known that the DLA's [Si/Fe] pattern deviates from the expected Milky Way \afe{} pattern, as the increased relative depletion of Fe (with respect to Si) onto dust will artificially enhance the measured [Si/Fe] as a function of metallicity \citep[e.g.][]{Rafelski12,Berg15}. As previously noted by \cite{Quiret16}, the subDLAs in Figure \ref{fig:AFe} appear to follow the same trend of [Si/Fe] as DLAs. The near-identical trends of DLAs and subDLA suggests the relative dust depletion pattern of Si to Fe is similar (to a precision within the scatter of \afe{}, i.e. $\approx\pm0.25$ dex), as both subDLAs and DLAs at [M/H]~$\gtrsim-0.5$ show [Si/Fe]~$\approx +0.5$ when \afe{} is (sub)solar in both the Milky Way and dwarf galaxies. Using dust depletion modelling from \cite{Jenkins09}, \cite{Quiret16} concluded that subDLAs show similar dust properties to DLAs, regardless of ionization effects \citep[contrary to expectations from previous studies, e.g.][]{Meiring09,Prochaska15}; agreeing with the implications of Figure \ref{fig:AFe}.

In Figure \ref{fig:AFe}, the scatter in \afe{} ($\approx\pm0.25$ dex) for both DLAs and subDLAs is on the order of the expected nucleosynthetic signature of the \afe{} enhancement (0.3 dex). Combined with the increasing \afe{} with metallicity with dust depletion, this washes out any potential underlying nucleosynthetic signature. Thus, \afe{} in subDLAs and DLAs cannot be used without an accurate dust correction to determine whether individual absorbers have surpassed the expected \afe{} knee. In attempts to correct for the depletion of Fe in DLAs, previous studies have used Zn as a proxy for Fe as Zn is volatile and the Milky Way shows a solar [Zn/Fe] for [M/H]~$\gtrsim-2$ \citep{Sneden88,Pettini97,Vladilo01,Nissen07}. As such, [$\alpha$/Zn] is typically measured to be (sub)solar across the full range of metallicities probed by DLAs \citep{Centurion00,Prochaska02II,Nissen04,DZavadsky06,Rafelski12,Berg15}. Recent studies of dwarf galaxies have shown that [Zn/Fe] is typically subsolar, and can be as low as $-1.0$ dex \citep{Shetrone03, Sbordone07,Berg15,Skuladottir17}, negating the use of [Zn/Fe] as a dust depletion metric in DLAs as DLAs likely probe a mix of both dwarf and Milky Way enrichment histories.  However, \cite{Berg15} argued that independent of assuming an intrinsically solar [Zn/Fe] for DLAs, the significant fraction of DLAs with low [$\alpha$/Zn] \citep{Rafelski12,Berg15} imply DLAs have undergone Type Ia supernovae enrichment. Given that Zn is not detected in the XQ-100 subDLAs, we cannot apply the same approach for the subDLAs. Higher signal-to-noise ratio spectra are required to study Zn and the nature of [$\alpha$/Fe] in subDLAs.

\subsection{[C/O] in subDLAs}
\label{sec:CO}
\begin{figure*}
\begin{center}
\includegraphics[scale=0.95]{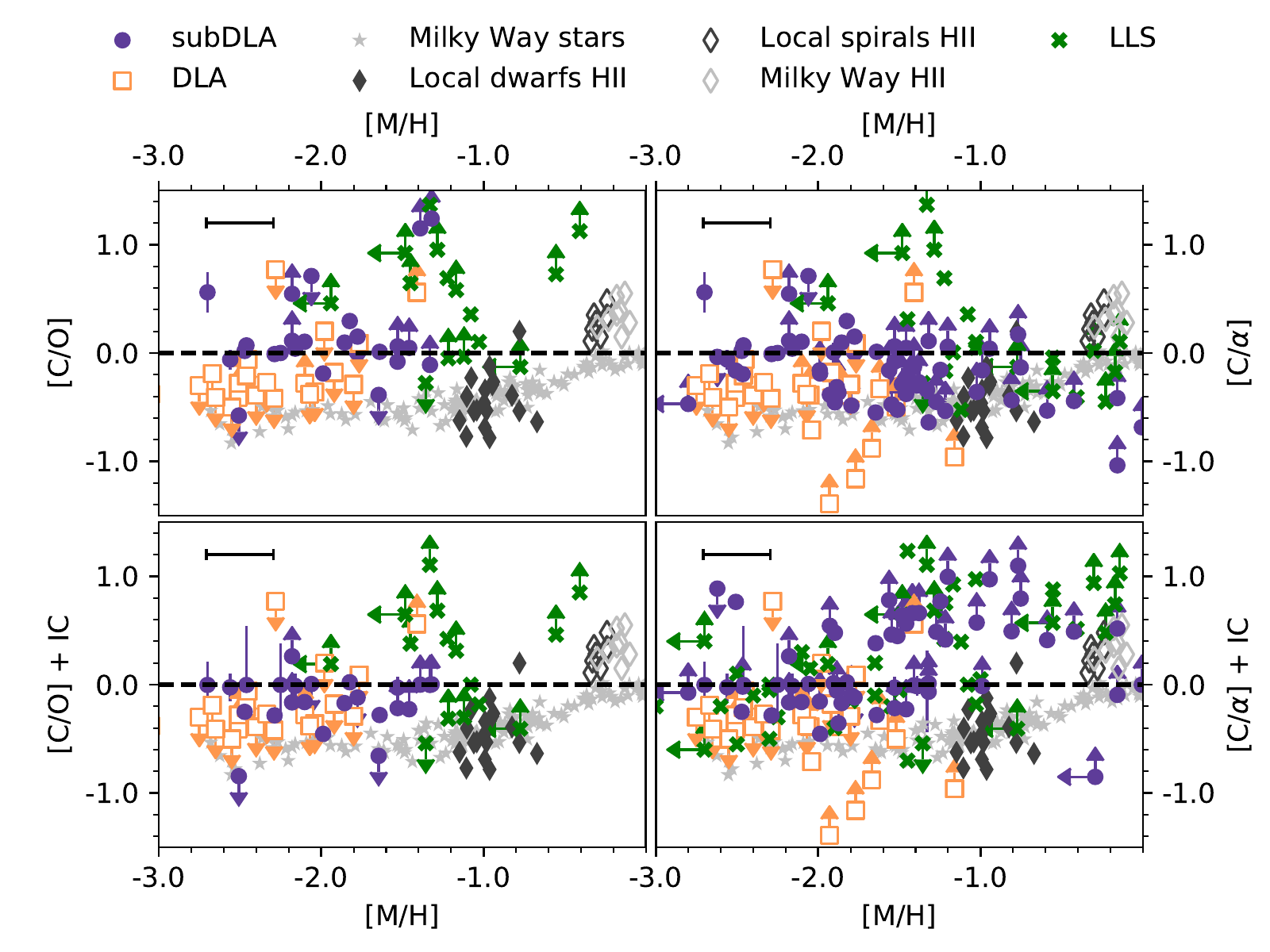}
\caption{[C/O] (left panels) and [C/$\alpha$] ratio (right panels) as a function of metallicity in subDLAs (purple circles), DLAs \citep[orange squares;][]{Cooke11,Rafelski14,Berg15, Cooke17}, LLSs \citep[green crosses][]{Prochaska15,Lehner16}, Milky Way stars \citep[light grey stars;][using the 3D non local thermodynamic equilibrium values]{Amarsi19} and H\ion{ii} regions (unfilled light grey diamonds), and H\ion{ii} regions in nearby spiral (unfilled dark grey diamonds) and dwarf galaxies (filled dark grey diamonds). The top panels show the measured values of the respective abundance ratios, while the lower panels have applied an ionization correction to the DLAs, subDLAs and LLSs. Errors in [C/$\alpha$] and [C/O] are typically smaller than the symbol size. The horizontal error bar in the top right of each panel shows the median error in metallicity for the subDLAs. Despite including the ionization correction; the bulk of subDLAs and LLS  above a metallicity of [M/H]~$\gtrsim-2$ appear to have near-solar [C/O] and [C/$\alpha$] at metallicities $-3\leq {\rm [M/H]} \leq -1$ whilst the measured abundance ratio in stars drops to a [C/O]$\approx-1$.}
\label{fig:CO}
\end{center}
\end{figure*}

[C/O] is a useful tracer of the nucleosynthetic contribution of high to intermediate mass stars, as [C/O] is expected to increase with increasing metallicities as the intermediate mass stars begin contributing C to the chemical evolution of the system \citep{DBerg16}. In low metallicity Milky Way stars and dwarfs galaxies ([M/H]~$\leq-1.5$), observations suggests [C/O] remains at a subsolar value. At metallicities [M/H]~$\lesssim-2.5$, there has been an observed increase in [C/O] with decreasing metallicity, that is thought to originate from massive zero-metallicity supernovae contributing to the production of carbon \citep{Akerman04,Spite05,DBerg16}, but this rise in [C/O] may also be a result of not including 3D or non-local thermodynamic equilibrium assumptions in stellar abundance modelling \citep[][see grey stars in Figure \ref{fig:CO}]{Nissen17, Amarsi19}. By varying the duration and efficiency of the star formation of dwarf galaxies, \cite{DBerg19} demonstrated that the rate of increase in [C/O] with metallicity could be explained by star formation burst cycles, with repeated nucleosynthetic contributions from massive stars (diluting [C/O]) followed by a delayed contribution from AGB winds (enhancing [C/O]). Such cycling in the star formation history could explain the return to solar [C/O] much more quickly in dwarf galaxies relative to the Milky Way. In DLAs, measuring the [C/O] ratio is only well constrained in the lowest metallicity systems, as the strong O\ion{i} 1302 \AA{} and C\ion{ii} 1334 \AA{} lines quickly saturate at [M/H]$\gtrsim-2$. For the most metal-poor DLAs, [C/O]~$\approx-0.4$ dex, and appears to agree with metal-poor Milky Way stars \citep{Pettini08,Cooke11,Cooke17}. Using Si in place of O as an $\alpha$-element tracer, previous work has demonstrated that [C/$\alpha$]$\approx0$ in LLSs at all metallicities, suggesting LLSs trace material that has been polluted by ejecta from galaxies \citep{Prochaska15,Lehner16}, and is consistent with the idea that LLSs gas is material associated with circumgalactic environments \citep{Fumagalli13, Fumagalli16}.

The top panels of Figure \ref{fig:CO} compare the metallicity evolution of the gas-phase [C/O] (left panels) and [C/$\alpha$] (right panels; using either O or Si as the $\alpha$-element tracer) in subDLAs (XQ-100 and HD-LLS subDLAs; purple circles), DLAs \citep[orange squares;][]{Cooke11,Berg15, Cooke17}, and LLSs \citep[green crosses;][]{Prochaska15,Lehner16}. [C/O] for Milky Way stars \citep[light grey stars;][which include 3D and non-local thermodynamic equilibrium effects]{Amarsi19}, as well as in H\ion{ii} regions in the Milky Way \citep[][light grey open diamonds]{GarciaRojas07}, neaby spirals \citep[][dark grey open diamons]{Esteban09}, and nearby dwarf galaxies \citep[][dark grey filled diamonds]{DBerg19} are shown in all four panels. The bottom panels show the effects of including ionization-corrections on the measured [C/O] and [C/$\alpha$] (for DLAs, subDLAs and LLSs). Note that we elected to continue to plot the ionization-corrected [C/O] and [C/$\alpha$] as a function of [M/H] in the bottom panel to emphasize the effect of ionization corrections. Using \ZIC{} in the bottom panels would shift the subDLAs points to $\approx0.3$ dex lower metallicities (i.e. the typical offets seen in Figure \ref{fig:Zcompare}.).

Independent of whether ionization corrections are included, subDLAs consistently exhibit [C/O]$\gtrsim-0.4$, similar to the typical [C/O] in metal-poor DLAs and $\gtrsim0.1$ dex higher than the Milky Way abundances of \cite{Amarsi19}. However, for the bulk of the subDLAs and DLAs, both the C and O lines become saturated with increasing metallicity, making [C/O] difficult to measure. In place of [C/O], the more commonly-measured [C/$\alpha$] (right panels of Figure \ref{fig:CO}) shows that subDLAs are consistent with LLS measurements \citep{Prochaska15, Lehner16}. Again, regardless of including ionization corrections, the bulk of subDLAs and LLS [C/$\alpha$] measurements are consistently above the Milky Way [C/O] trend by at least 0.1 dex (without ionization corrections) or 0.7 dex (with ionization corrections). Such large discrepancies in [C/$\alpha$] could be potentially be explained by dust depletion, as Si is more refractory than C. As discussed in Sections \ref{sec:meanZ} and \ref{sec:aFe}, we currently do not have sufficient detections of elements to constrain the amount of dust depletion in individual subDLAs, but it appears that subDLAs follow similar amounts of depletion as DLAs. Assuming a typical depletion pattern from DLAs would only decrease [C/$\alpha$] by 0.1--0.3 dex \citep{Vladilo11,Berg15,deCia18}, which is insufficient to reach the typical Milky Way values upon including ionization corrections. Given the results of \cite{DBerg19} and the assumption that DLAs and subDLAs have similar relative depletion patterns as DLAs, the bulk of LLS \citep[as suggested by][]{Lehner16} and subDLAs are consistent with being enriched by material ejected from several burst cycles in the star formation of galaxies. 

\section{Conclusion}
This paper presents the ionic column density and abundance measurements  in the gas-phase for 20 ionic species (Table \ref{tab:ionSum}; $\approx4$ ionic species are detected on average per subDLA system) in a blind, \HI{}-selected sample of subDLAs from the XQ-100 survey.  Using the measured column densities for all ionic species as constraints (mainly C\ion{iv}, Si\ion{ii}, Mg\ion{ii}, Si\ion{iv}, Al\ion{ii}, Fe\ion{ii}, C\ion{ii}, and O\ion{i}; in order of decreasing detection frequency), we estimate the ionization corrected gas-phase metallicity (\ZIC{}) of the subDLAs derived from Markov Chain Monte Carlo analysis based on a large grid of \textsc{Cloudy} ionization models. No dust corrections have been included in the \ZIC{} measurements. We expect similar statistical offsets in \ZIC{} from including a \cite{Jenkins09} dust depletion model in the Markov Chain Monte Carlo analysis found by \cite{Fumagalli16} (median \ZIC{} offset of $0.00^{+0.10}_{-0.06}$; Section \ref{sec:dustAssump}).  Comparing with gas-phase metal abundances of other \HI{}-selected subDLAs \citep[i.e. XQ-100 and HD-LLS;][]{Prochaska15}, we investigated the redshift evolution of the \HI{}-weighted mean gas-phase metallicity (\meanZ{}; Figure \ref{fig:zZ}), the redshift evolution of the cosmic mass density of metals (\omegaM{}; Figure \ref{fig:omegaM}), and the metallicity evolution of \afe{} (Figure \ref{fig:AFe}) and [C/O] (Figure \ref{fig:CO}) probed by both subDLAs and DLAs. We find that: (\emph{i}) subDLAs appear to be systematically more metal poor by 0.2--1.0 dex than DLAs when ionization corrections are included across redshifts $3.0\leq$~\zabs{}~$\leq4.3$ in contrast to \zabs{}~$\lesssim2$ studies \citep{Peroux07, DZavadsky09,Som13,Quiret16}, (\emph{ii}) the observed metallicity evolution of gas-phase \afe{} in subDLAs traces that of DLAs almost identically (within $\approx\pm0.25$ dex of scatter), and (\emph{iii}) the gas-phase [C/O] in subDLAs remains constantly higher than the Milky Way value by, on average, $\gtrsim0.7$ dex across a large range in metallicities ($-3\lesssim$~[M/H]~$\lesssim-1$). These results are robust to ionization corrections derived from Markov Chain Monte Carlo modelling using several different \textsc{Cloudy} ionization grids.  The near-identical trend of \afe{} in subDLAs and DLAs is suggestive that the relative dust depletion patterns of DLAs and subDLAs is similar. However, the total amount of the dust depletion for each ion in subDLAs may be different. As seen in lower redshifts systems, the relatively low metallicity gas in subDLAs with respect to DLAs, and the near-solar [C/O] is suggestive of subDLAs typically tracing CGM environments that host carbon-rich ejecta \citep{Fumagalli16,Lehner16, Prochaska17,Kacprzak19}. The tentative redshift evolution in the contribution to \omegaM{} from logN(\HI{})~$\lesssim19.5$ absorbers supports the idea that the CGM is being enriched with metals more quickly than higher column density environments typically associated with the ISM. Given that ionization corrections are critical for accurate measurements in subDLAs, more ionization modelling in future is necessary to accurately model the metallicity of multiphase gas seen along QSO sightlines. Accurate chemical abundances measurements are required to constrain upcoming cosmological simulations of the gaseous environments of galaxies probed by QSO absorption line systems \citep{Peeples19,Vandevoort19}.

\section*{Acknowledgments}
We are grateful to Ryan Cooke for providing unpublished abundance measurements to compute the ionization corrections for the metal-poor DLAs. This project has received funding from the European Research Council (ERC) under the European Union's Horizon 2020 research and innovation programme (grant agreement No 757535). This work has been supported by Fondazione Cariplo, grant No 2018-2329. SL was funded by FONDECYT grant number 1191232.
\bibliography{bibref}

\appendix
\section{Supplementary material}
\label{sec:App}

\begin{figure*}
\begin{center}
\begin{subfigure}{\textwidth}
\includegraphics[width=0.95\textwidth]{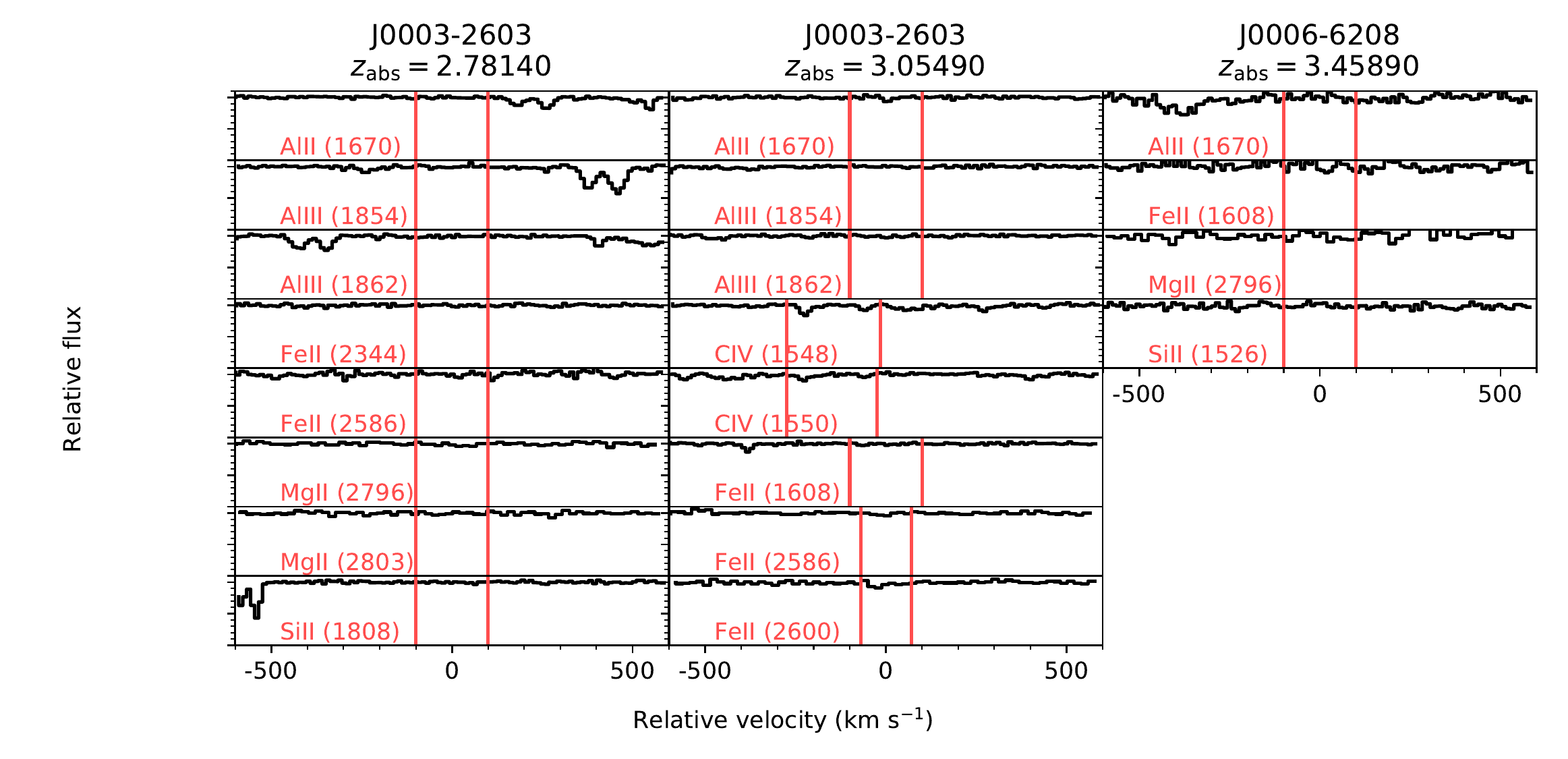}
\end{subfigure}
\caption[]{The metal line profiles for species in the XQ-100 subDLA sample. Only metal species with at least one detection are shown. Each column shows all the metal line profiles, with measured column densities, for each subDLA. Every panel has the same y-axis scaling, with large ticks at a relative flux of 0, 0.5, and 1 (labels not shown for readability) and smaller ticks for every 0.1 increment. The vertical red lines in each panel show the velocity bounds used for the respective column density computation. }
\label{fig:velprof}
\end{center}
\end{figure*}

\begin{figure*}
\ContinuedFloat
\begin{center}
\begin{subfigure}{\textwidth}
\includegraphics[width=0.95\textwidth]{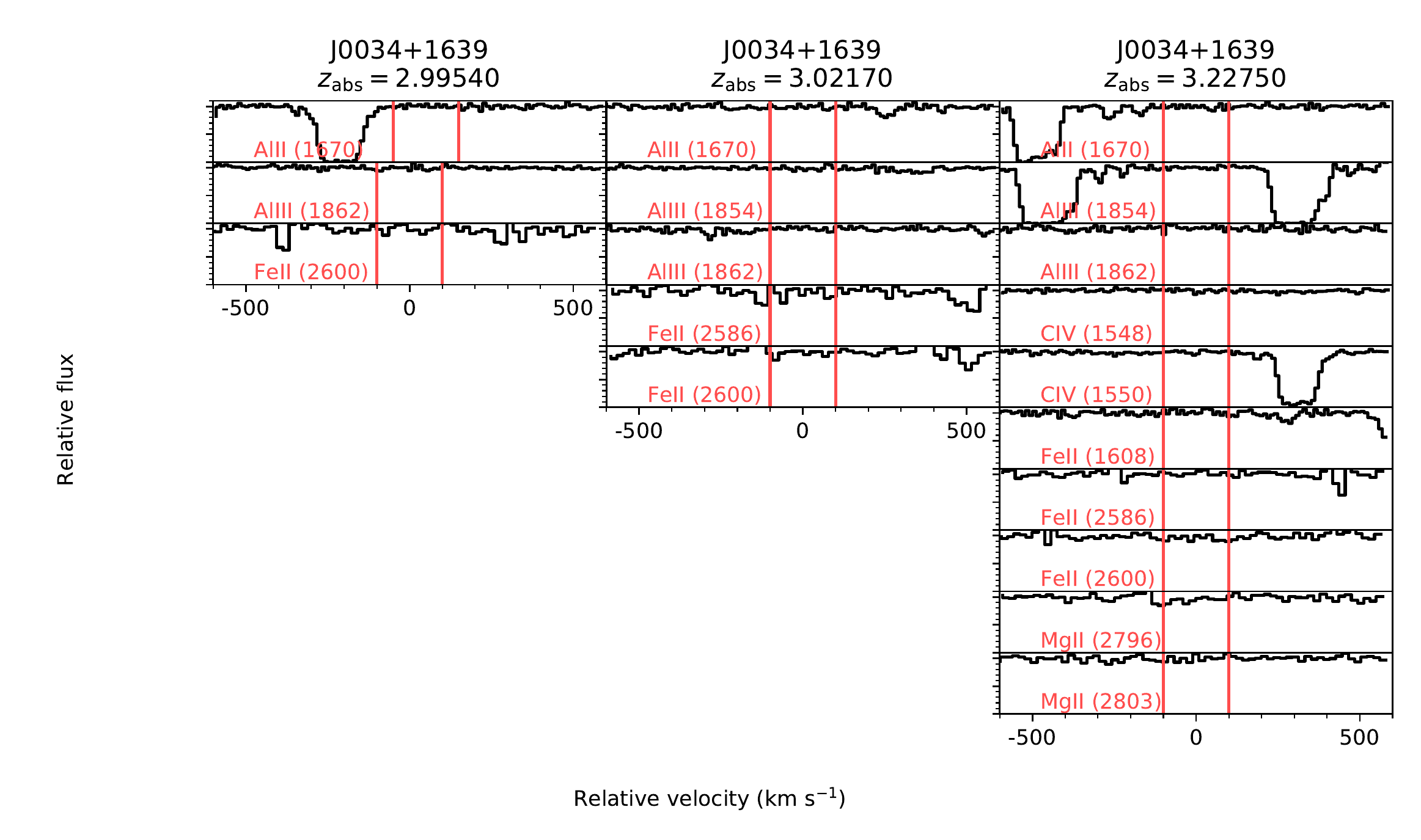}
\end{subfigure}
\caption{(cont'd)}
\end{center}
\end{figure*}

\begin{figure*}
\ContinuedFloat
\begin{center}
\begin{subfigure}{\textwidth}
\includegraphics[width=0.95\textwidth]{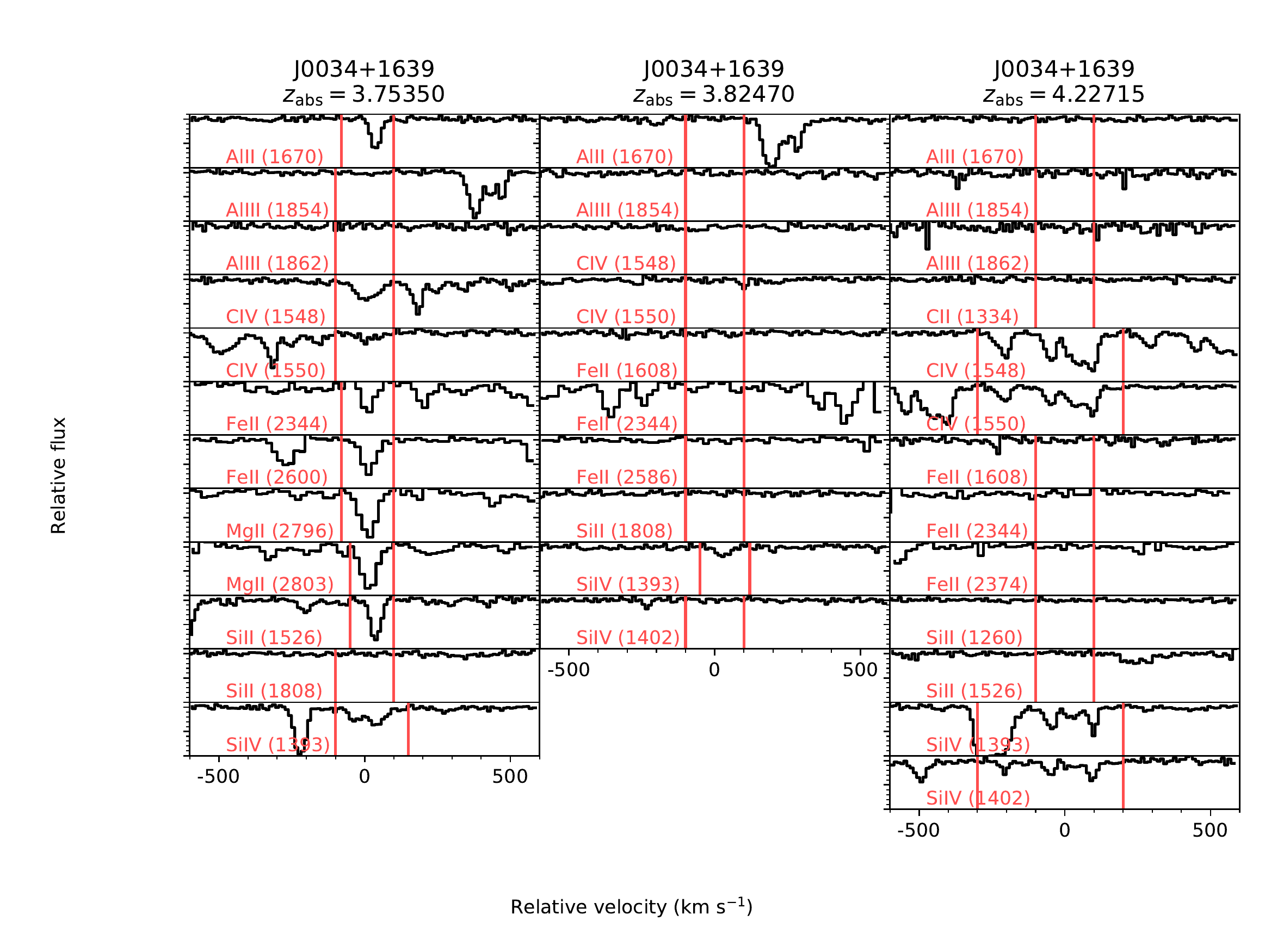}
\end{subfigure}
\caption{(cont'd)}
\end{center}
\end{figure*}

\begin{figure*}
\ContinuedFloat
\begin{center}
\begin{subfigure}{\textwidth}
\includegraphics[width=0.95\textwidth]{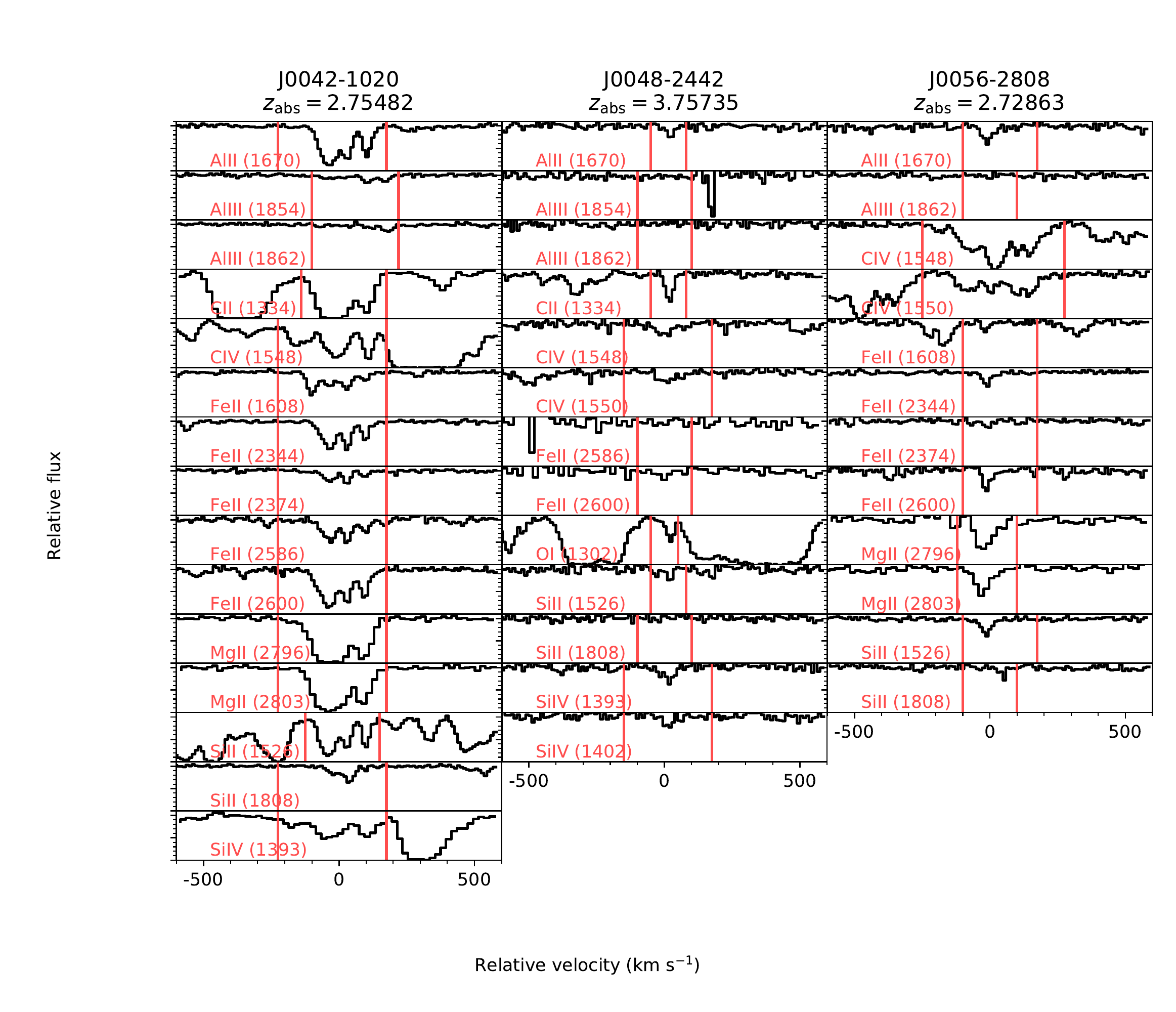}
\end{subfigure}
\caption{(cont'd)}
\end{center}
\end{figure*}

\begin{figure*}
\ContinuedFloat
\begin{center}
\begin{subfigure}{\textwidth}
\includegraphics[width=0.95\textwidth]{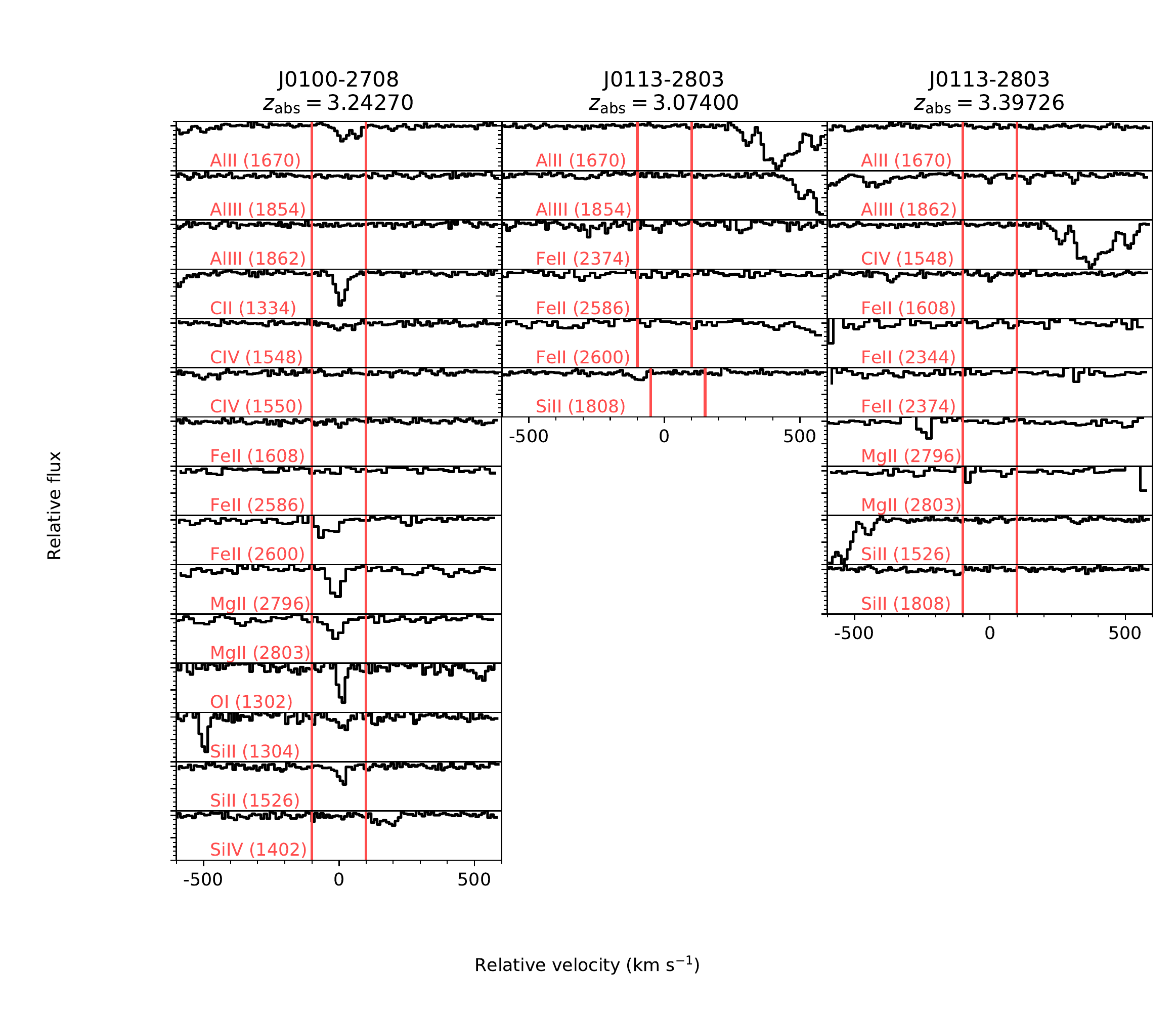}
\end{subfigure}
\caption{(cont'd)}
\end{center}
\end{figure*}

\begin{figure*}
\ContinuedFloat
\begin{center}
\begin{subfigure}{\textwidth}
\includegraphics[width=0.95\textwidth]{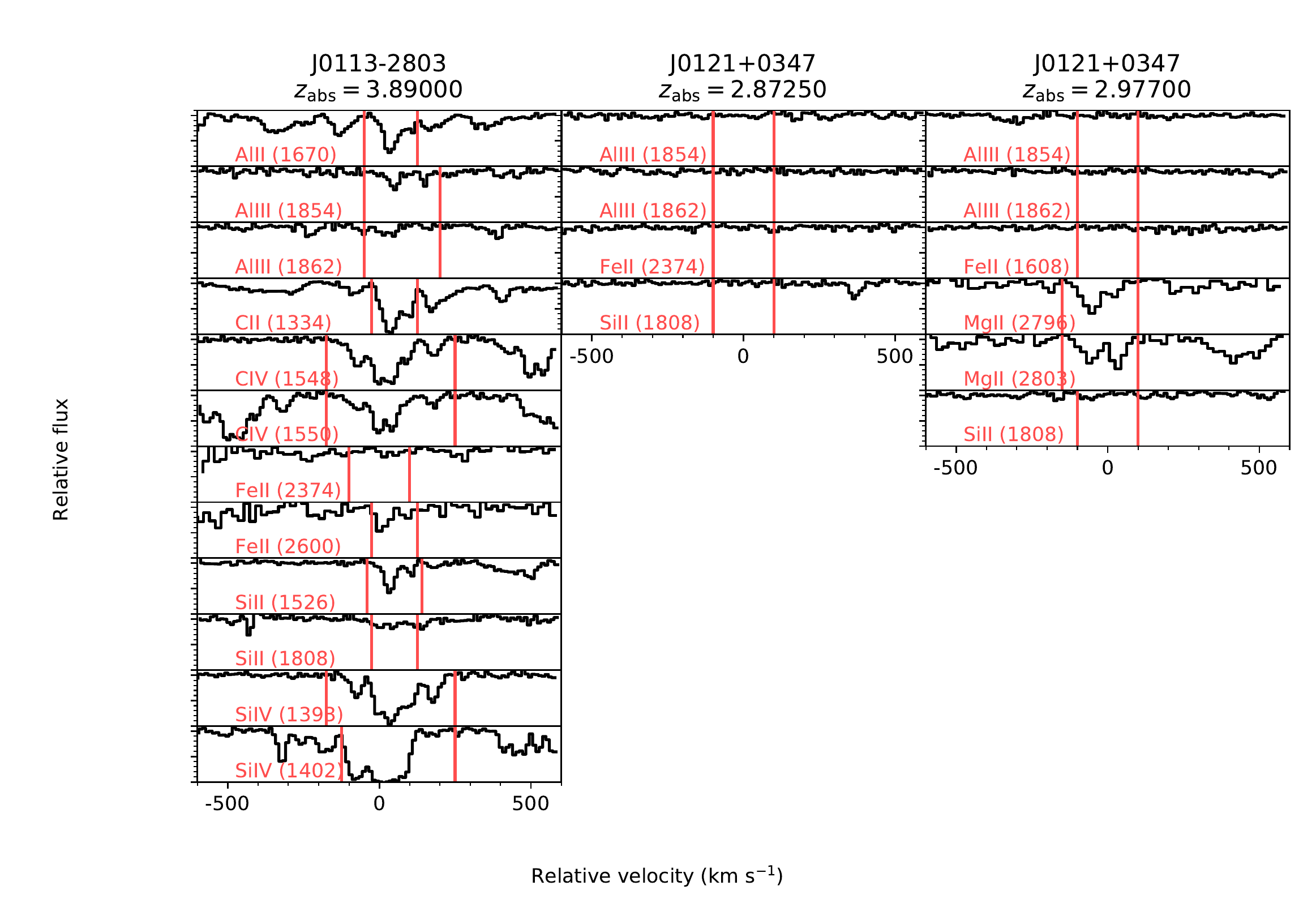}
\end{subfigure}
\caption{(cont'd)}
\end{center}
\end{figure*}

\begin{figure*}
\ContinuedFloat
\begin{center}
\begin{subfigure}{\textwidth}
\includegraphics[width=0.95\textwidth]{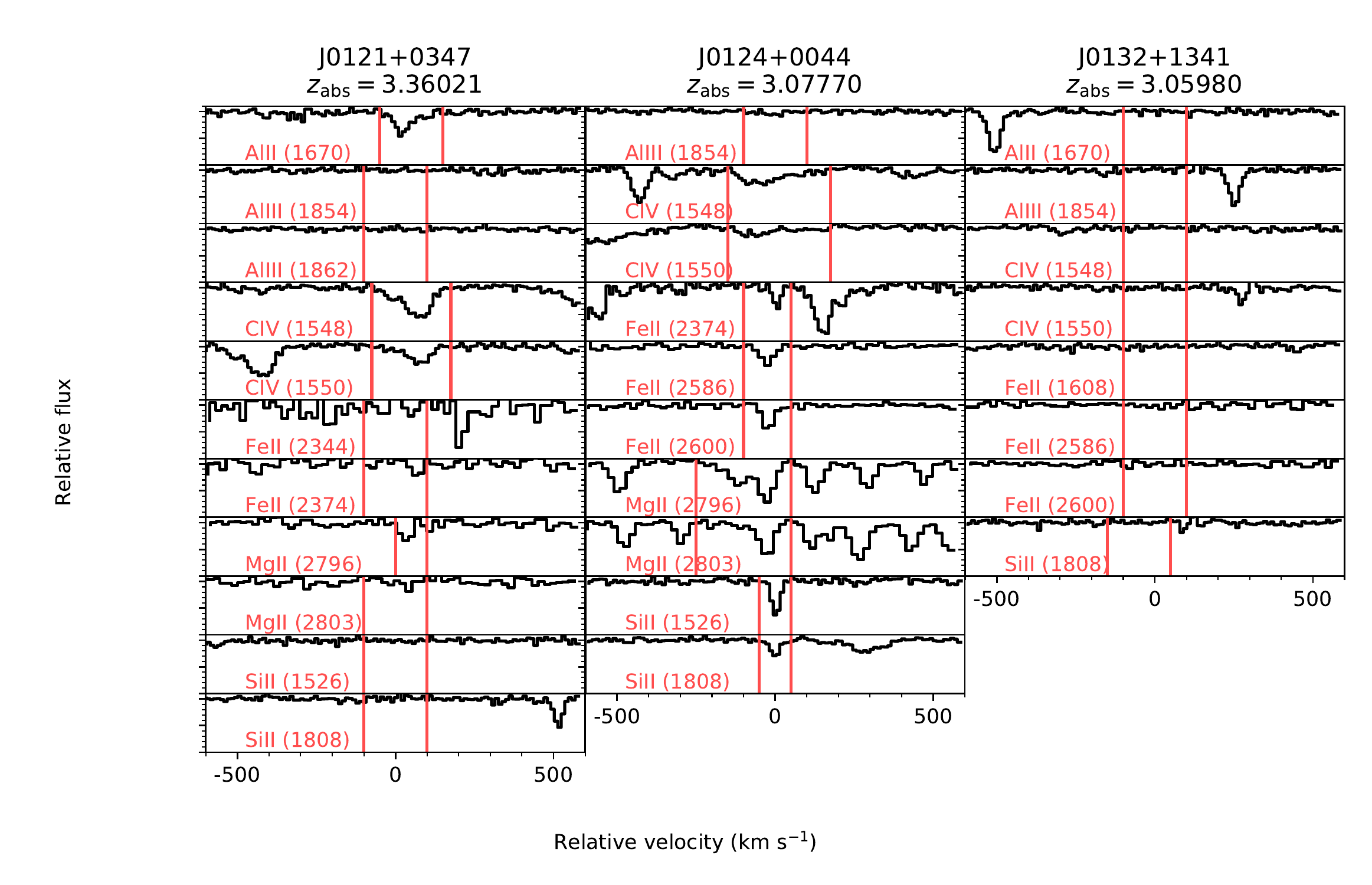}
\end{subfigure}
\caption{(cont'd)}
\end{center}
\end{figure*}

\begin{figure*}
\ContinuedFloat
\begin{center}
\begin{subfigure}{\textwidth}
\includegraphics[width=0.95\textwidth]{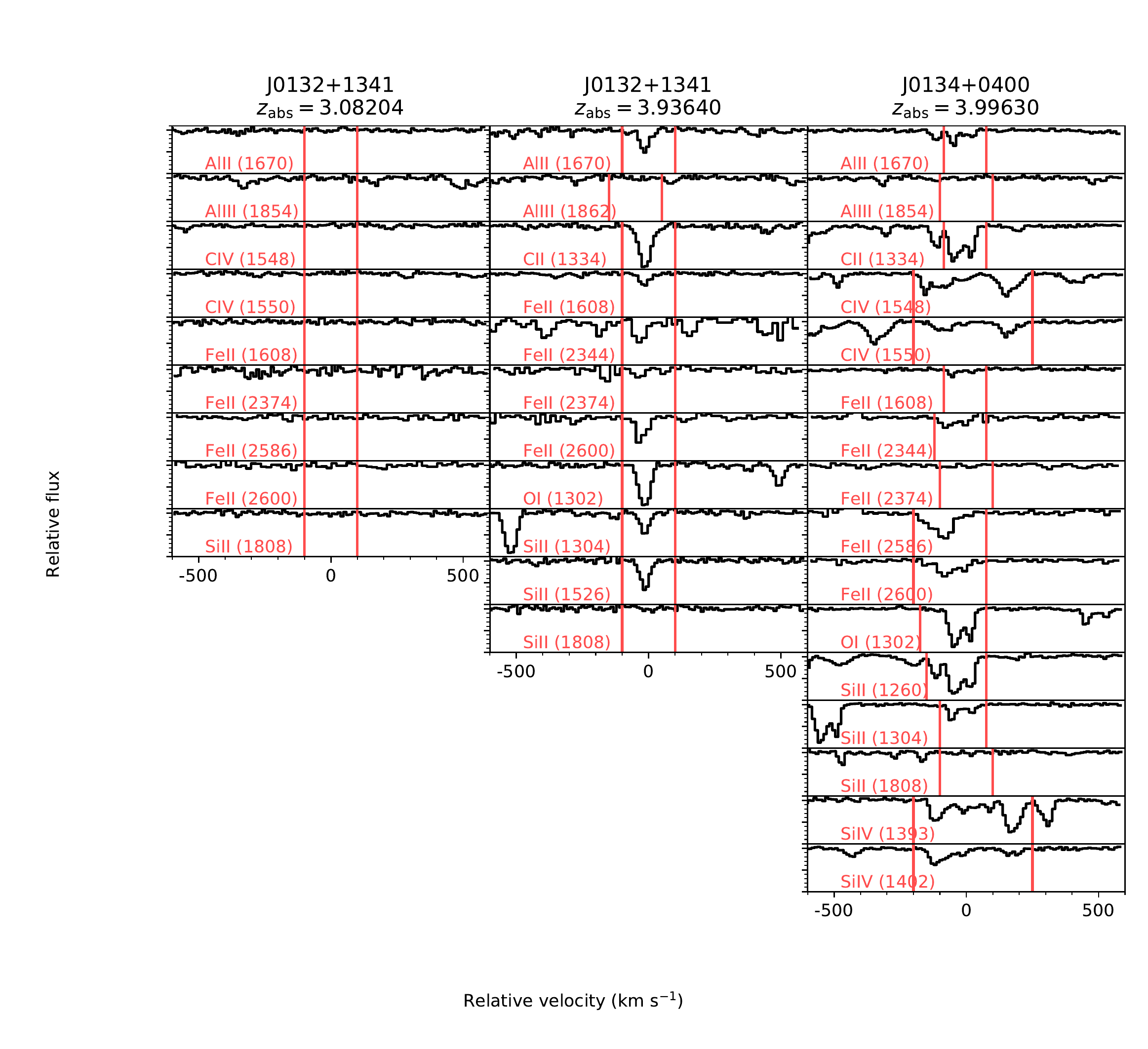}
\end{subfigure}
\caption{(cont'd)}
\end{center}
\end{figure*}

\begin{figure*}
\ContinuedFloat
\begin{center}
\begin{subfigure}{\textwidth}
\includegraphics[width=0.95\textwidth]{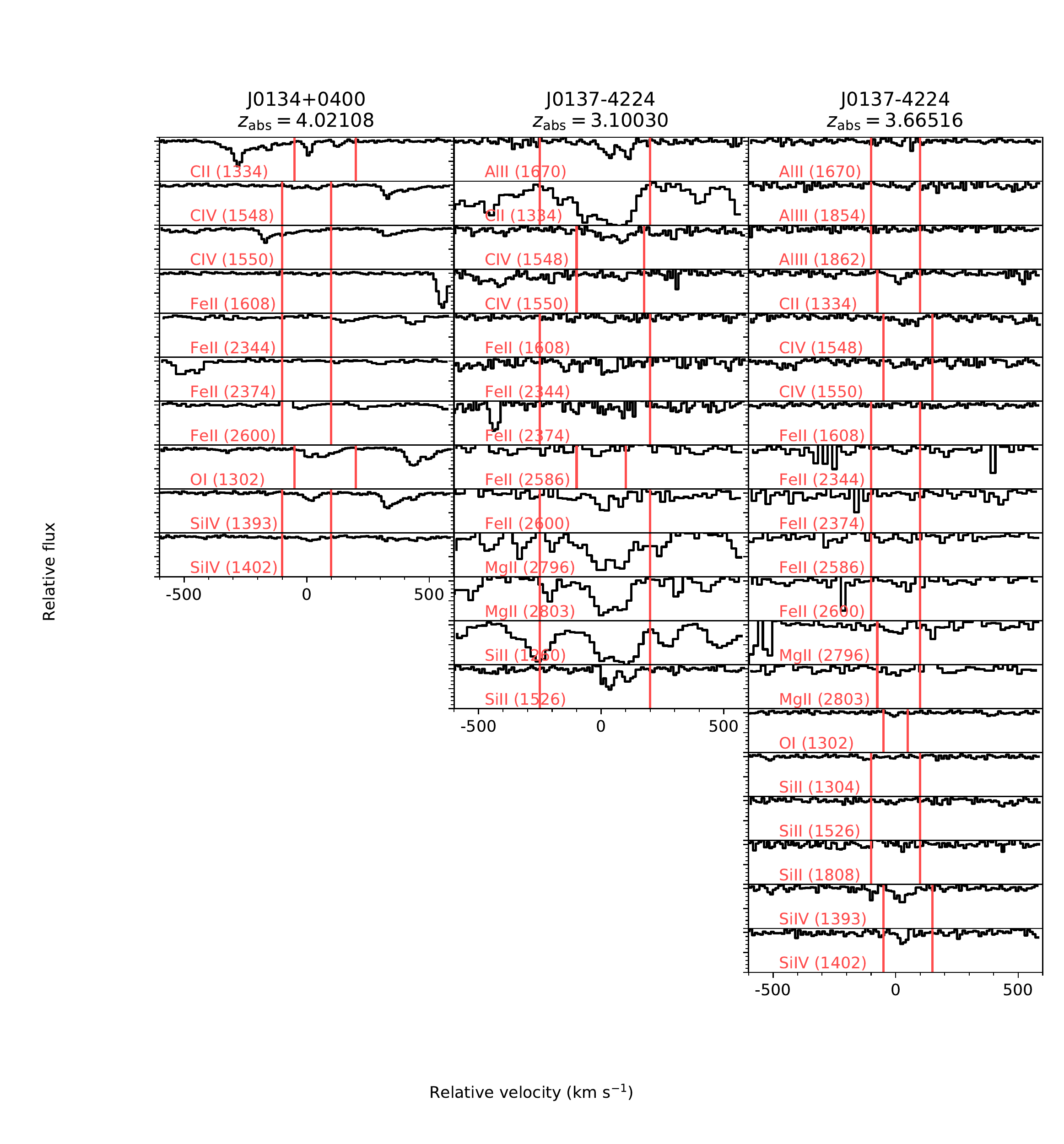}
\end{subfigure}
\caption{(cont'd)}
\end{center}
\end{figure*}

\begin{figure*}
\ContinuedFloat
\begin{center}
\begin{subfigure}{\textwidth}
\includegraphics[width=0.95\textwidth]{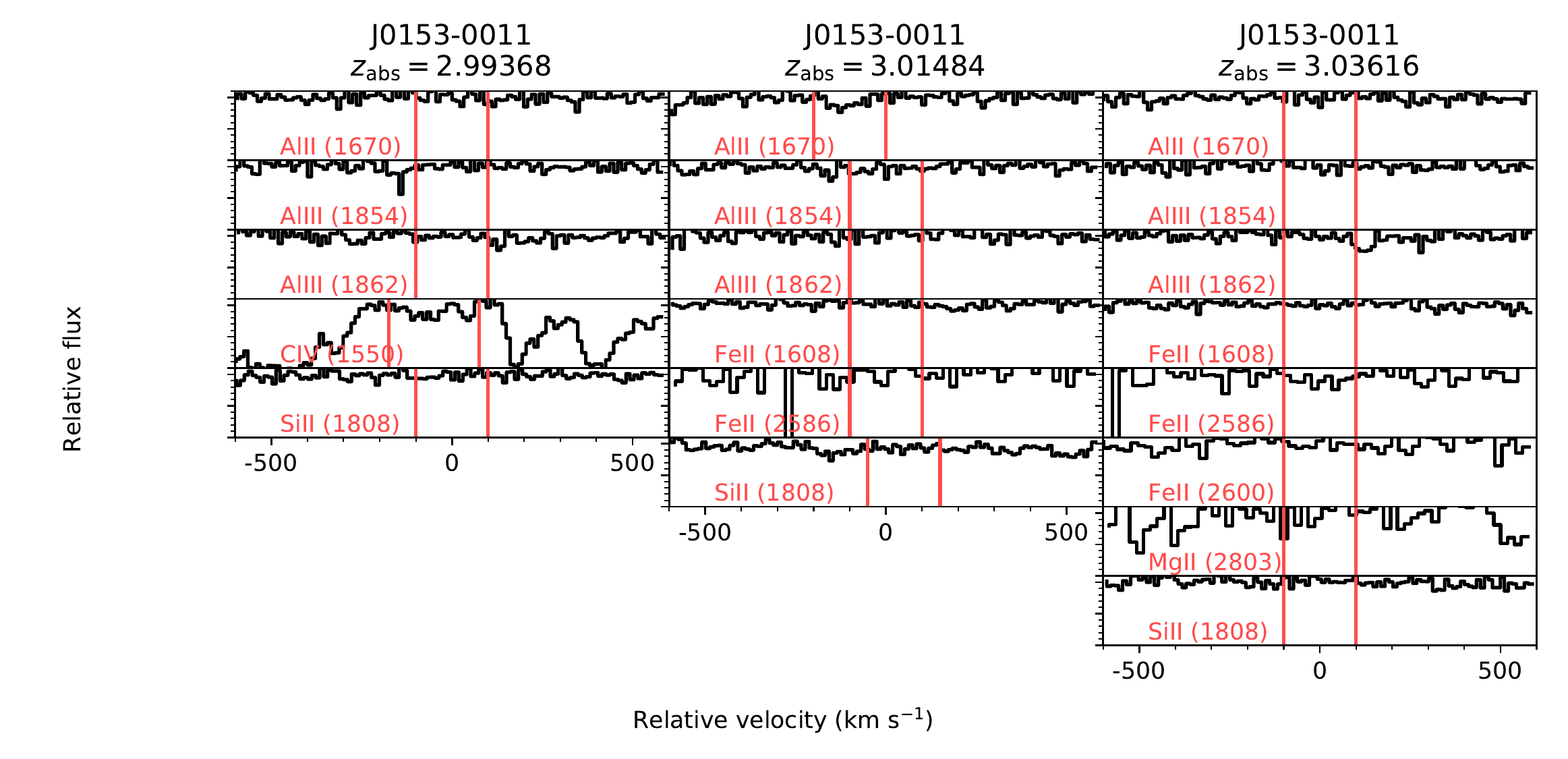}
\end{subfigure}
\caption{(cont'd)}
\end{center}
\end{figure*}

\begin{figure*}
\ContinuedFloat
\begin{center}
\begin{subfigure}{\textwidth}
\includegraphics[width=0.95\textwidth]{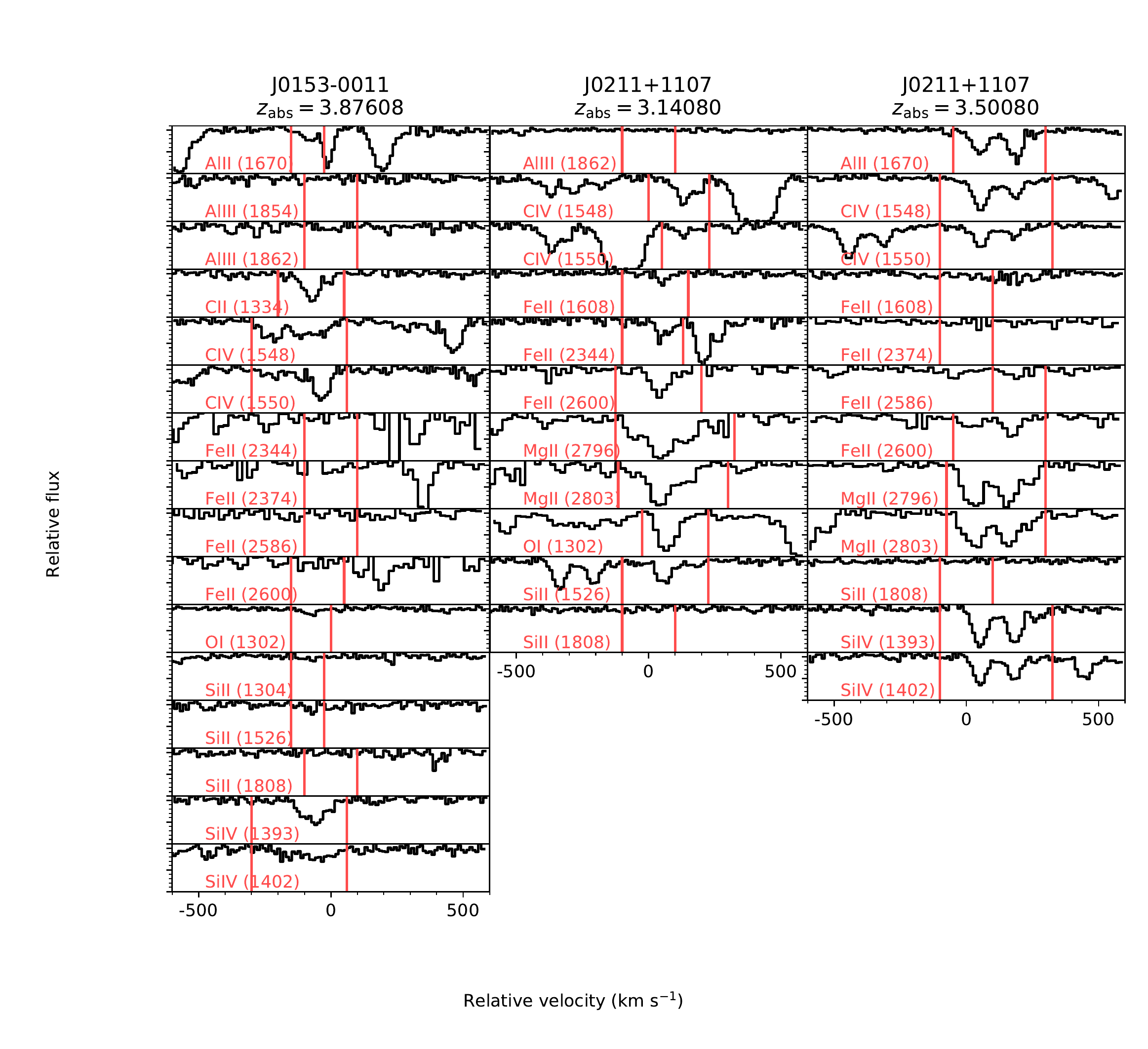}
\end{subfigure}
\caption{(cont'd)}
\end{center}
\end{figure*}

\begin{figure*}
\ContinuedFloat
\begin{center}
\begin{subfigure}{\textwidth}
\includegraphics[width=0.95\textwidth]{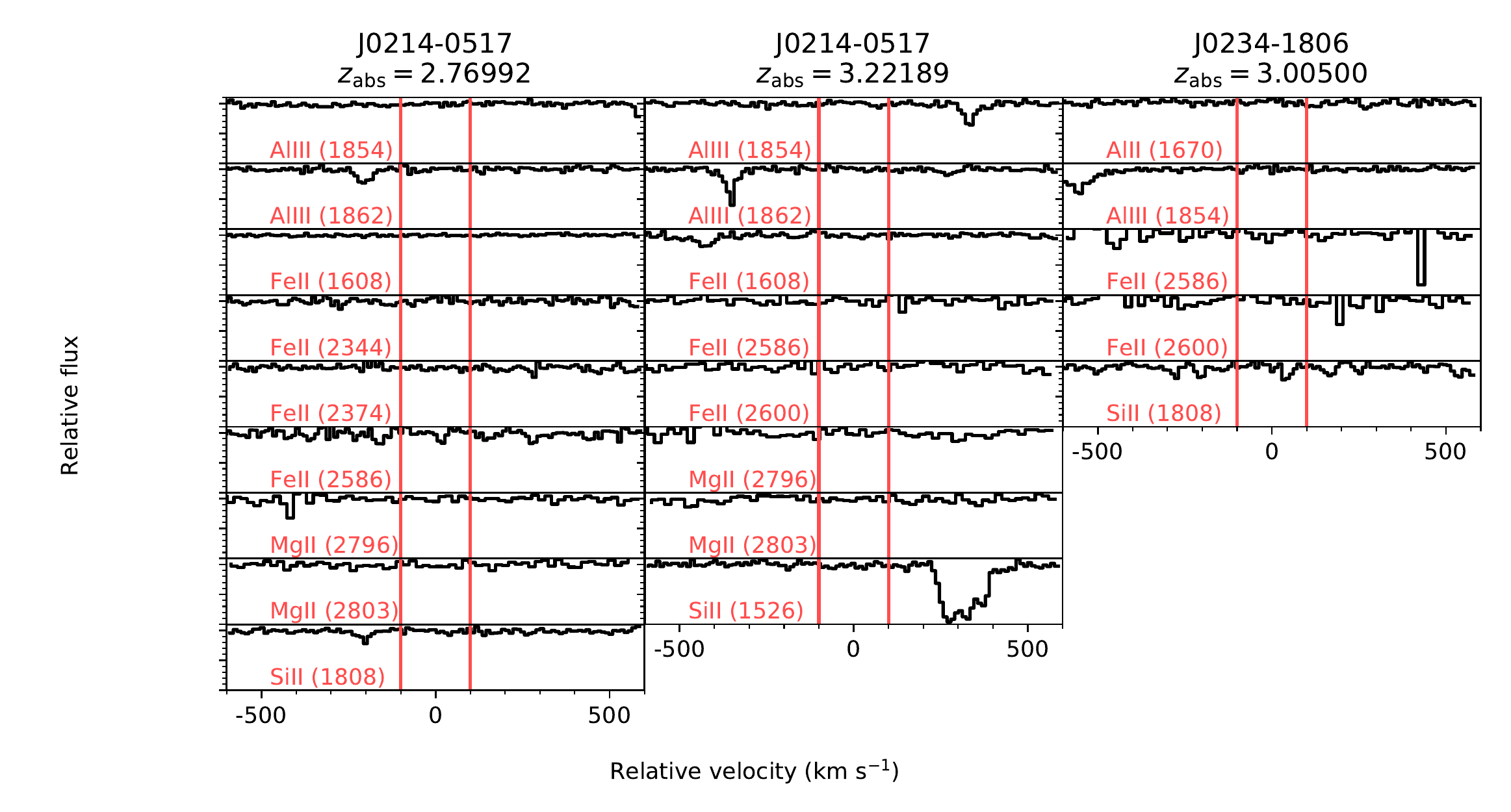}
\end{subfigure}
\caption{(cont'd)}
\end{center}
\end{figure*}

\begin{figure*}
\ContinuedFloat
\begin{center}
\begin{subfigure}{\textwidth}
\includegraphics[width=0.95\textwidth]{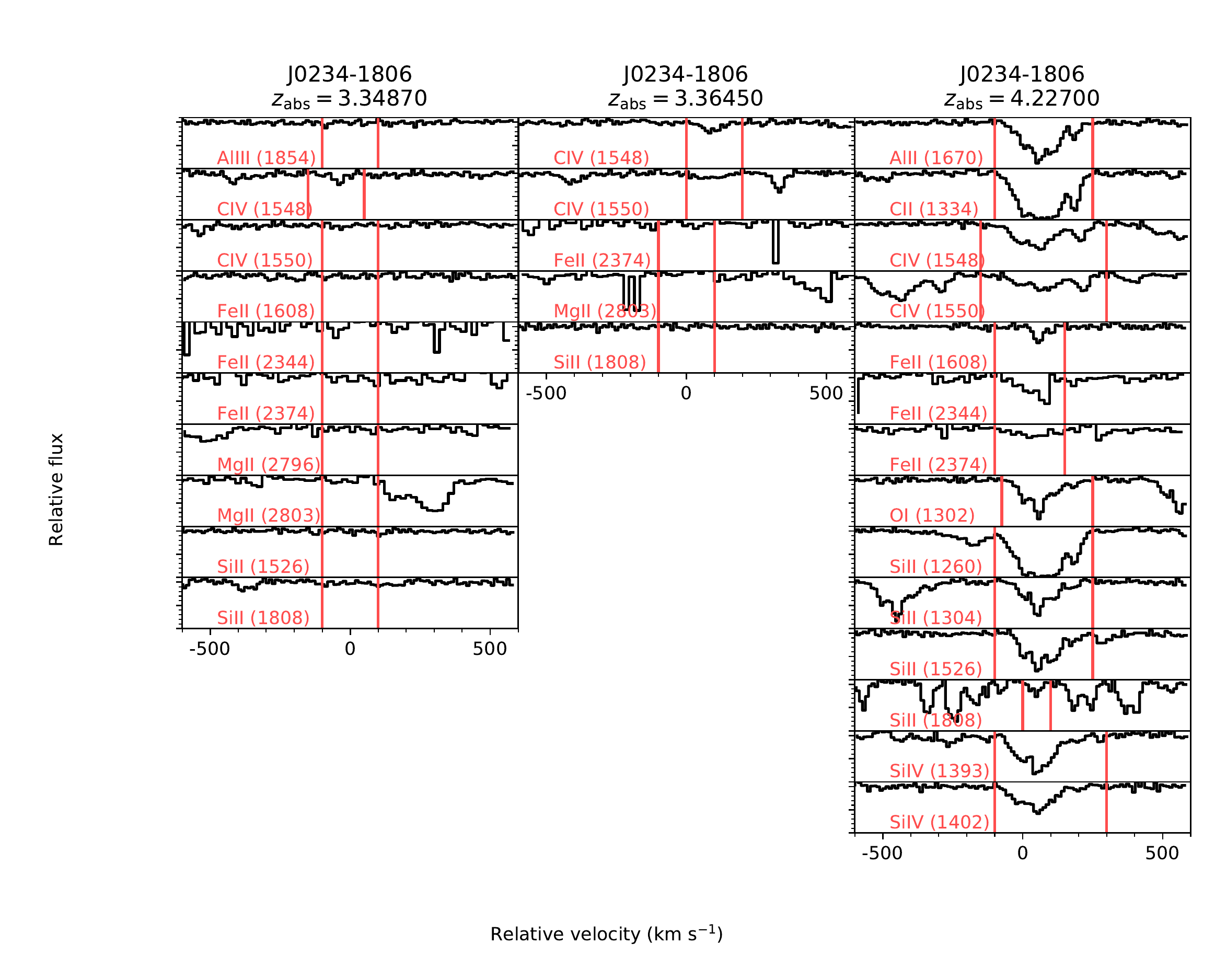}
\end{subfigure}
\caption{(cont'd)}
\end{center}
\end{figure*}

\begin{figure*}
\ContinuedFloat
\begin{center}
\begin{subfigure}{\textwidth}
\includegraphics[width=0.95\textwidth]{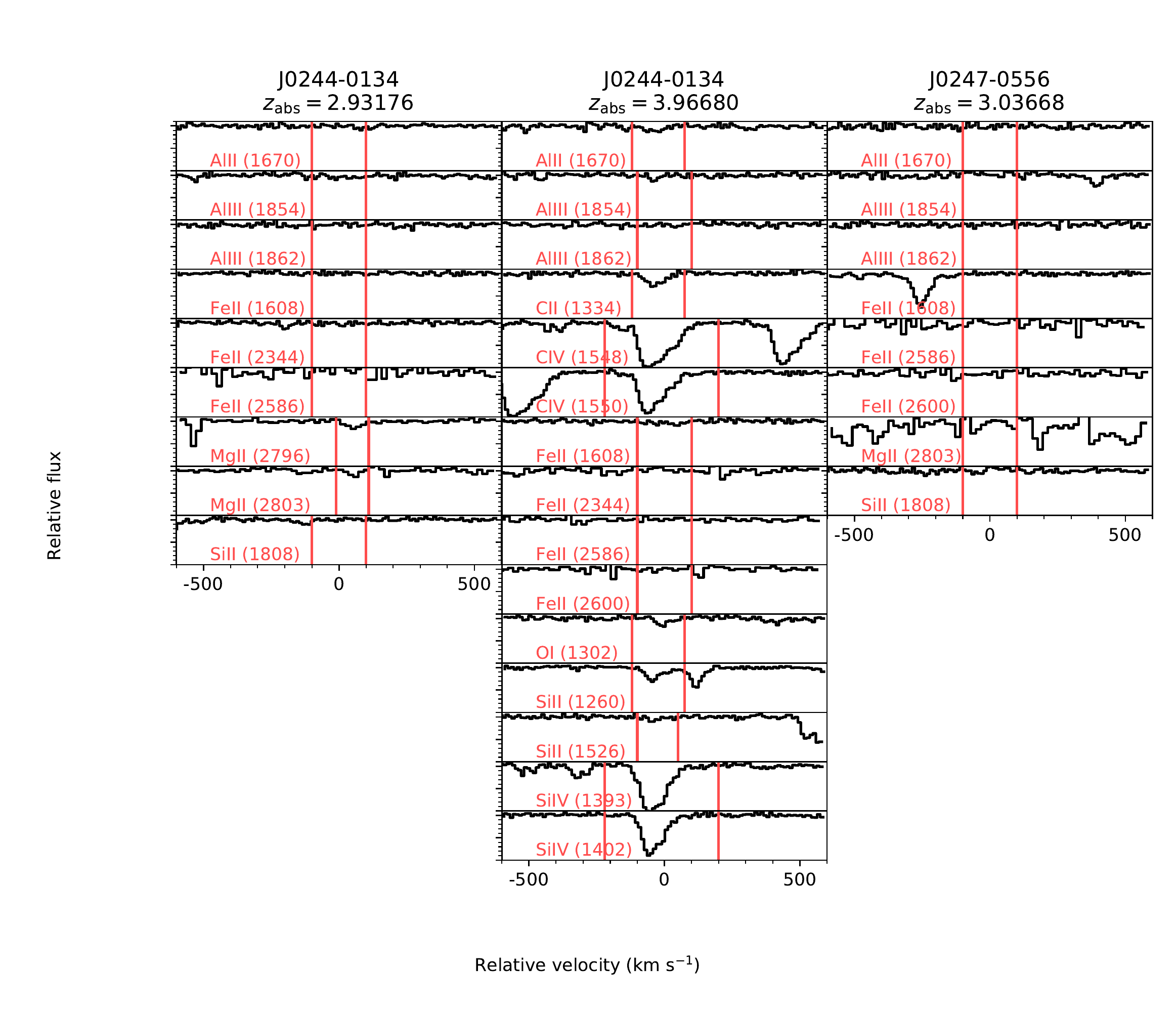}
\end{subfigure}
\caption{(cont'd)}
\end{center}
\end{figure*}

\begin{figure*}
\ContinuedFloat
\begin{center}
\begin{subfigure}{\textwidth}
\includegraphics[width=0.95\textwidth]{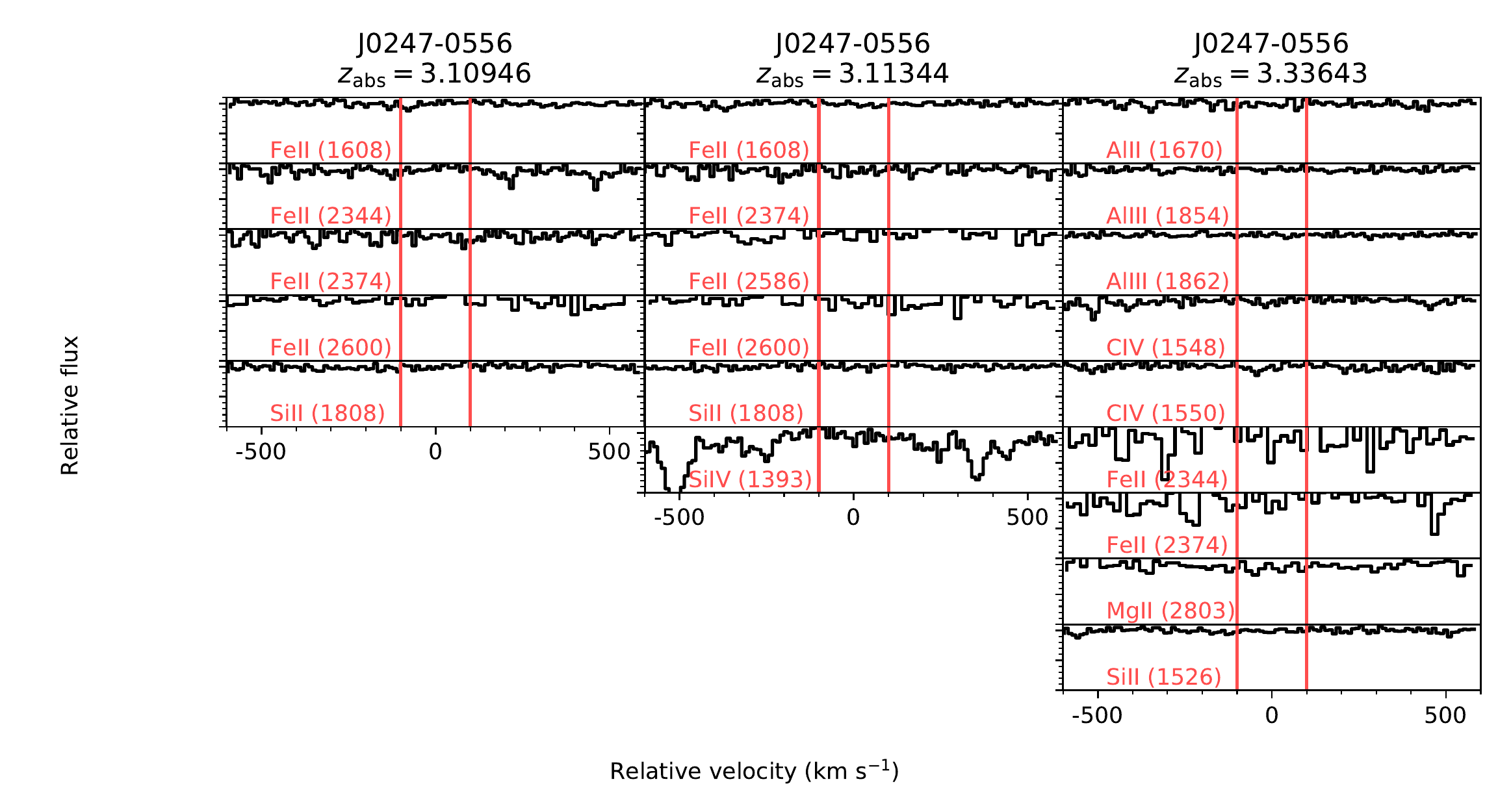}
\end{subfigure}
\caption{(cont'd)}
\end{center}
\end{figure*}

\begin{figure*}
\ContinuedFloat
\begin{center}
\begin{subfigure}{\textwidth}
\includegraphics[width=0.95\textwidth]{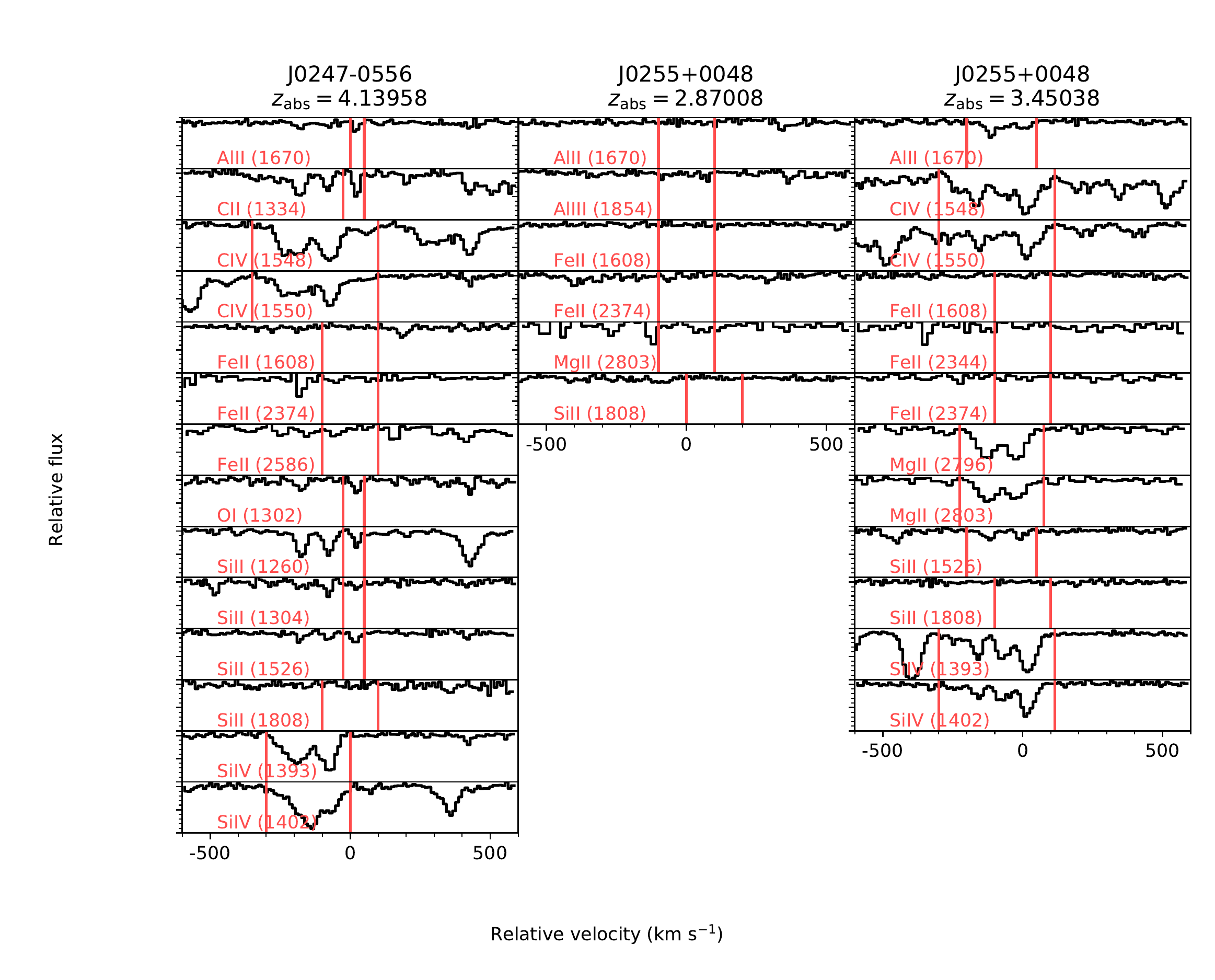}
\end{subfigure}
\caption{(cont'd)}
\end{center}
\end{figure*}

\begin{figure*}
\ContinuedFloat
\begin{center}
\begin{subfigure}{\textwidth}
\includegraphics[width=0.95\textwidth]{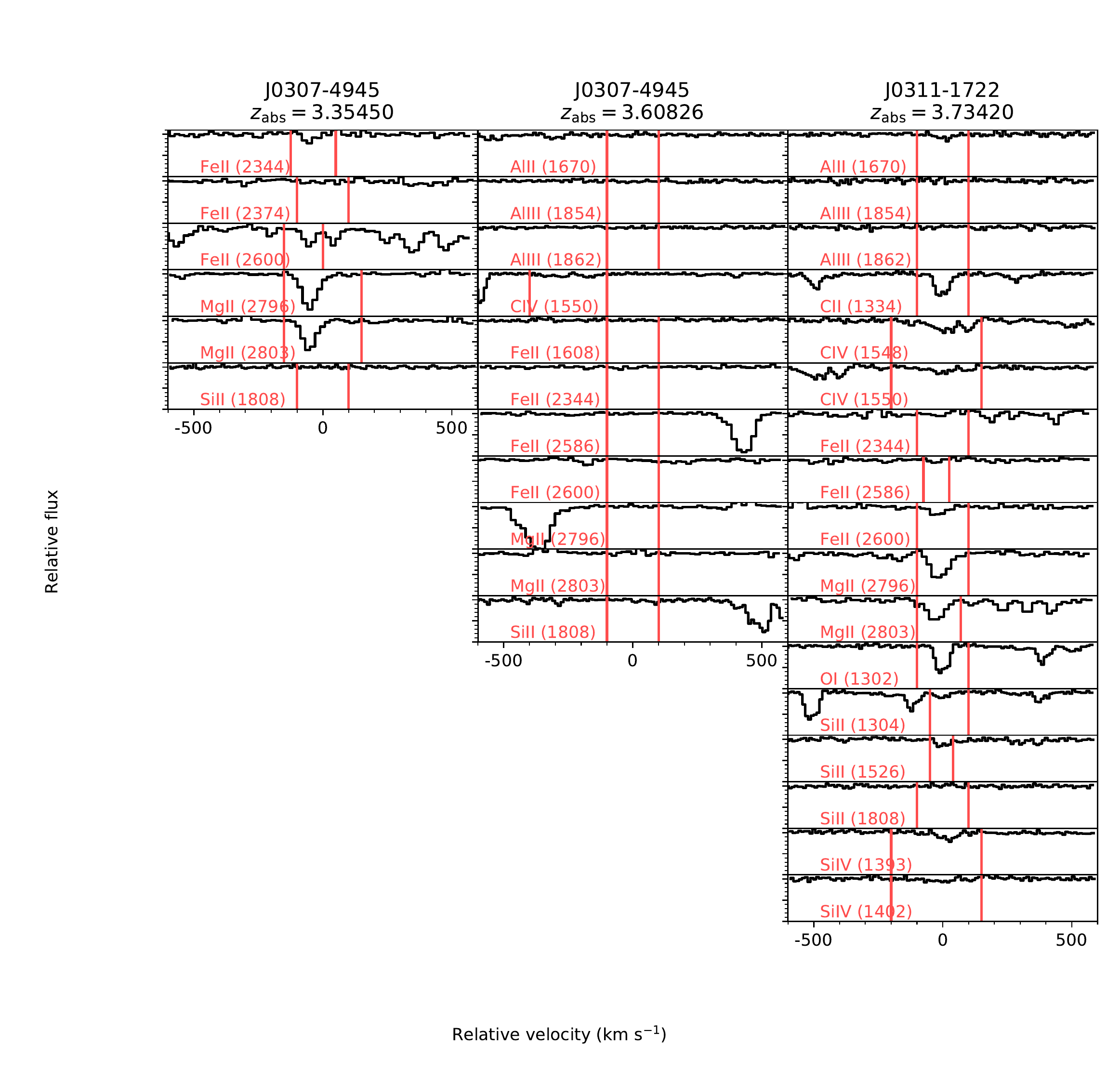}
\end{subfigure}
\caption{(cont'd)}
\end{center}
\end{figure*}

\begin{figure*}
\ContinuedFloat
\begin{center}
\begin{subfigure}{\textwidth}
\includegraphics[width=0.95\textwidth]{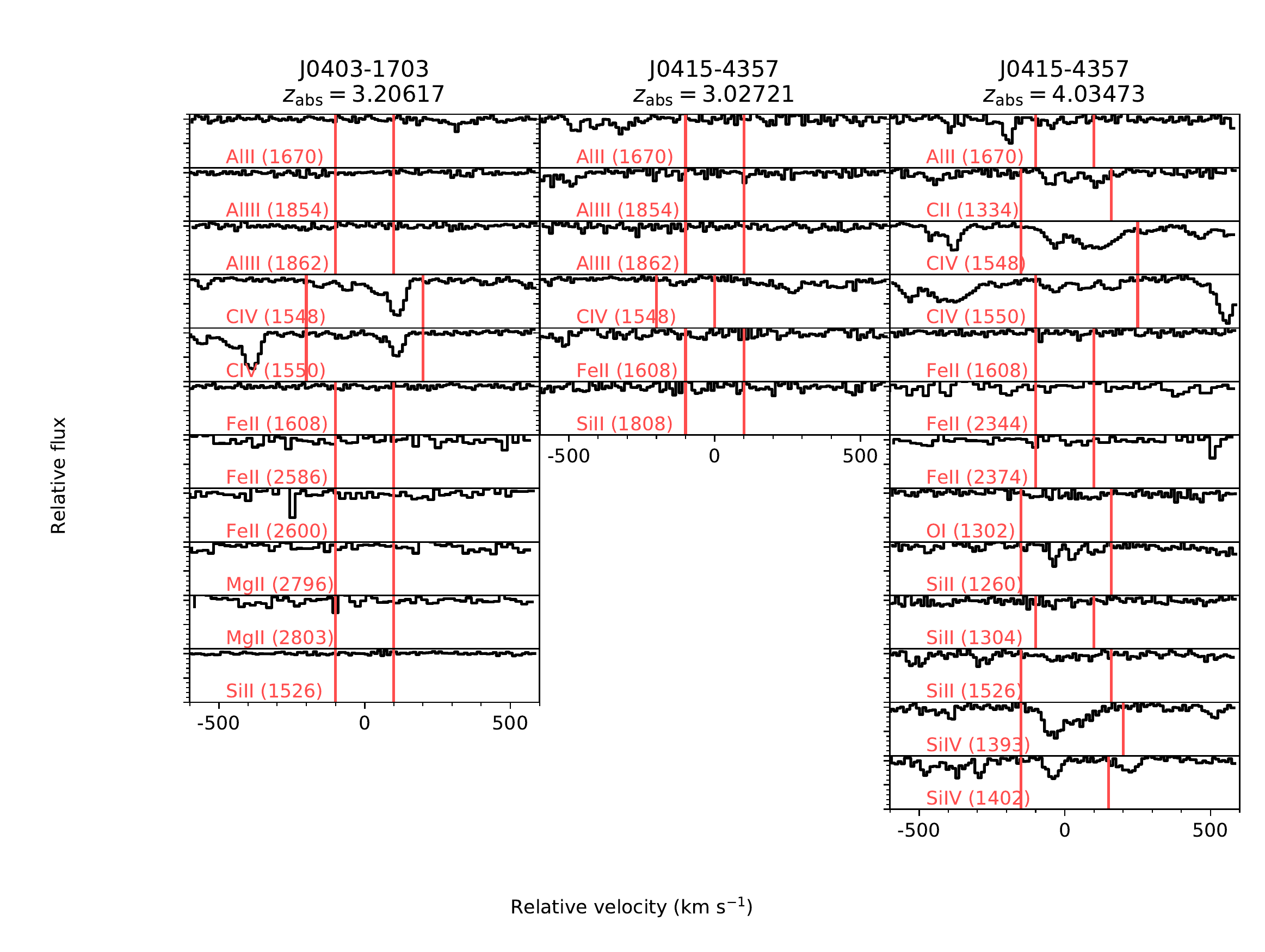}
\end{subfigure}
\caption{(cont'd)}
\end{center}
\end{figure*}

\begin{figure*}
\ContinuedFloat
\begin{center}
\begin{subfigure}{\textwidth}
\includegraphics[width=0.95\textwidth]{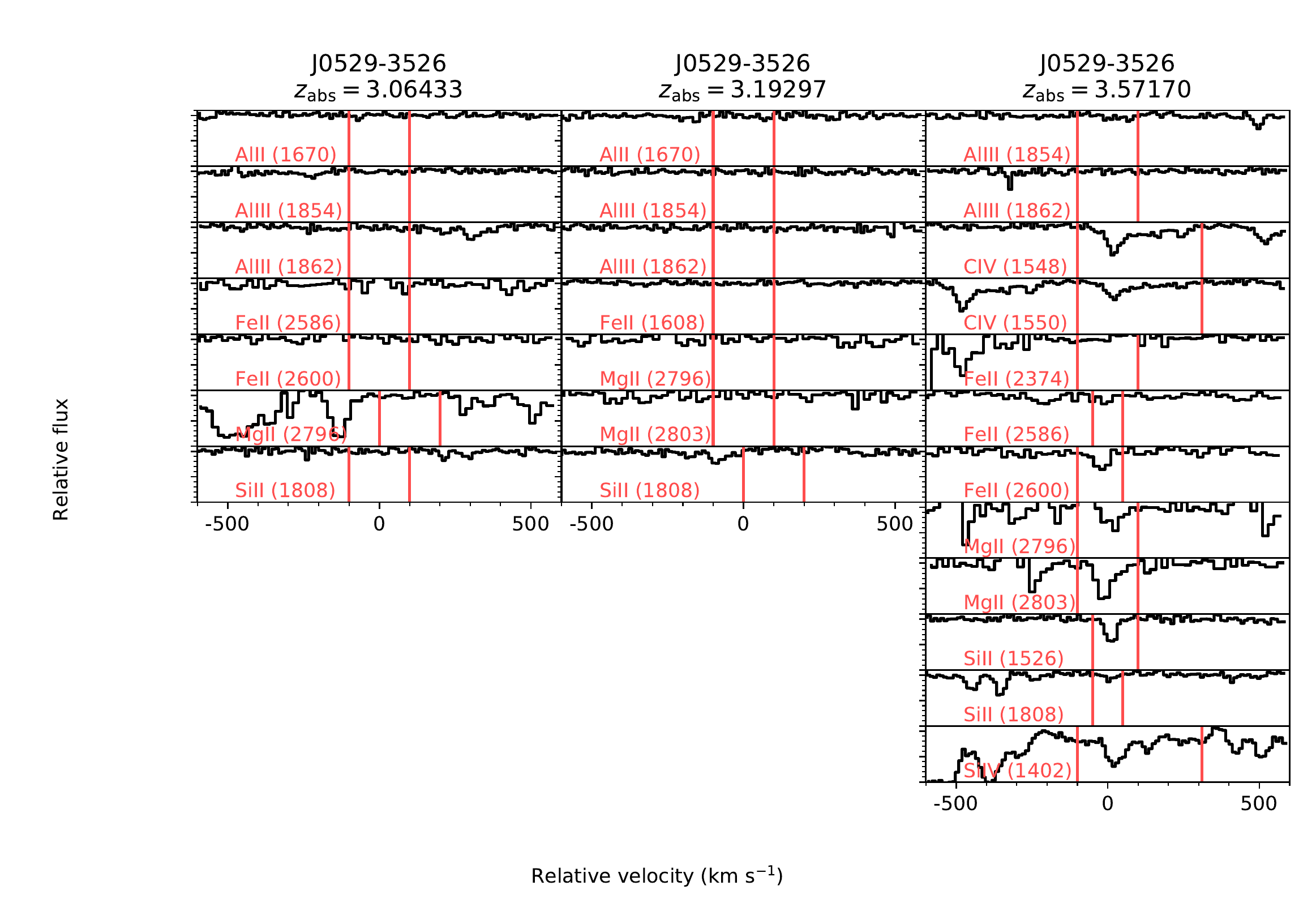}
\end{subfigure}
\caption{(cont'd)}
\end{center}
\end{figure*}

\begin{figure*}
\ContinuedFloat
\begin{center}
\begin{subfigure}{\textwidth}
\includegraphics[width=0.95\textwidth]{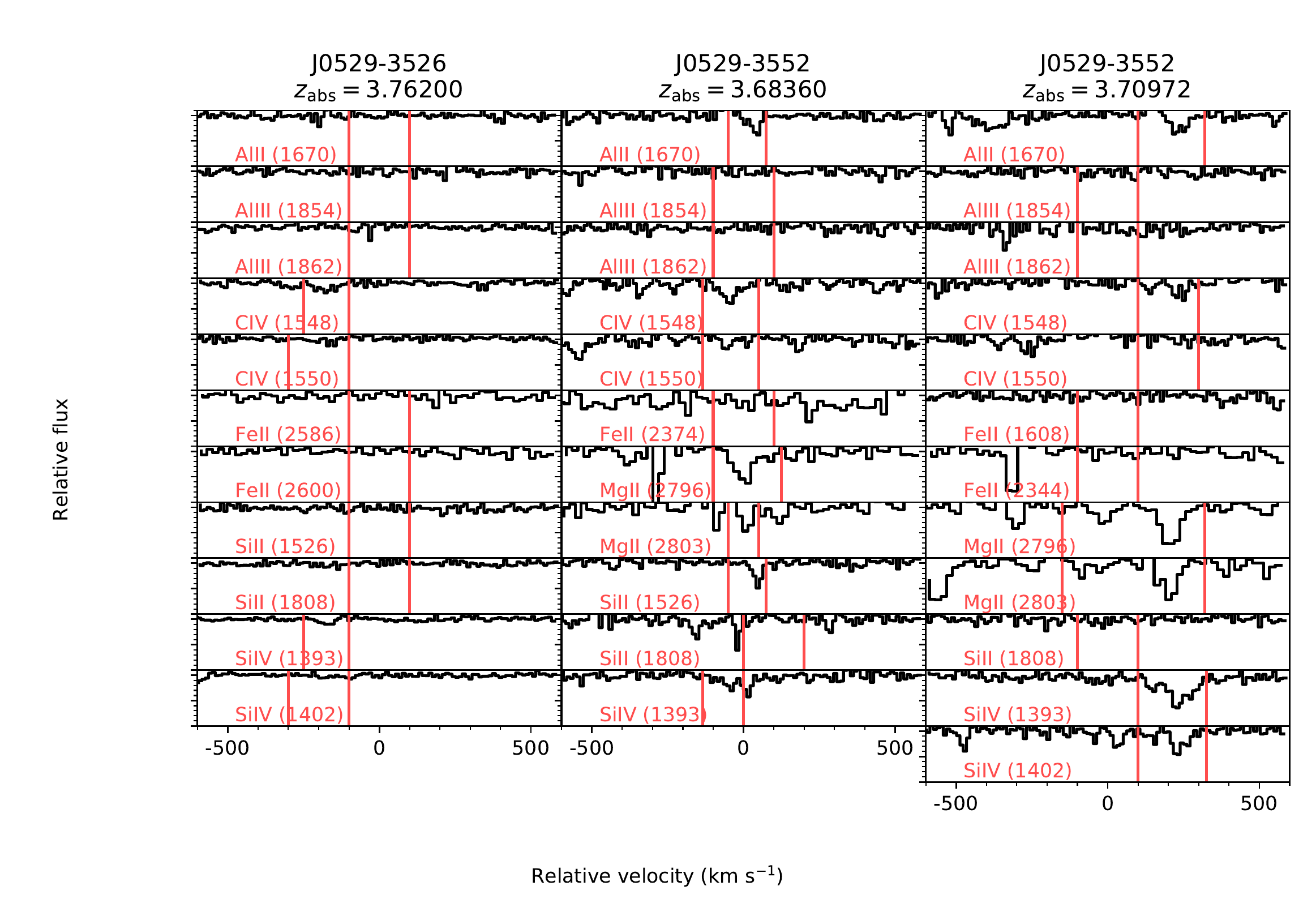}
\end{subfigure}
\caption{(cont'd)}
\end{center}
\end{figure*}

\begin{figure*}
\ContinuedFloat
\begin{center}
\begin{subfigure}{\textwidth}
\includegraphics[width=0.95\textwidth]{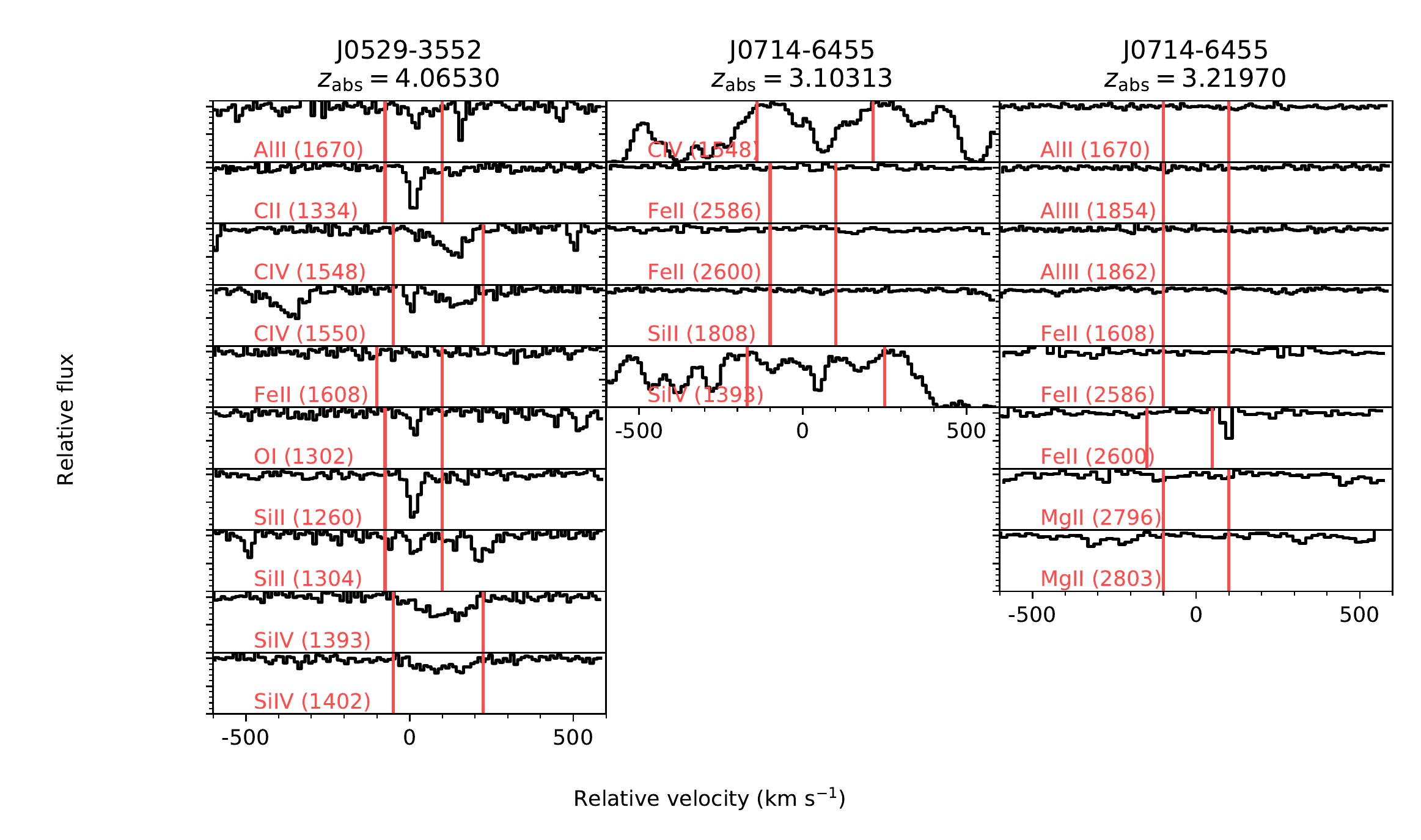}
\end{subfigure}
\caption{(cont'd)}
\end{center}
\end{figure*}

\begin{figure*}
\ContinuedFloat
\begin{center}
\begin{subfigure}{\textwidth}
\includegraphics[width=0.95\textwidth]{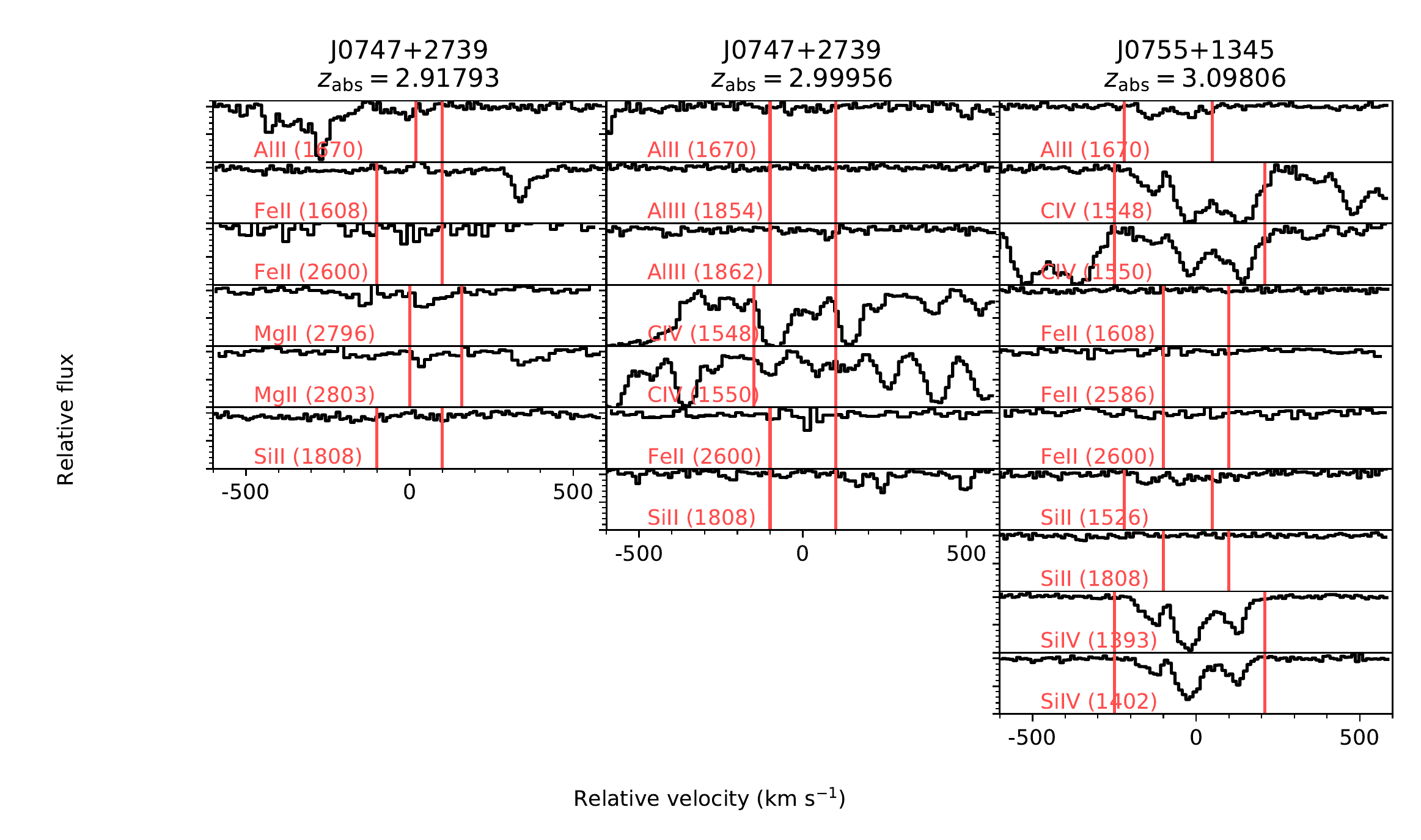}
\end{subfigure}
\caption{(cont'd)}
\end{center}
\end{figure*}

\begin{figure*}
\ContinuedFloat
\begin{center}
\begin{subfigure}{\textwidth}
\includegraphics[width=0.95\textwidth]{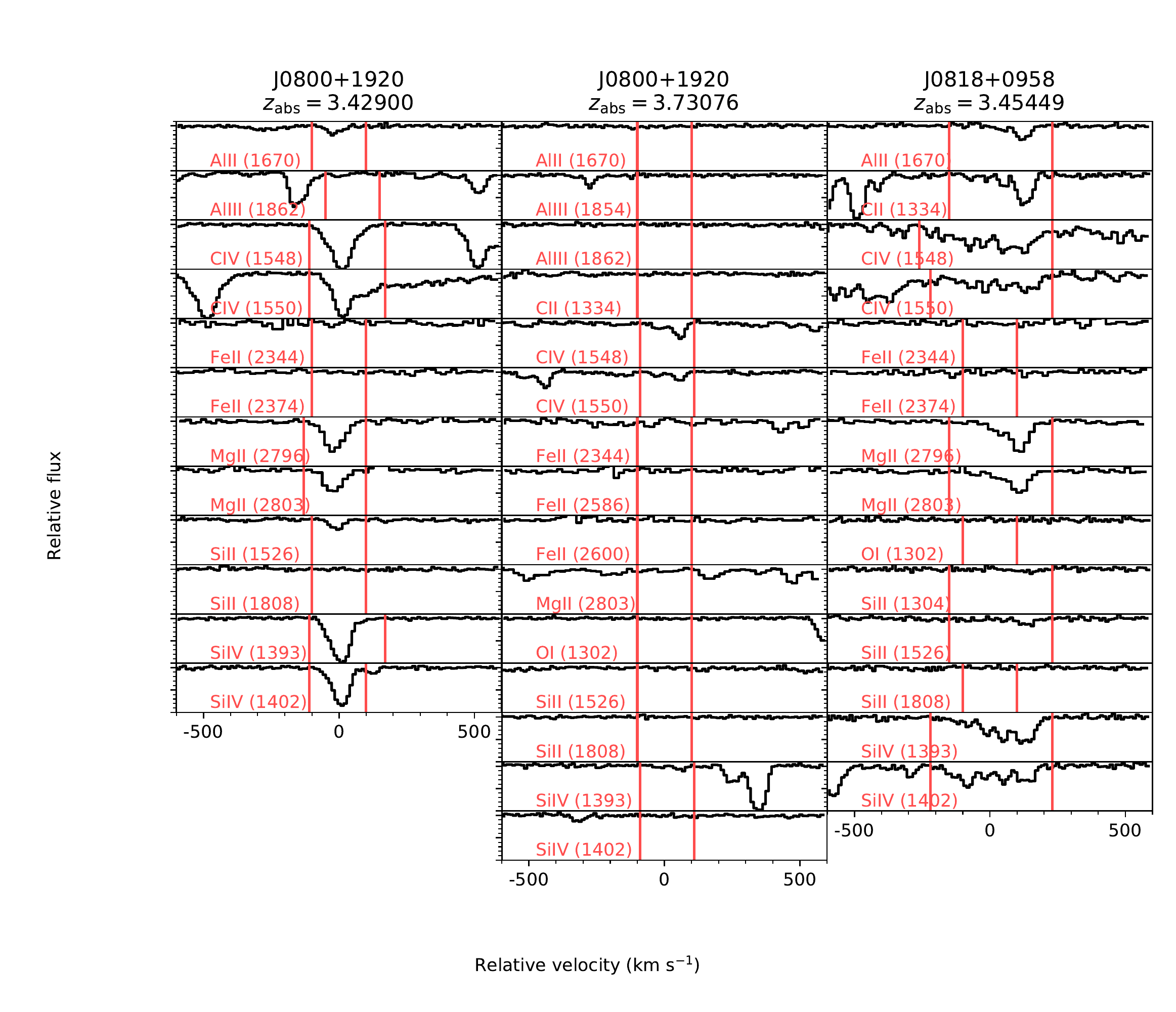}
\end{subfigure}
\caption{(cont'd)}
\end{center}
\end{figure*}

\begin{figure*}
\ContinuedFloat
\begin{center}
\begin{subfigure}{\textwidth}
\includegraphics[width=0.95\textwidth]{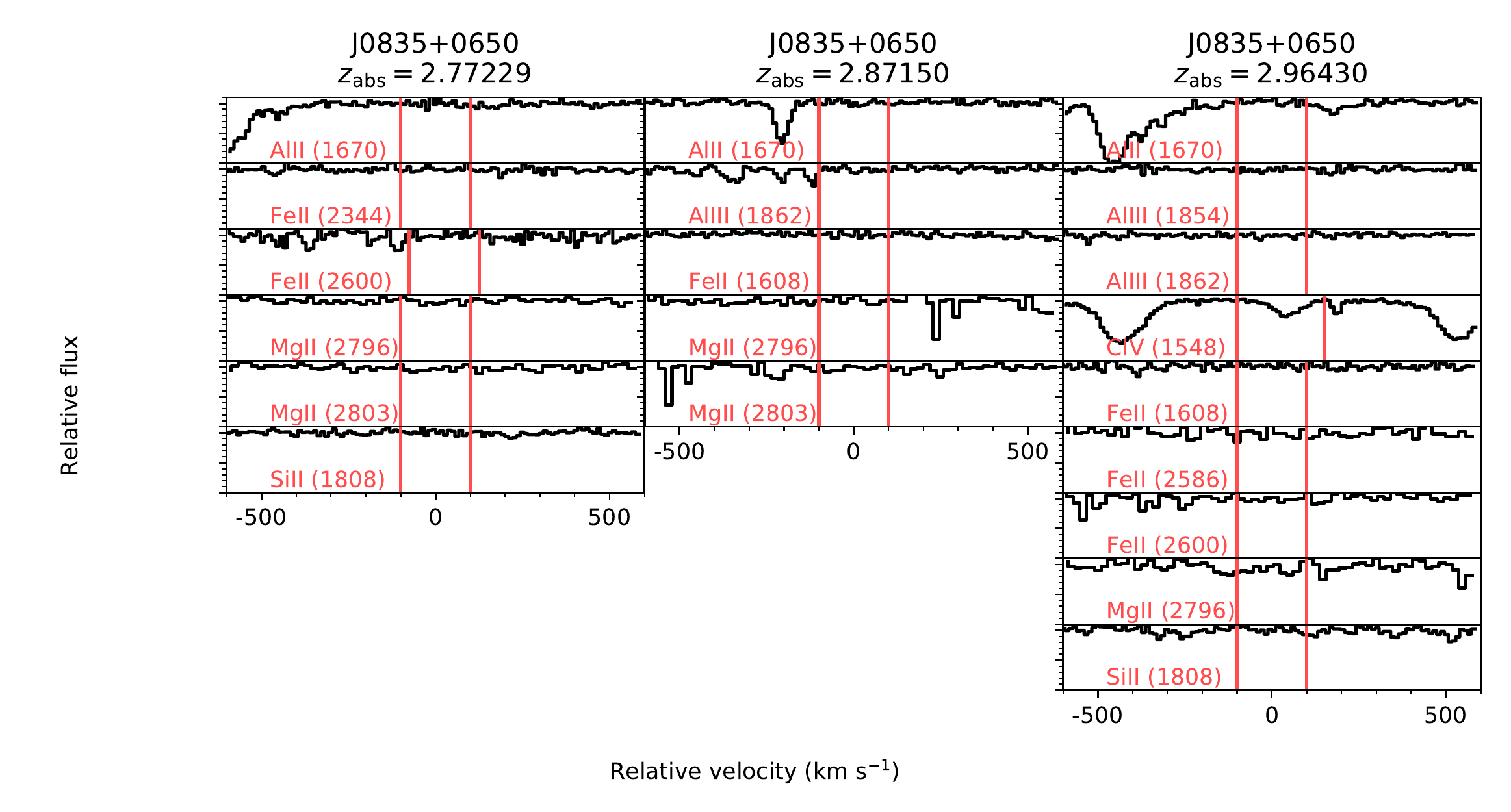}
\end{subfigure}
\caption{(cont'd)}
\end{center}
\end{figure*}

\begin{figure*}
\ContinuedFloat
\begin{center}
\begin{subfigure}{\textwidth}
\includegraphics[width=0.95\textwidth]{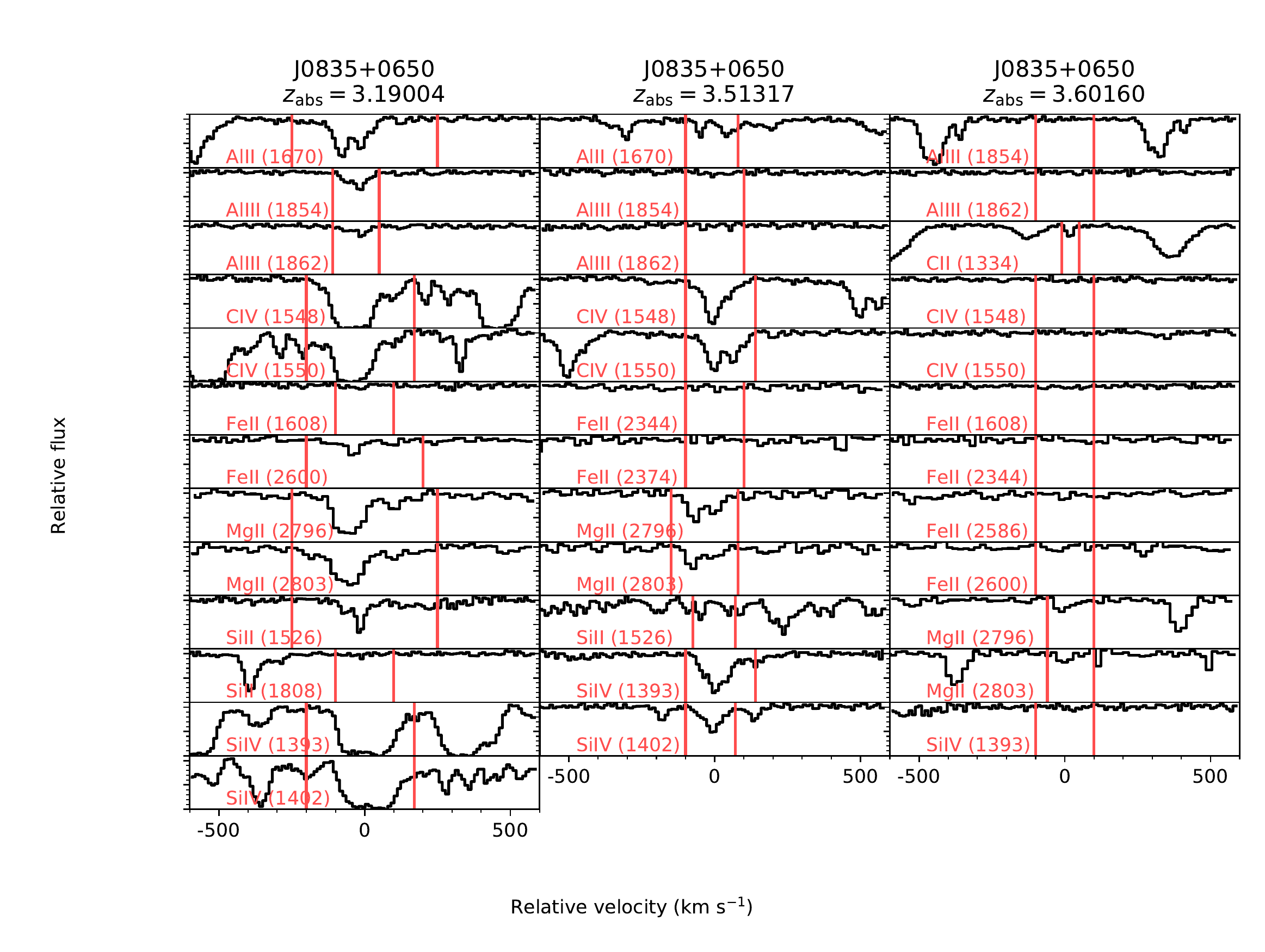}
\end{subfigure}
\caption{(cont'd)}
\end{center}
\end{figure*}

\begin{figure*}
\ContinuedFloat
\begin{center}
\begin{subfigure}{\textwidth}
\includegraphics[width=0.95\textwidth]{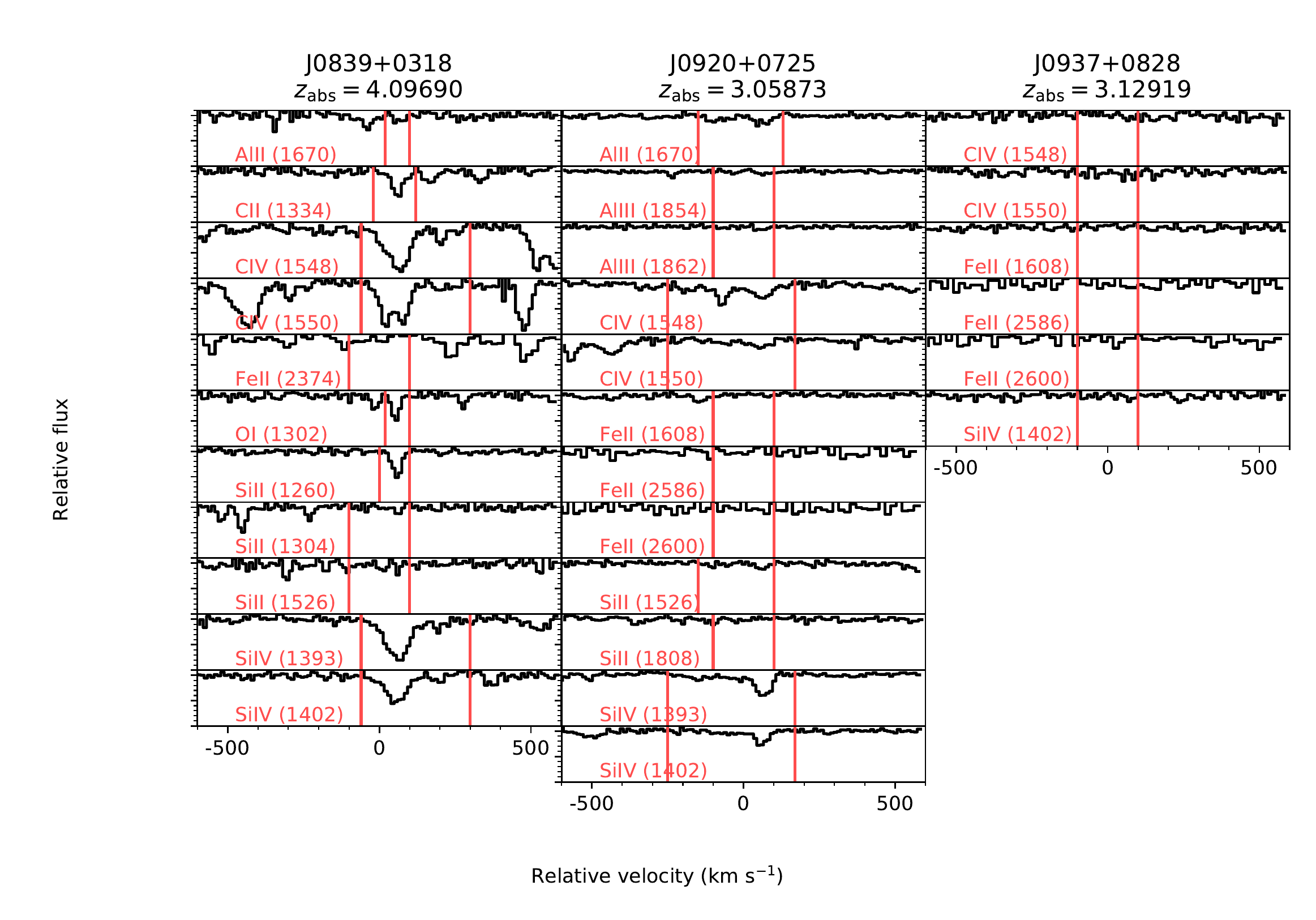}
\end{subfigure}
\caption{(cont'd)}
\end{center}
\end{figure*}

\begin{figure*}
\ContinuedFloat
\begin{center}
\begin{subfigure}{\textwidth}
\includegraphics[width=0.95\textwidth]{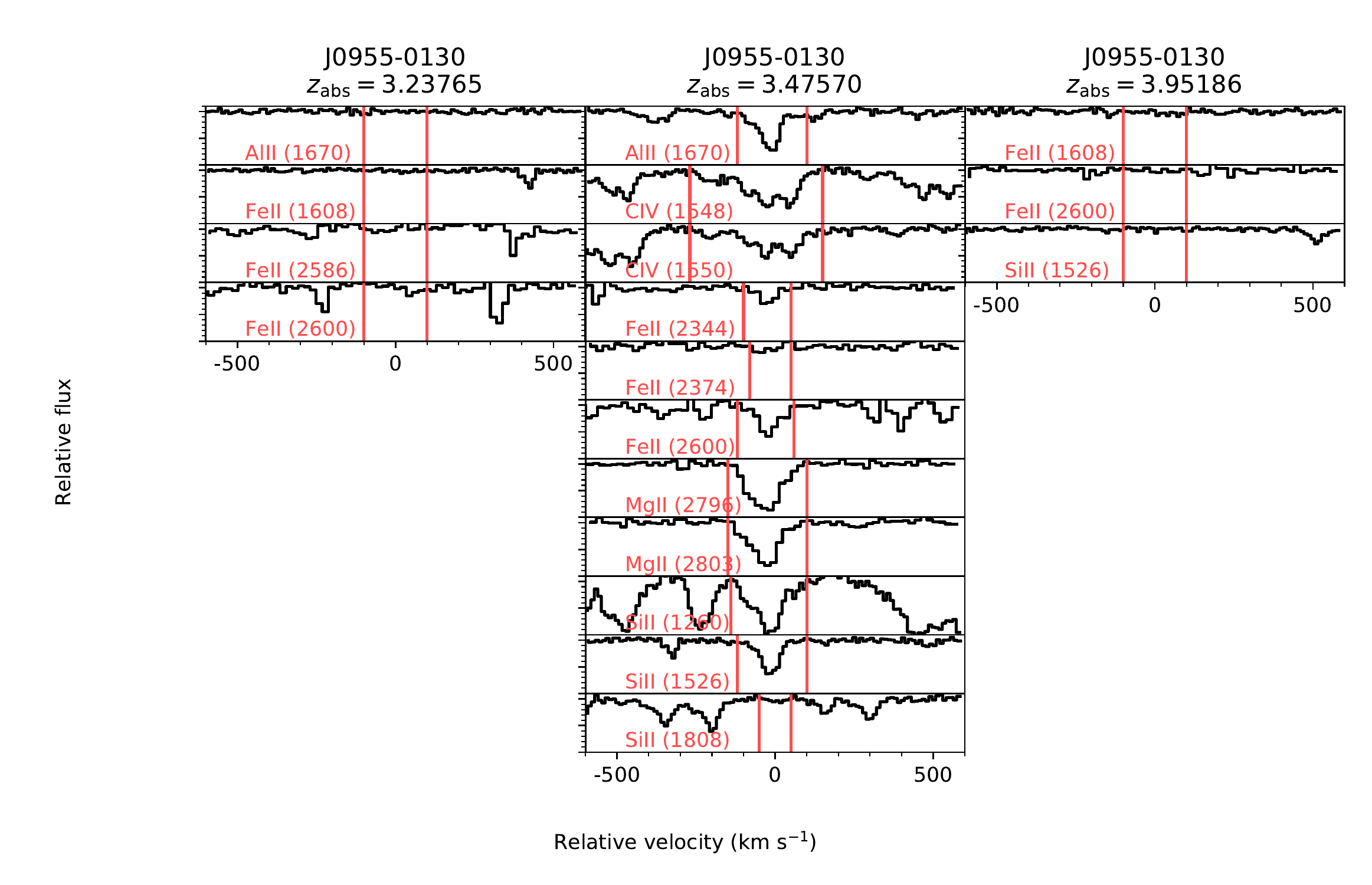}
\end{subfigure}
\caption{(cont'd)}
\end{center}
\end{figure*}

\begin{figure*}
\ContinuedFloat
\begin{center}
\begin{subfigure}{\textwidth}
\includegraphics[width=0.95\textwidth]{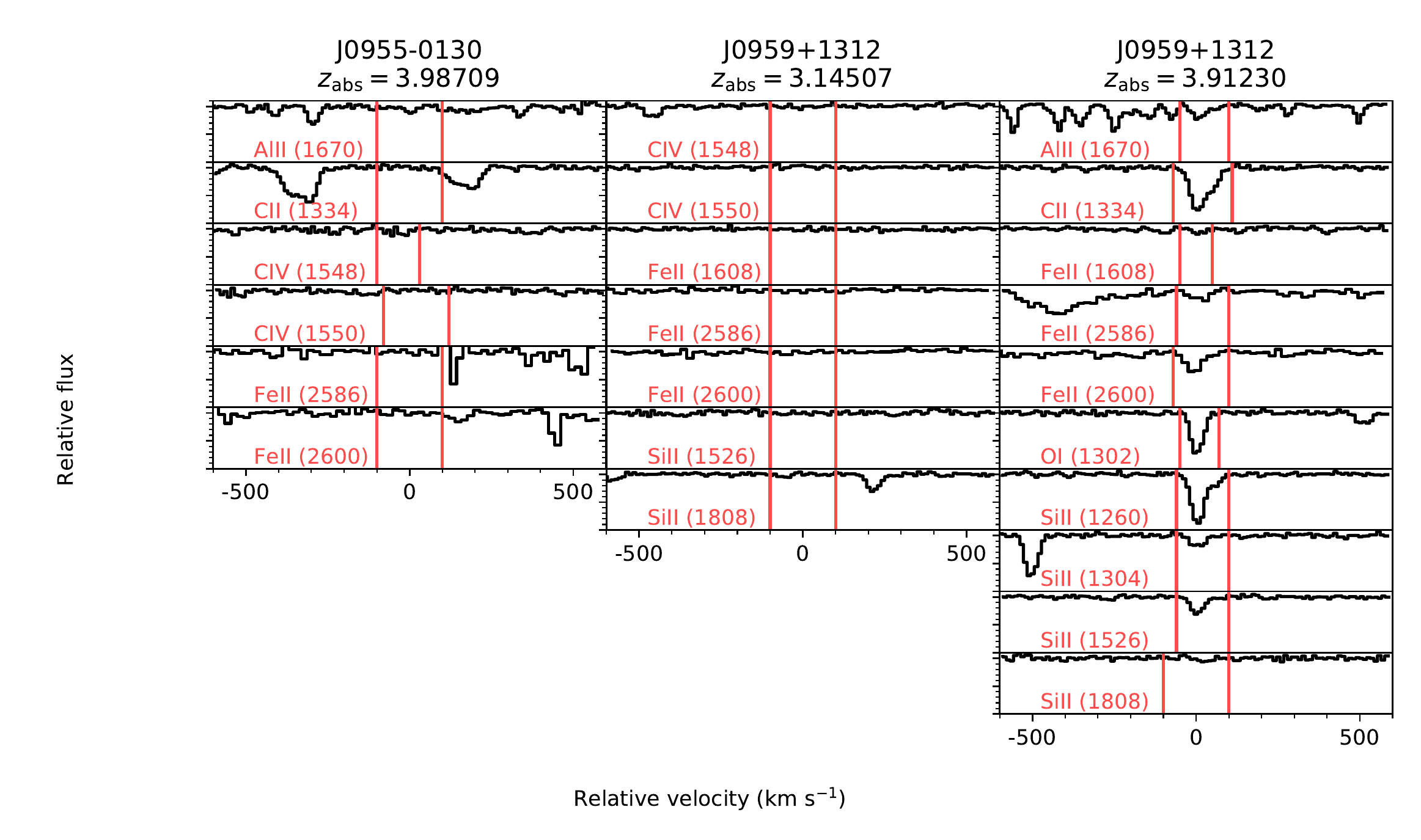}
\end{subfigure}
\caption{(cont'd)}
\end{center}
\end{figure*}

\begin{figure*}
\ContinuedFloat
\begin{center}
\begin{subfigure}{\textwidth}
\includegraphics[width=0.95\textwidth]{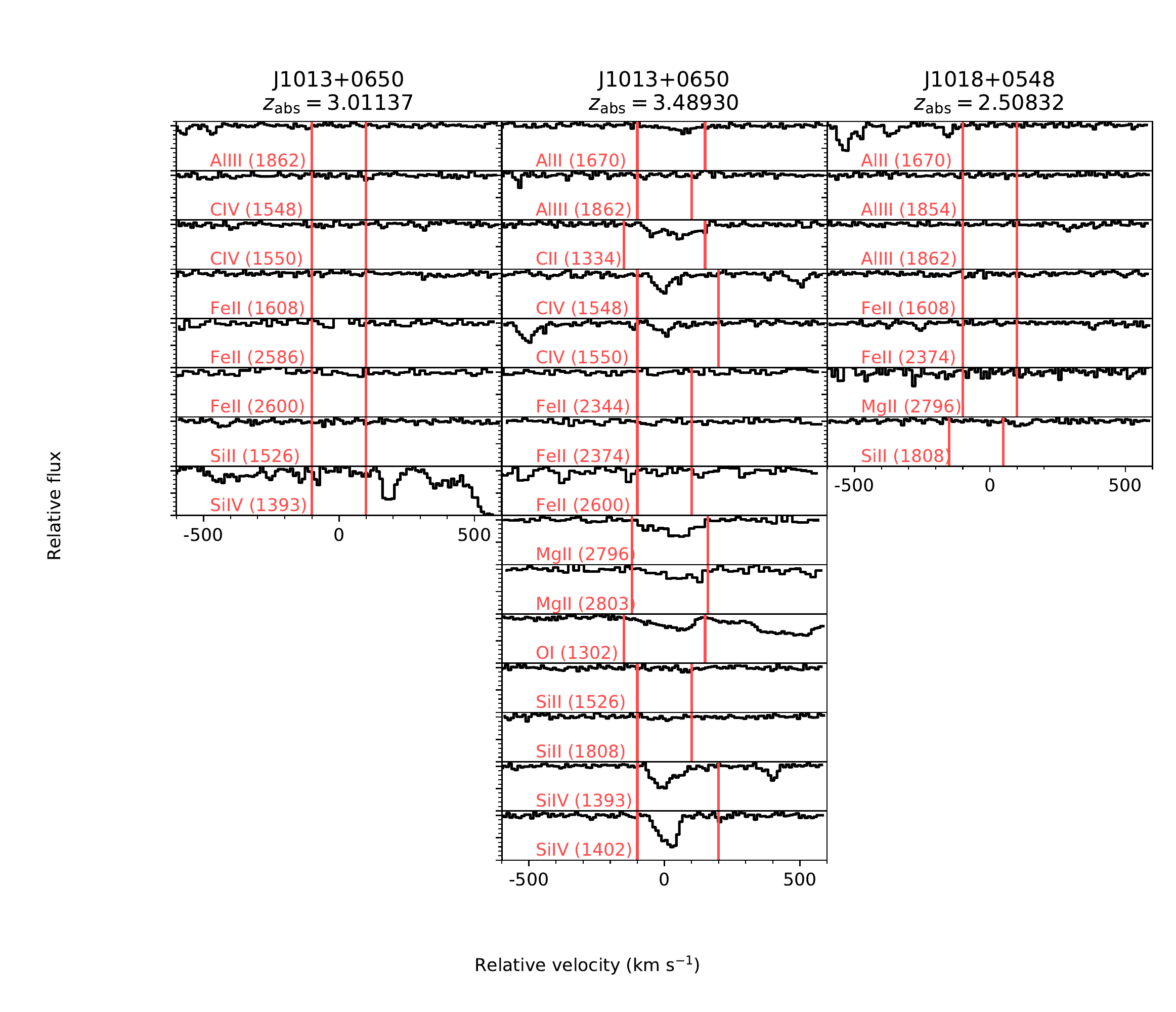}
\end{subfigure}
\caption{(cont'd)}
\end{center}
\end{figure*}

\begin{figure*}
\ContinuedFloat
\begin{center}
\begin{subfigure}{\textwidth}
\includegraphics[width=0.95\textwidth]{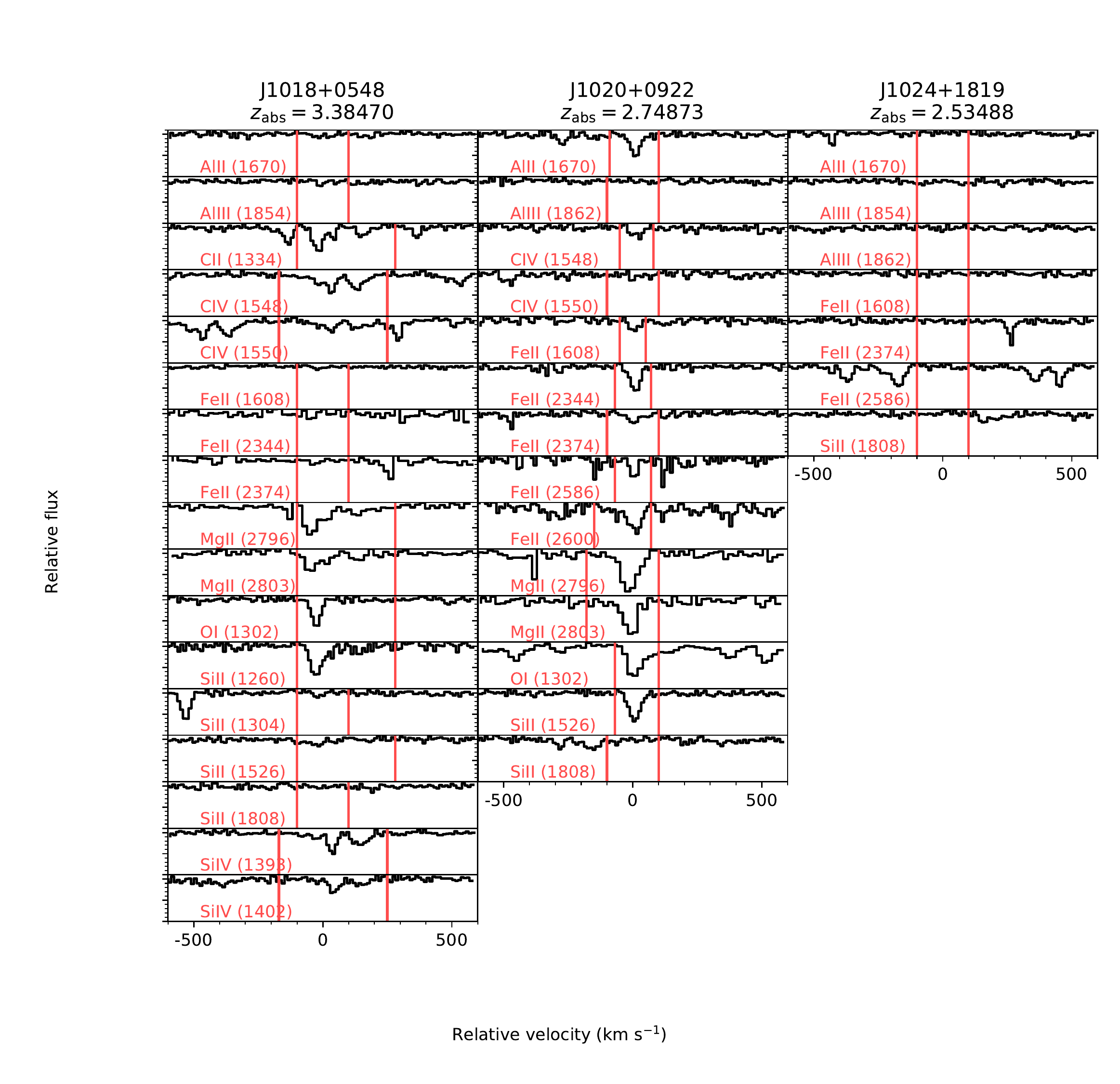}
\end{subfigure}
\caption{(cont'd)}
\end{center}
\end{figure*}

\begin{figure*}
\ContinuedFloat
\begin{center}
\begin{subfigure}{\textwidth}
\includegraphics[width=0.95\textwidth]{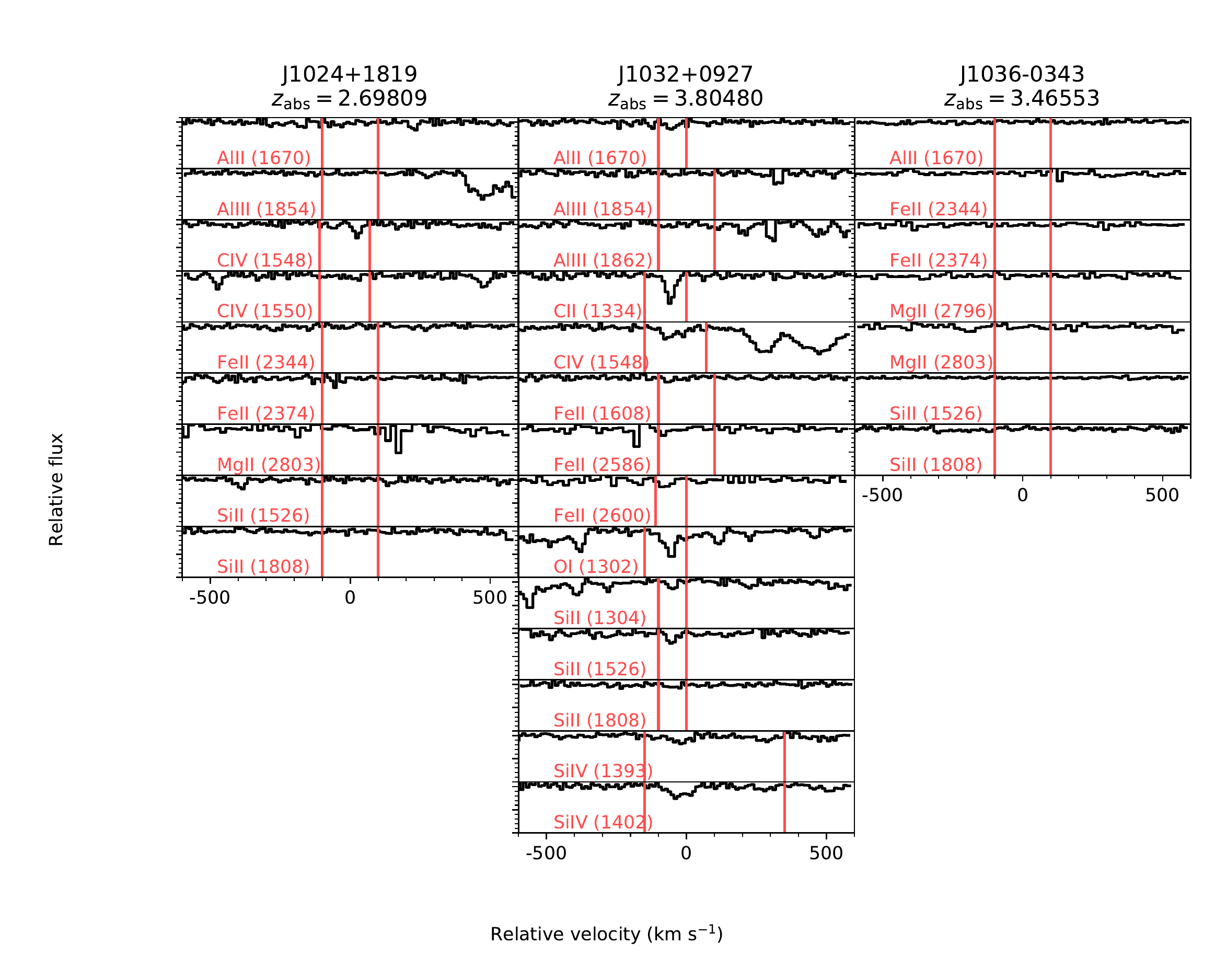}
\end{subfigure}
\caption{(cont'd)}
\end{center}
\end{figure*}

\begin{figure*}
\ContinuedFloat
\begin{center}
\begin{subfigure}{\textwidth}
\includegraphics[width=0.95\textwidth]{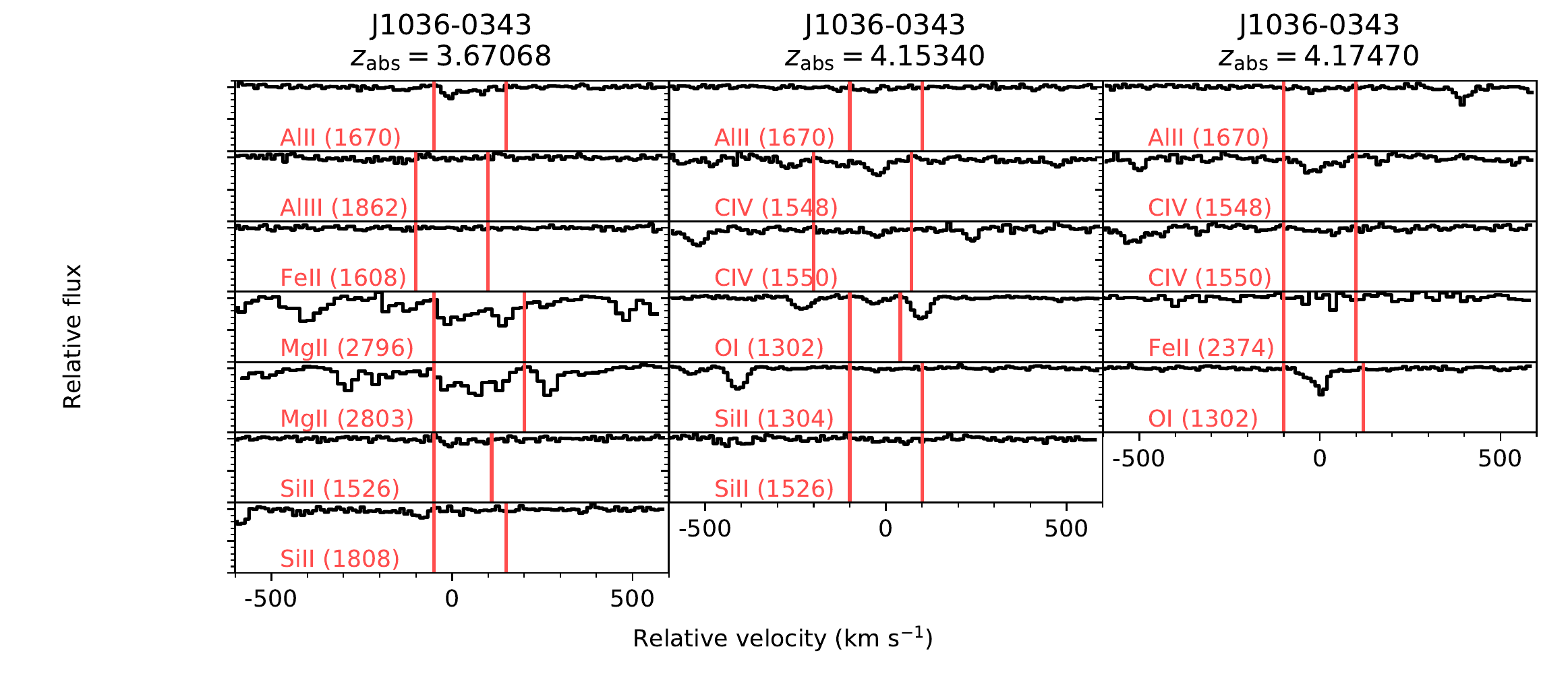}
\end{subfigure}
\caption{(cont'd)}
\end{center}
\end{figure*}

\begin{figure*}
\ContinuedFloat
\begin{center}
\begin{subfigure}{\textwidth}
\includegraphics[width=0.95\textwidth]{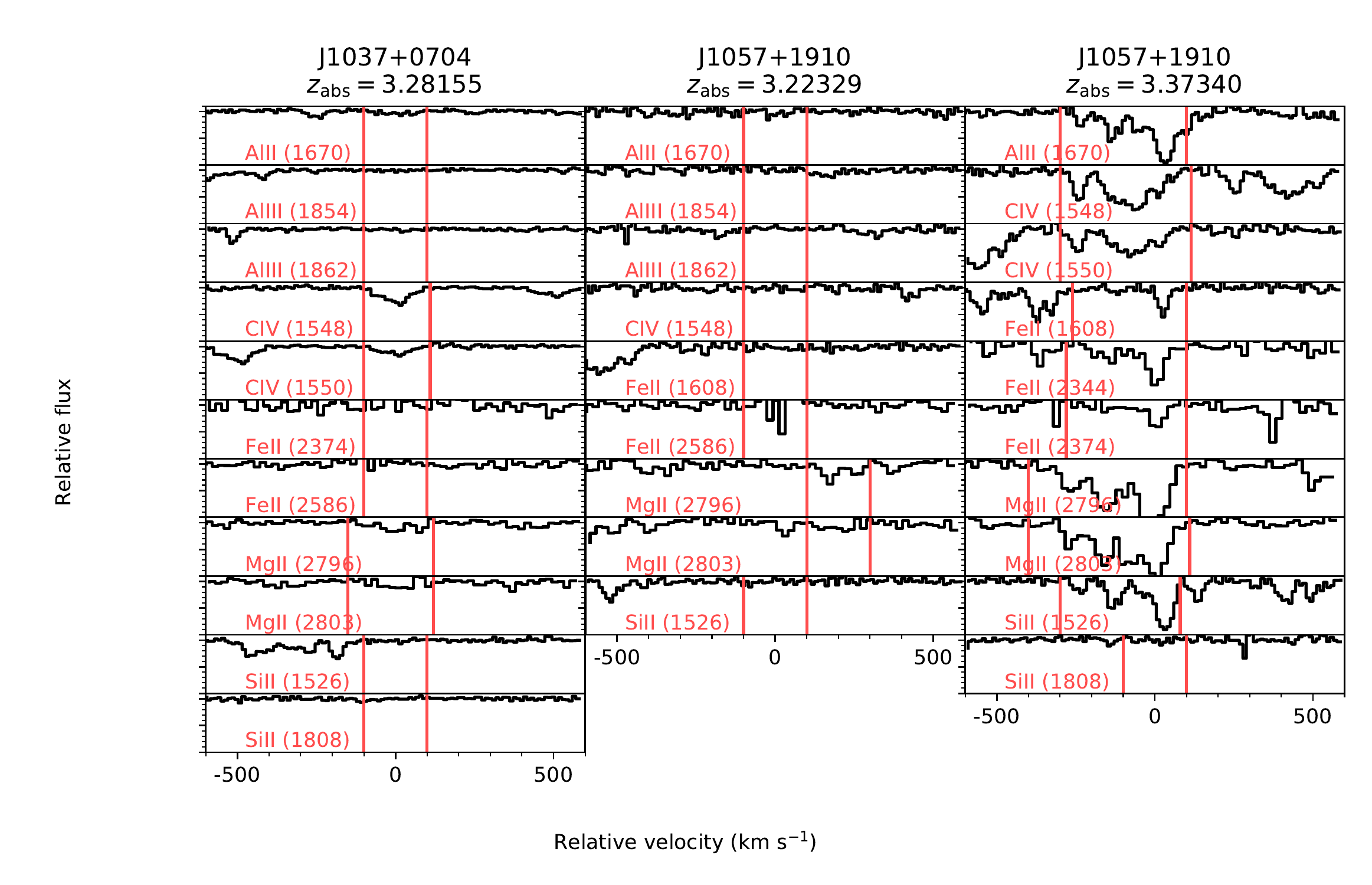}
\end{subfigure}
\caption{(cont'd)}
\end{center}
\end{figure*}

\begin{figure*}
\ContinuedFloat
\begin{center}
\begin{subfigure}{\textwidth}
\includegraphics[width=0.95\textwidth]{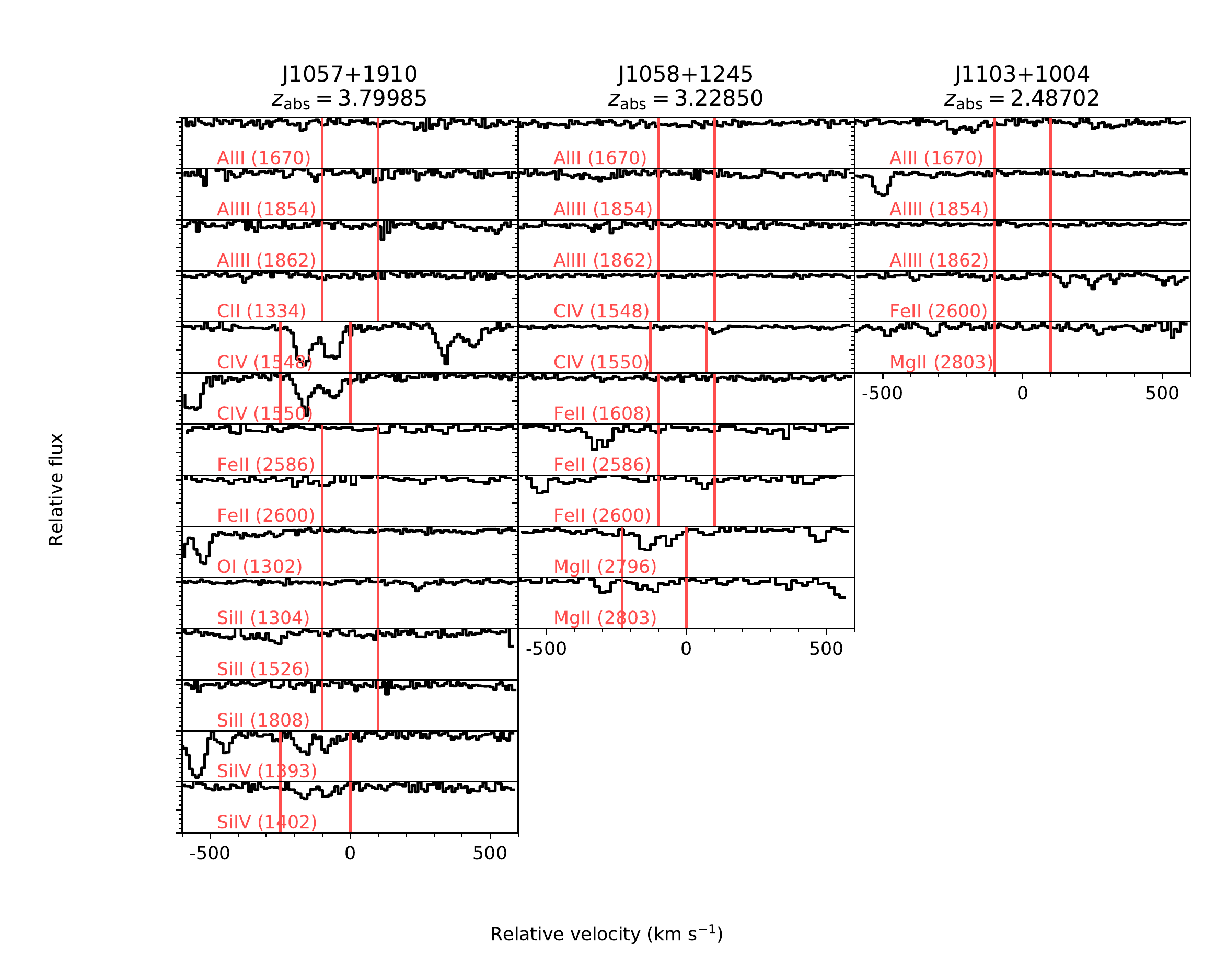}
\end{subfigure}
\caption{(cont'd)}
\end{center}
\end{figure*}

\begin{figure*}
\ContinuedFloat
\begin{center}
\begin{subfigure}{\textwidth}
\includegraphics[width=0.95\textwidth]{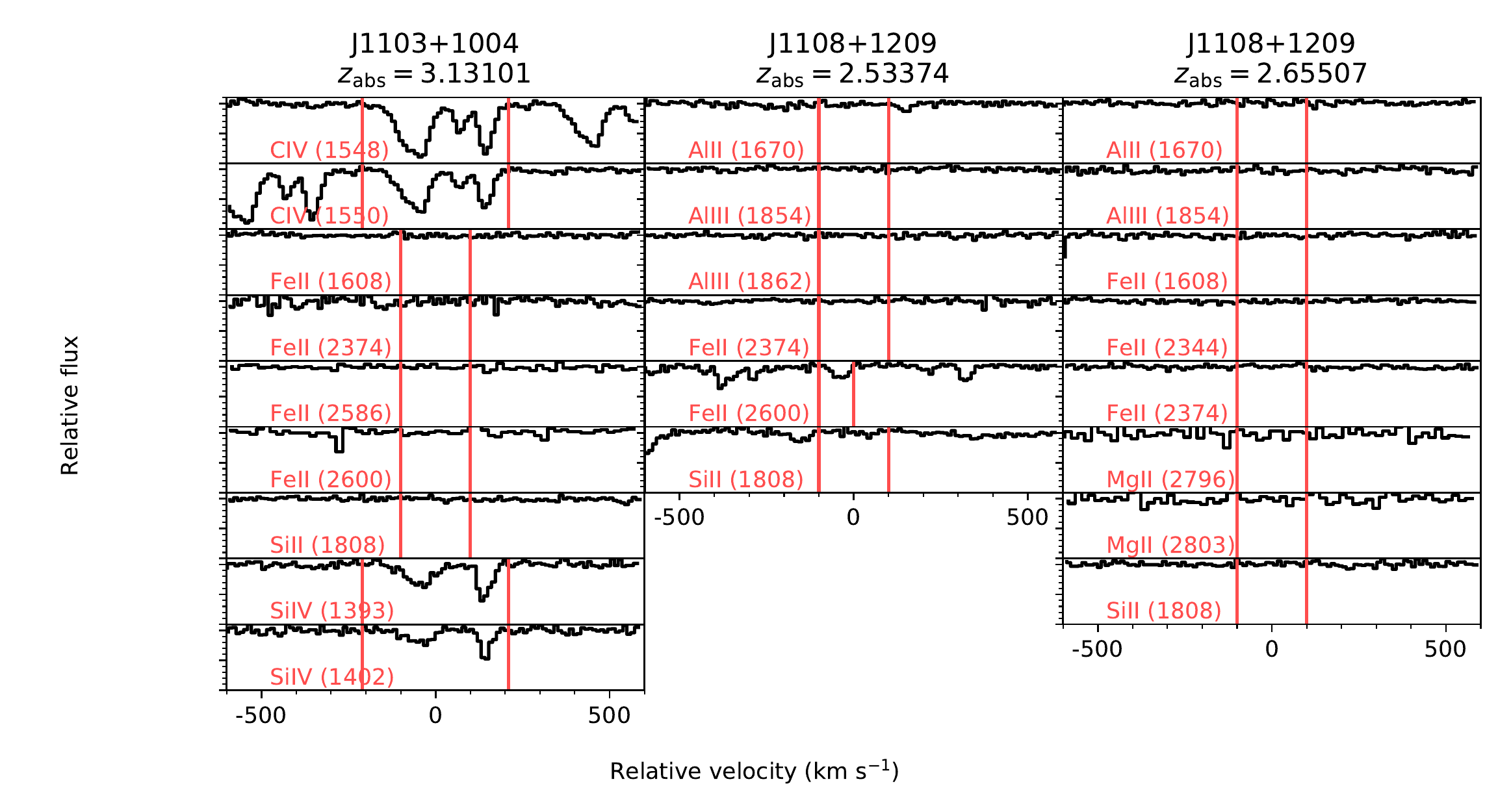}
\end{subfigure}
\caption{(cont'd)}
\end{center}
\end{figure*}

\begin{figure*}
\ContinuedFloat
\begin{center}
\begin{subfigure}{\textwidth}
\includegraphics[width=0.95\textwidth]{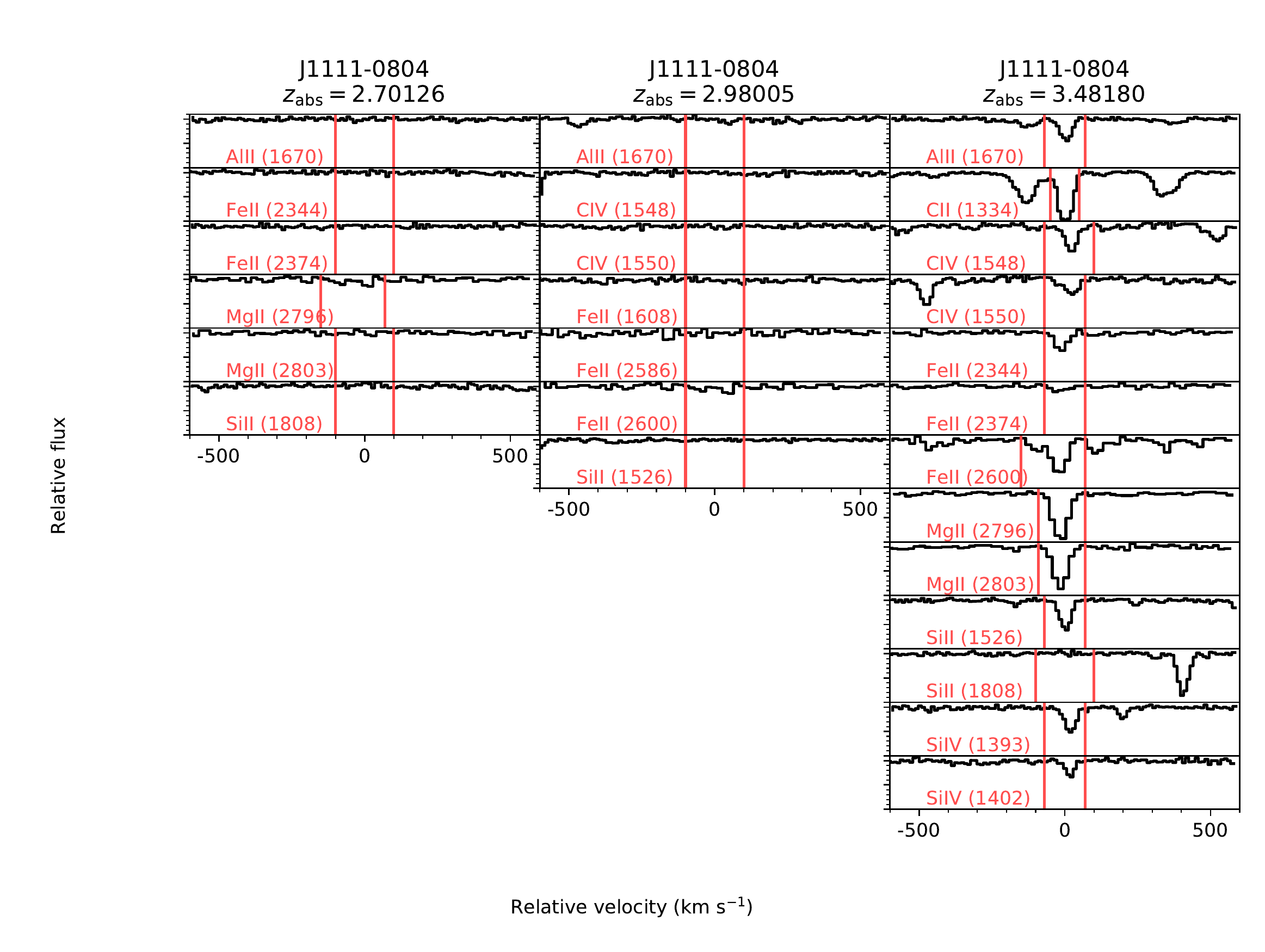}
\end{subfigure}
\caption{(cont'd)}
\end{center}
\end{figure*}

\begin{figure*}
\ContinuedFloat
\begin{center}
\begin{subfigure}{\textwidth}
\includegraphics[width=0.95\textwidth]{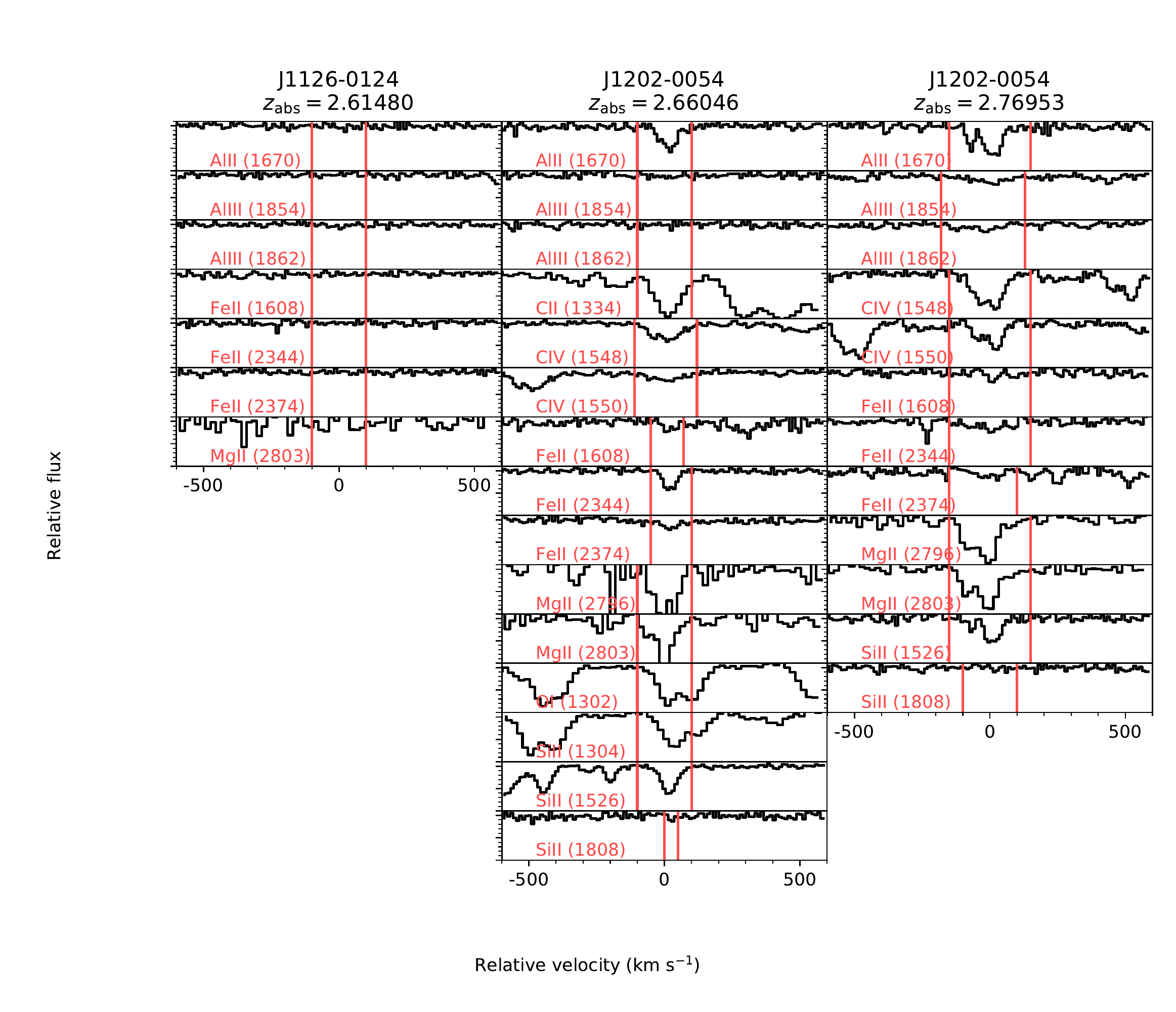}
\end{subfigure}
\caption{(cont'd)}
\end{center}
\end{figure*}

\begin{figure*}
\ContinuedFloat
\begin{center}
\begin{subfigure}{\textwidth}
\includegraphics[width=0.95\textwidth]{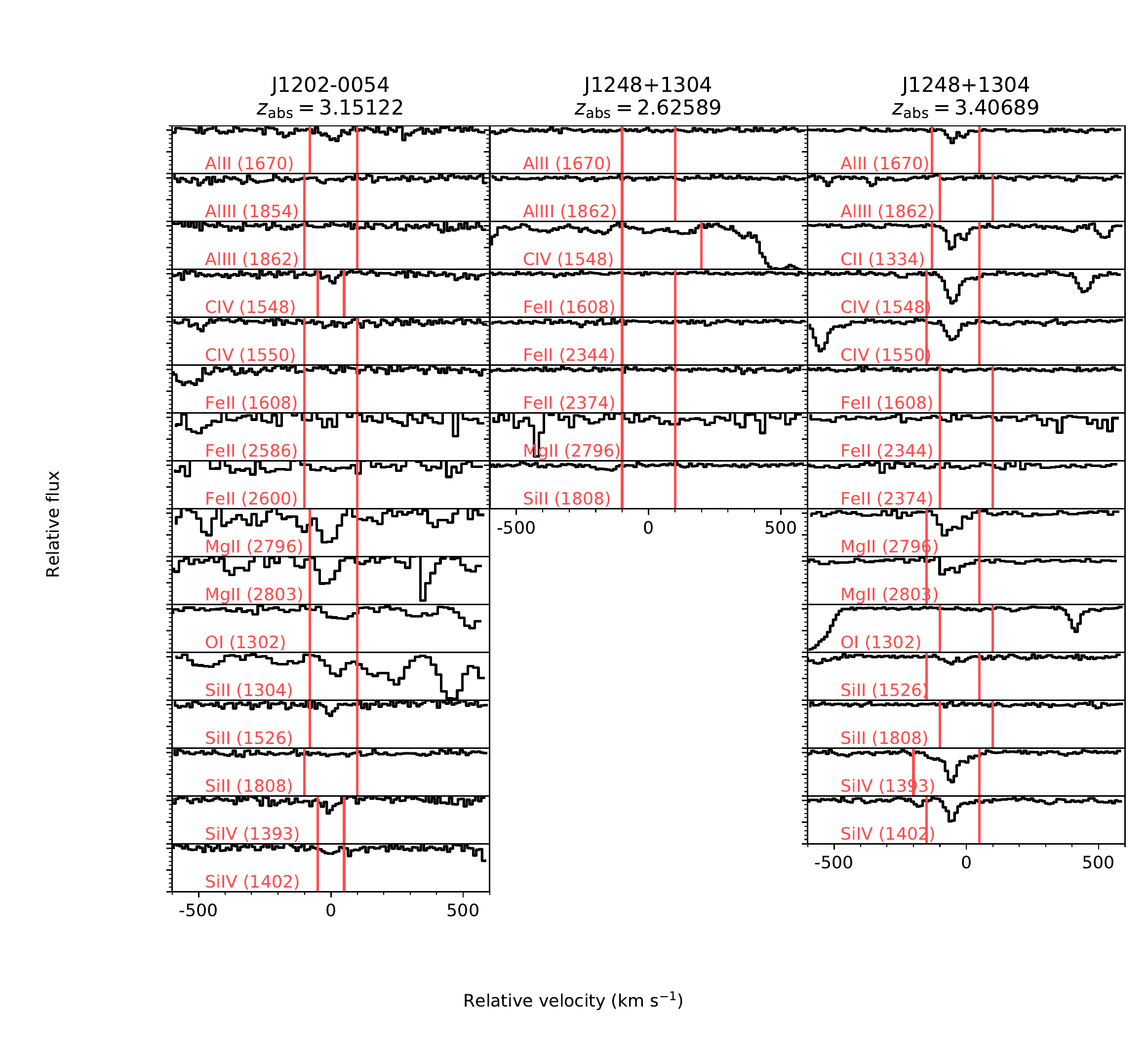}
\end{subfigure}
\caption{(cont'd)}
\end{center}
\end{figure*}

\begin{figure*}
\ContinuedFloat
\begin{center}
\begin{subfigure}{\textwidth}
\includegraphics[width=0.95\textwidth]{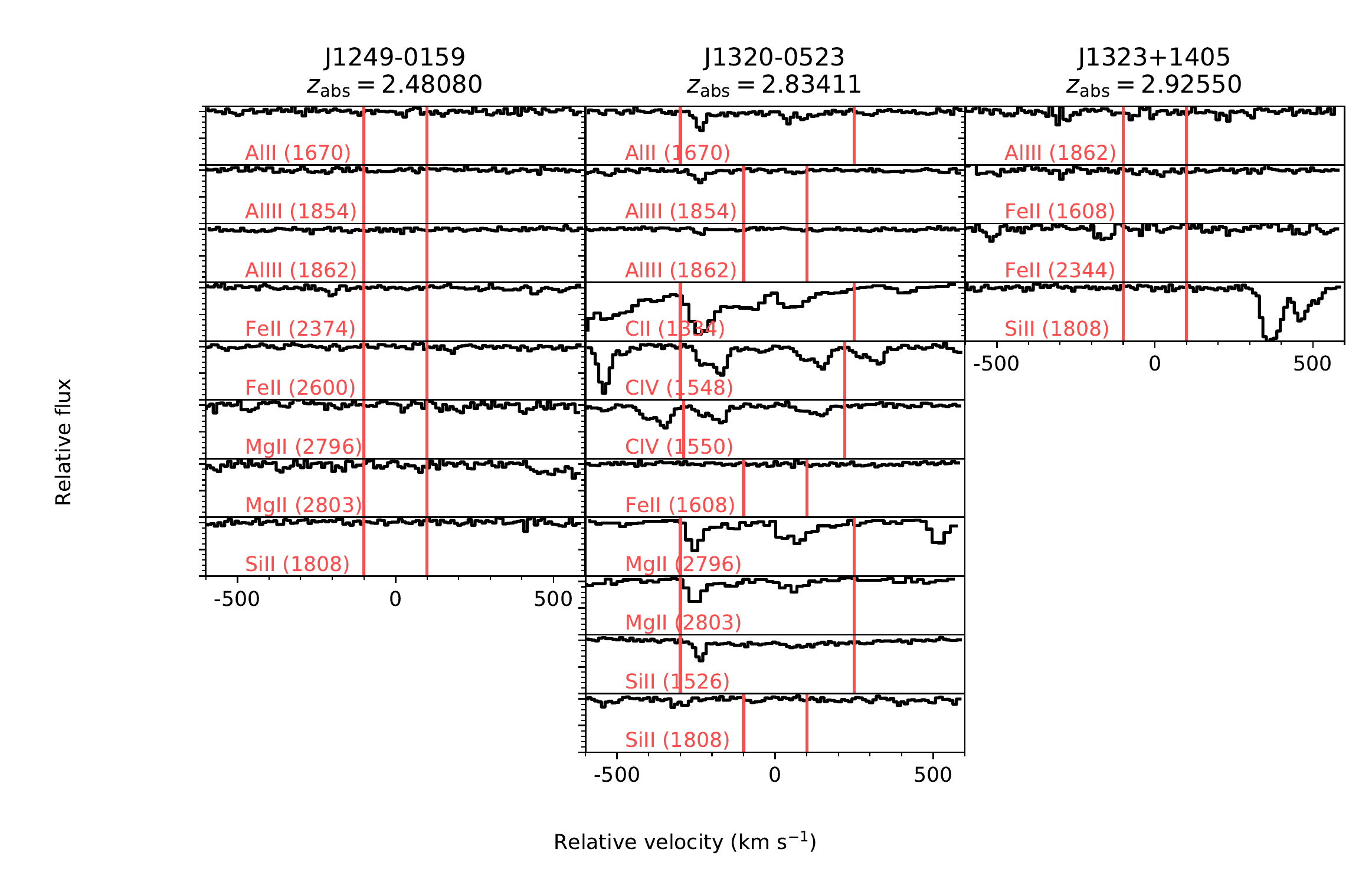}
\end{subfigure}
\caption{(cont'd)}
\end{center}
\end{figure*}

\begin{figure*}
\ContinuedFloat
\begin{center}
\begin{subfigure}{\textwidth}
\includegraphics[width=0.95\textwidth]{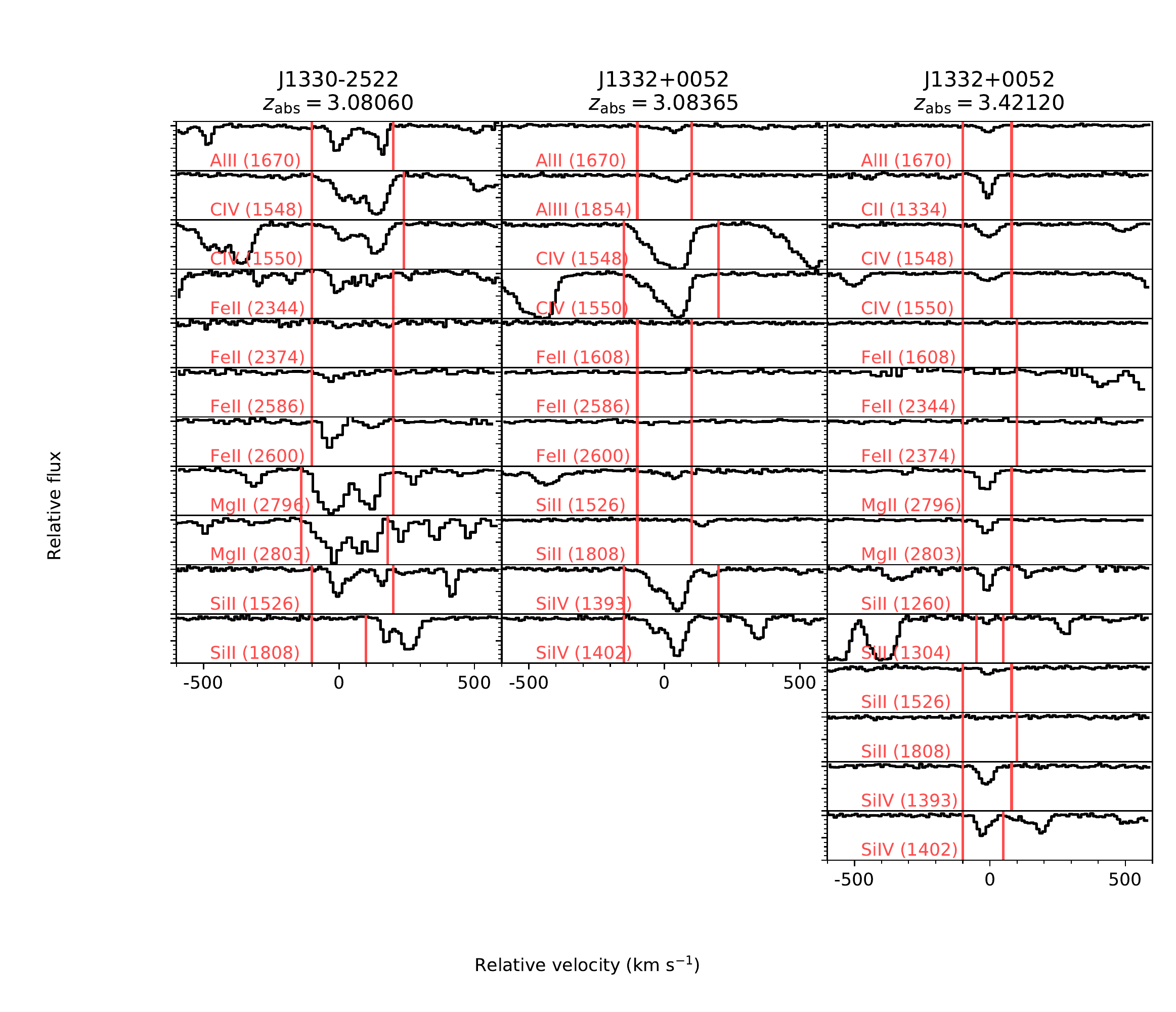}
\end{subfigure}
\caption{(cont'd)}
\end{center}
\end{figure*}

\begin{figure*}
\ContinuedFloat
\begin{center}
\begin{subfigure}{\textwidth}
\includegraphics[width=0.95\textwidth]{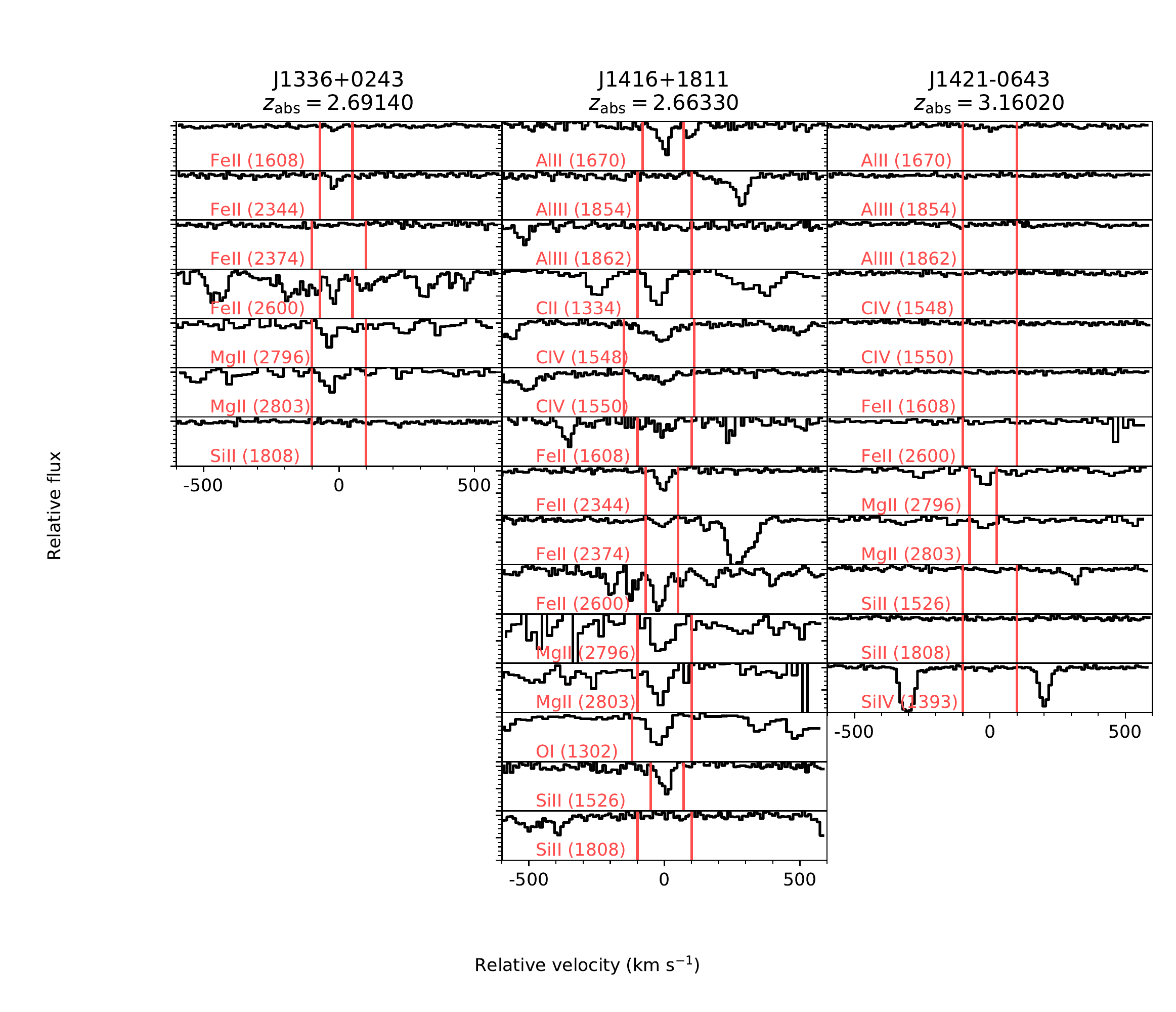}
\end{subfigure}
\caption{(cont'd)}
\end{center}
\end{figure*}

\begin{figure*}
\ContinuedFloat
\begin{center}
\begin{subfigure}{\textwidth}
\includegraphics[width=0.95\textwidth]{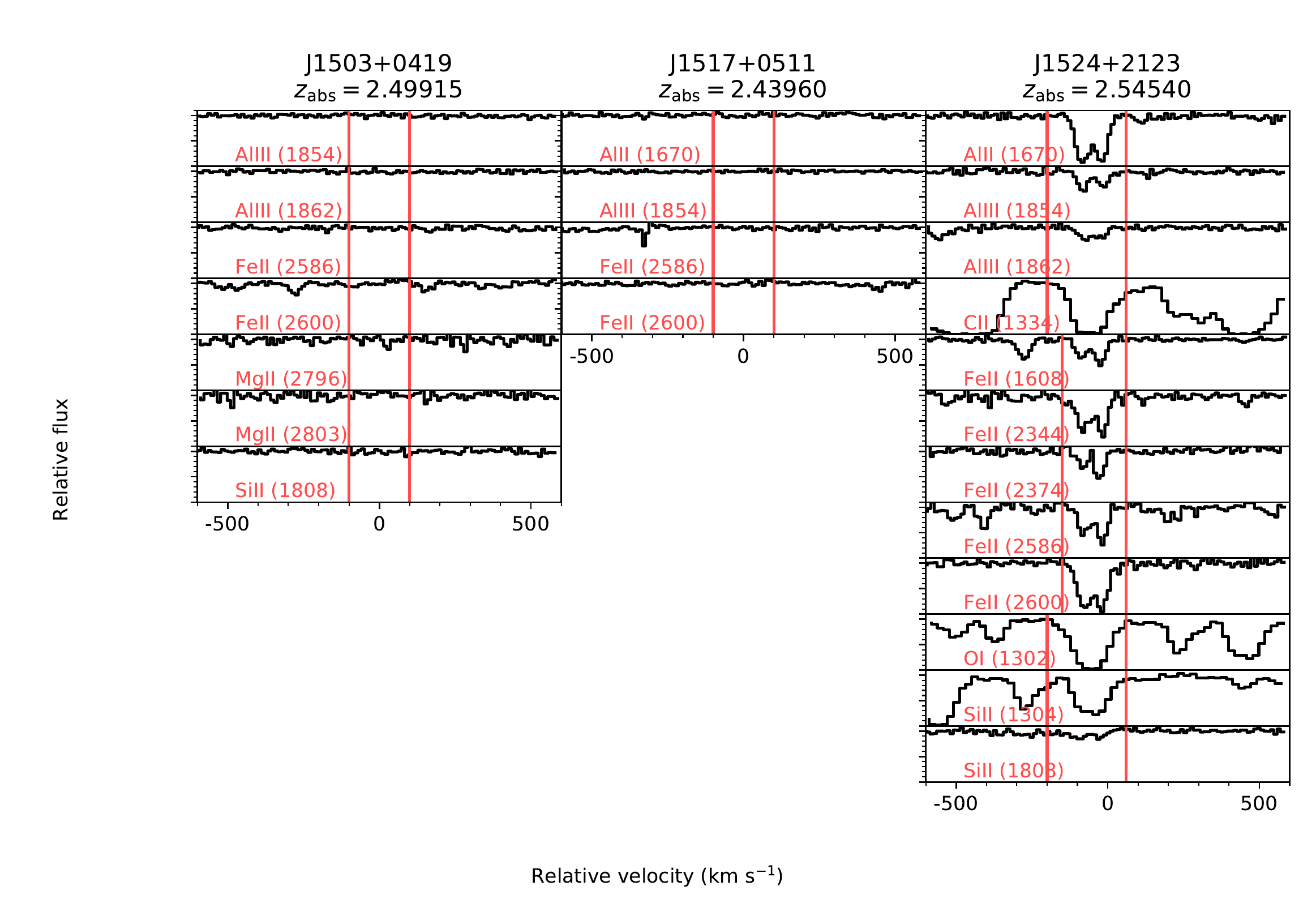}
\end{subfigure}
\caption{(cont'd)}
\end{center}
\end{figure*}

\begin{figure*}
\ContinuedFloat
\begin{center}
\begin{subfigure}{\textwidth}
\includegraphics[width=0.95\textwidth]{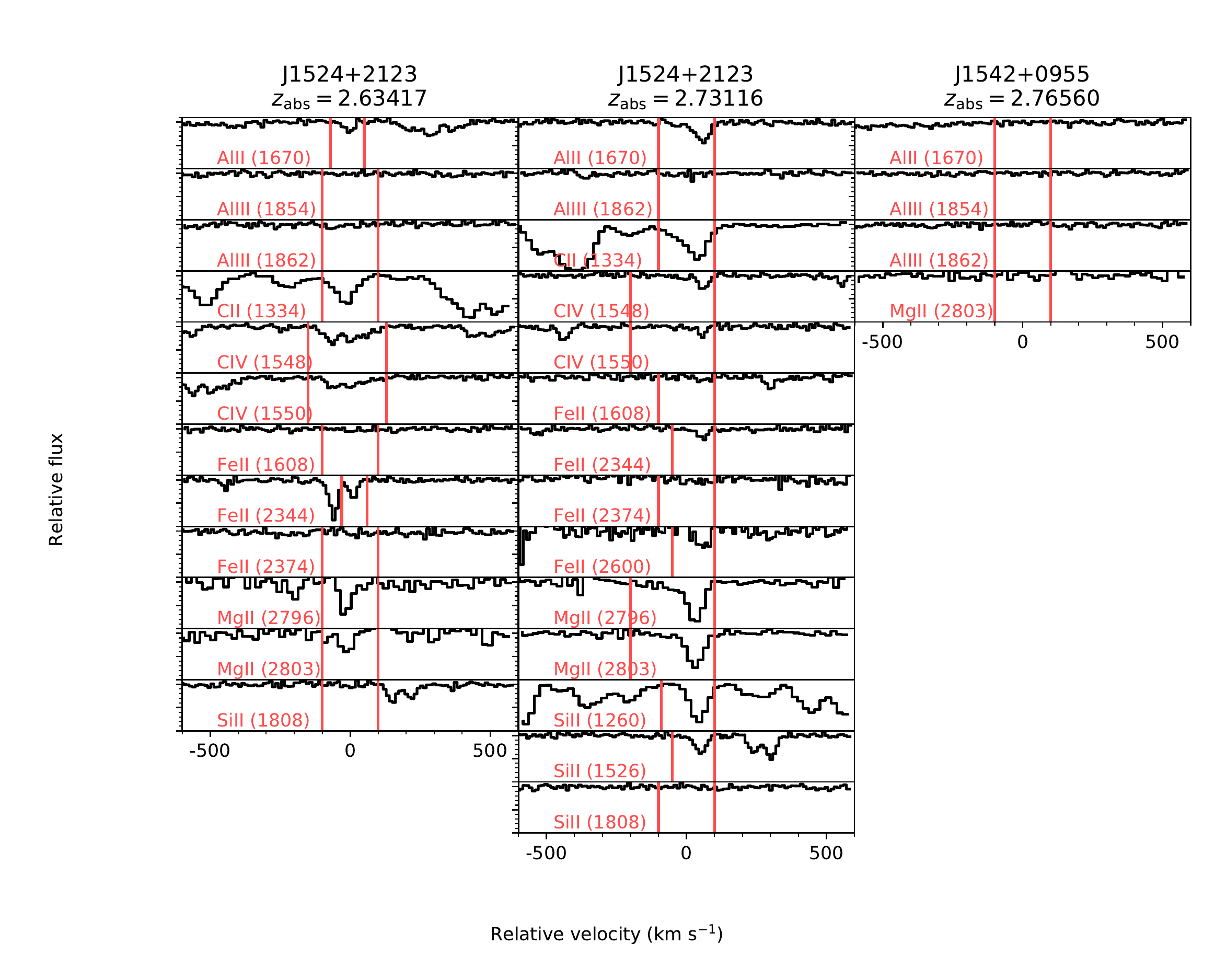}
\end{subfigure}
\caption{(cont'd)}
\end{center}
\end{figure*}

\begin{figure*}
\ContinuedFloat
\begin{center}
\begin{subfigure}{\textwidth}
\includegraphics[width=0.95\textwidth]{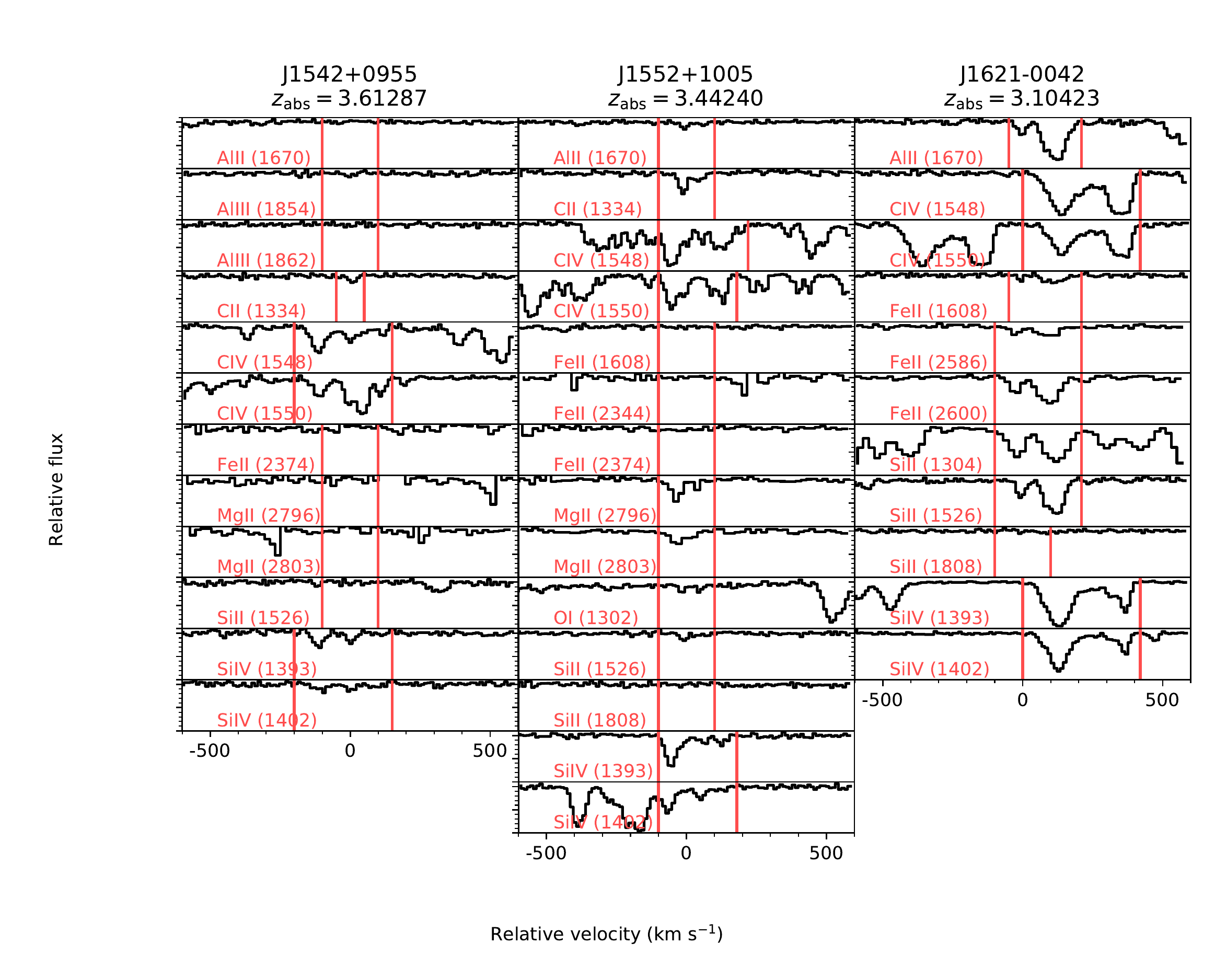}
\end{subfigure}
\caption{(cont'd)}
\end{center}
\end{figure*}

\begin{figure*}
\ContinuedFloat
\begin{center}
\begin{subfigure}{\textwidth}
\includegraphics[width=0.95\textwidth]{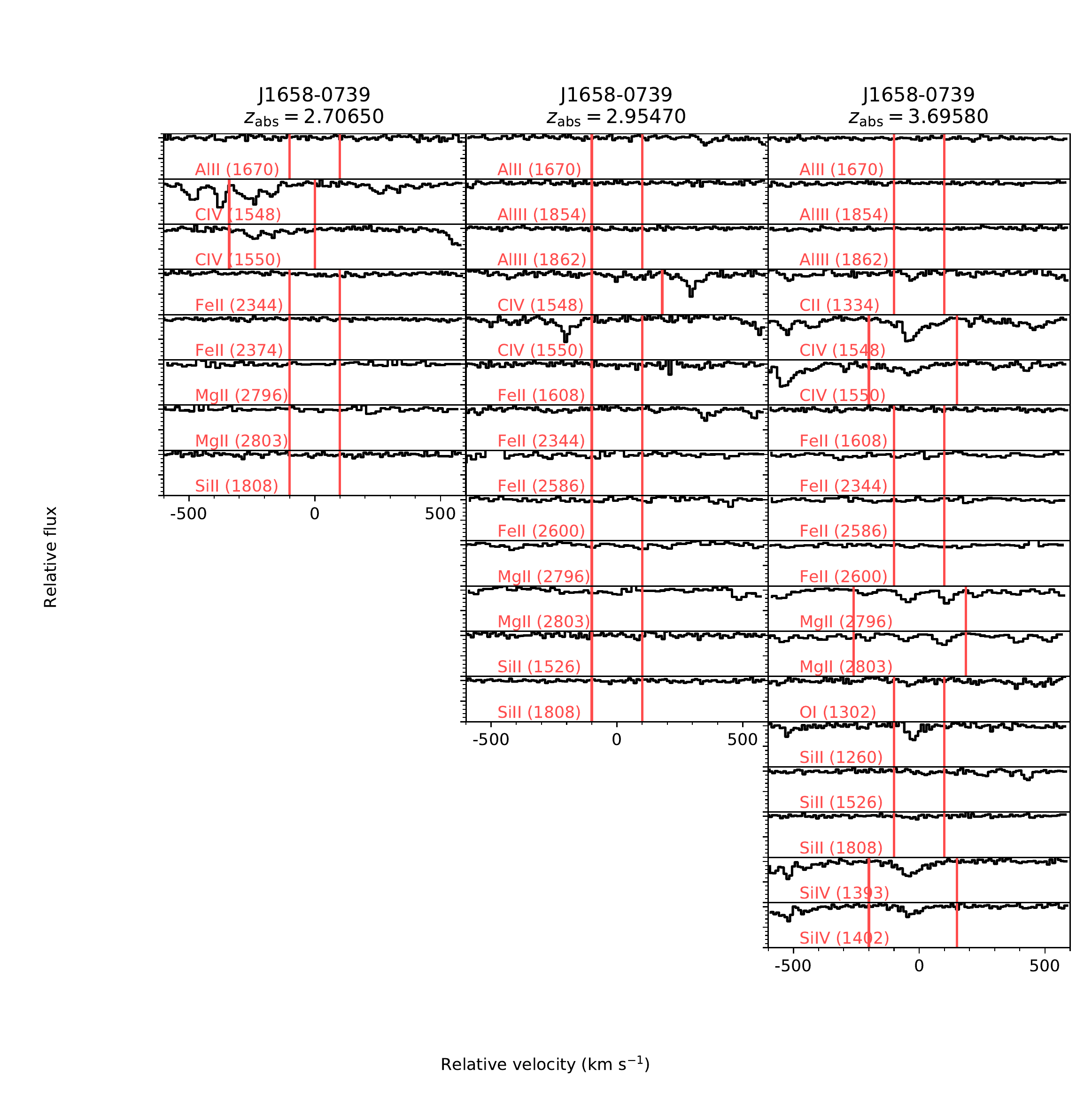}
\end{subfigure}
\caption{(cont'd)}
\end{center}
\end{figure*}

\begin{figure*}
\ContinuedFloat
\begin{center}
\begin{subfigure}{\textwidth}
\includegraphics[width=0.95\textwidth]{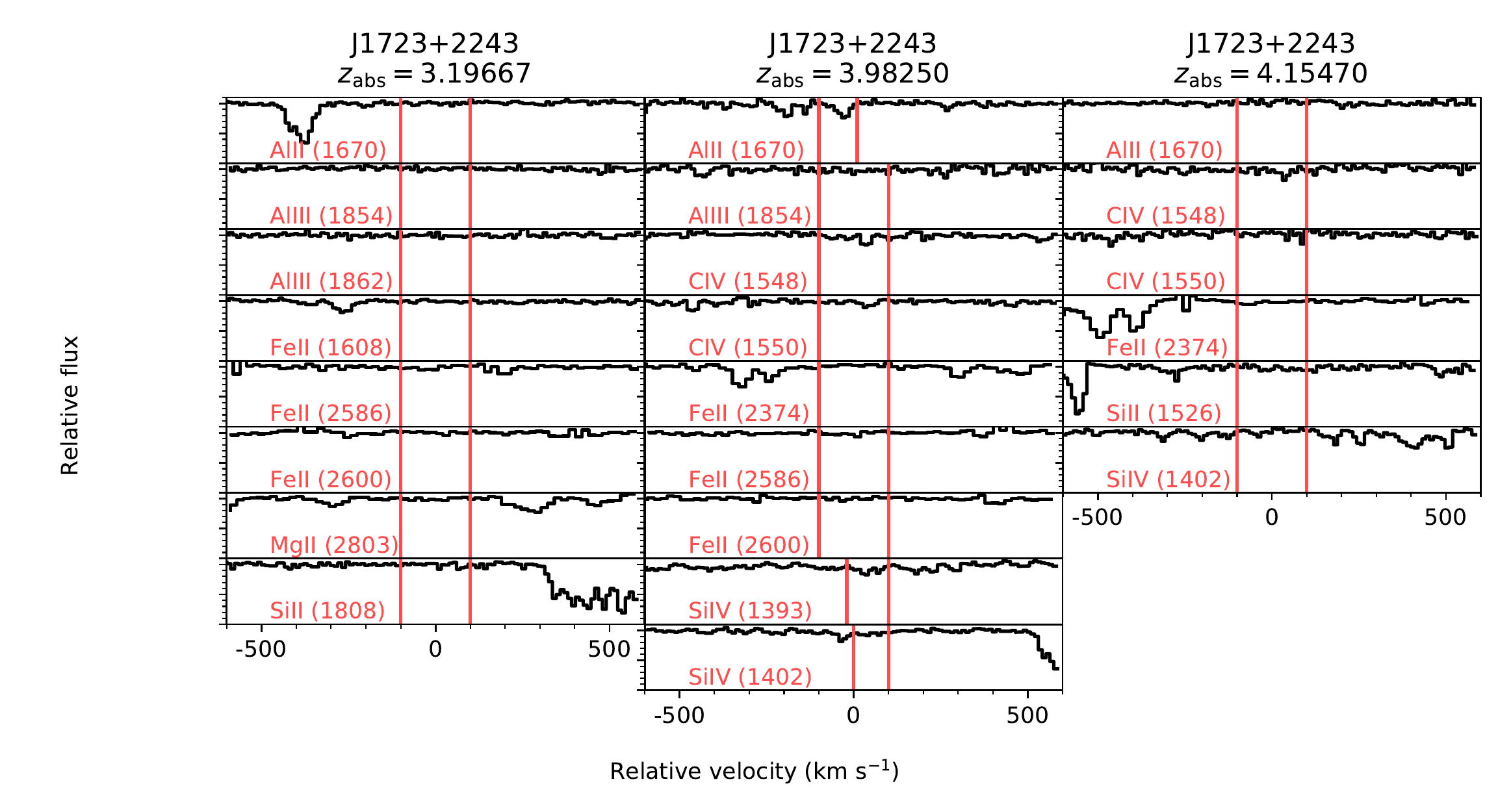}
\end{subfigure}
\caption{(cont'd)}
\end{center}
\end{figure*}

\begin{figure*}
\ContinuedFloat
\begin{center}
\begin{subfigure}{\textwidth}
\includegraphics[width=0.95\textwidth]{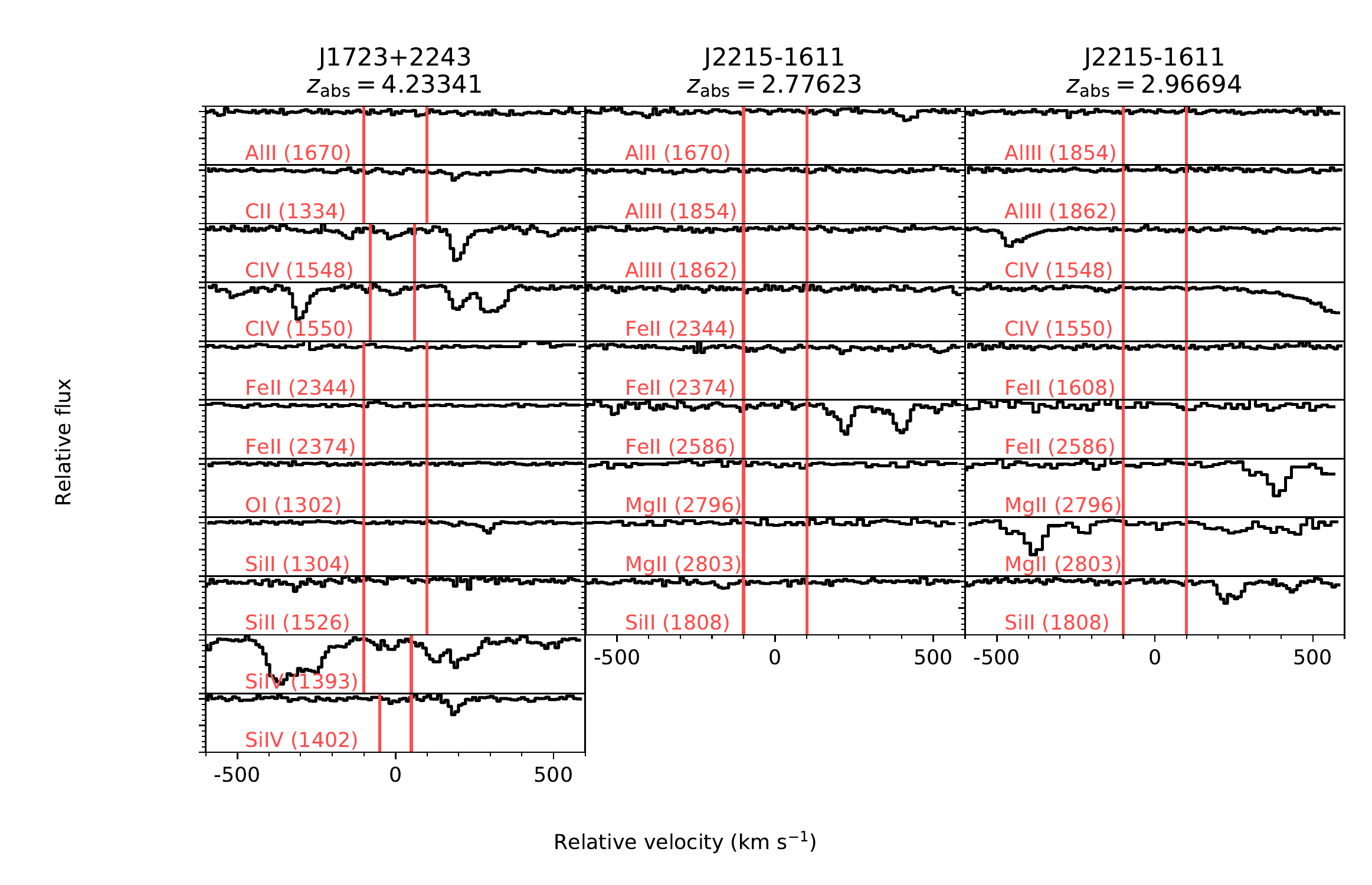}
\end{subfigure}
\caption{(cont'd)}
\end{center}
\end{figure*}

\begin{figure*}
\ContinuedFloat
\begin{center}
\begin{subfigure}{\textwidth}
\includegraphics[width=0.95\textwidth]{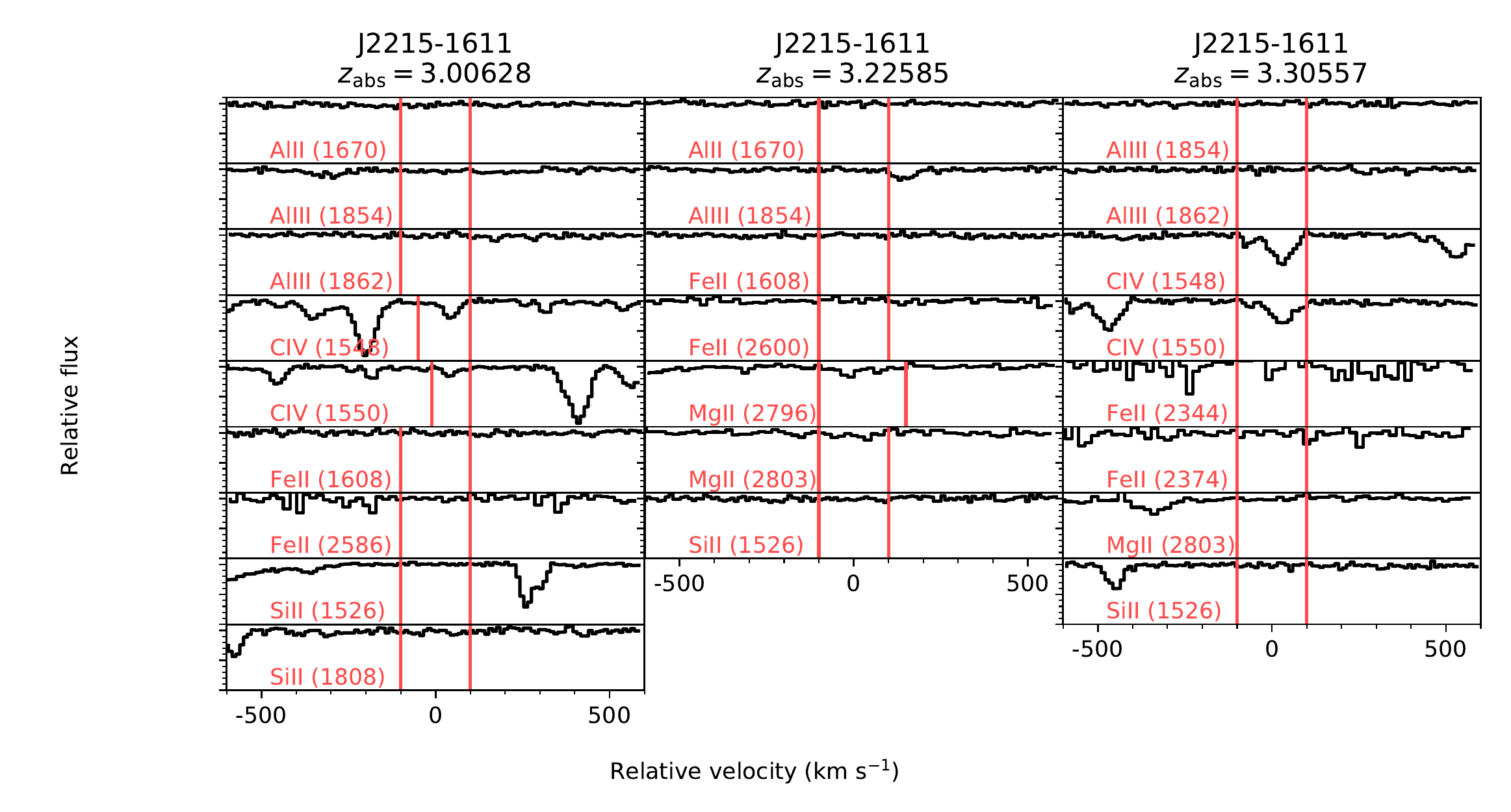}
\end{subfigure}
\caption{(cont'd)}
\end{center}
\end{figure*}

\begin{figure*}
\ContinuedFloat
\begin{center}
\begin{subfigure}{\textwidth}
\includegraphics[width=0.95\textwidth]{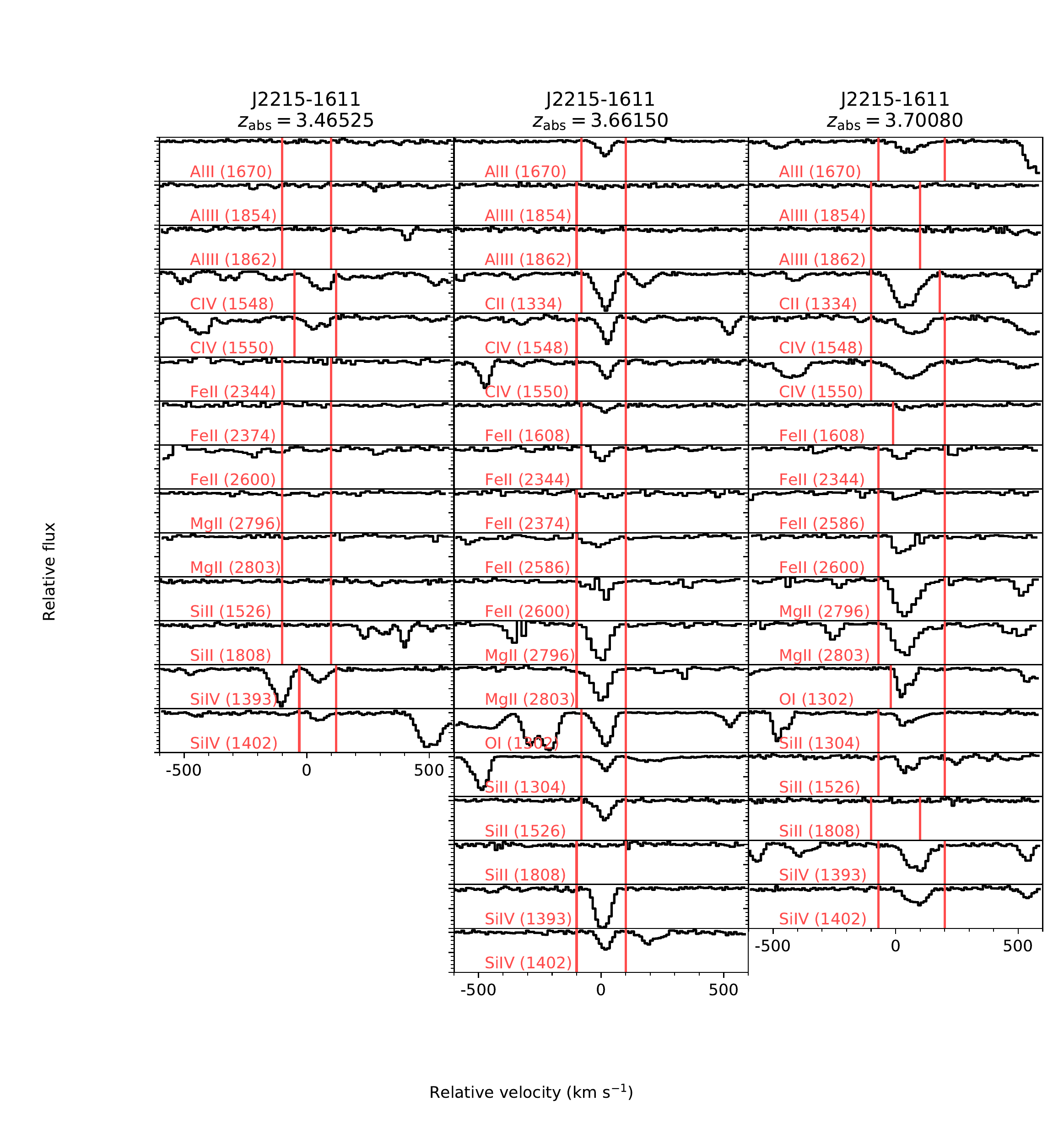}
\end{subfigure}
\caption{(cont'd)}
\end{center}
\end{figure*}

\clearpage

\begin{figure*}
\ContinuedFloat
\begin{center}
\begin{subfigure}{\textwidth}
\includegraphics[width=0.95\textwidth]{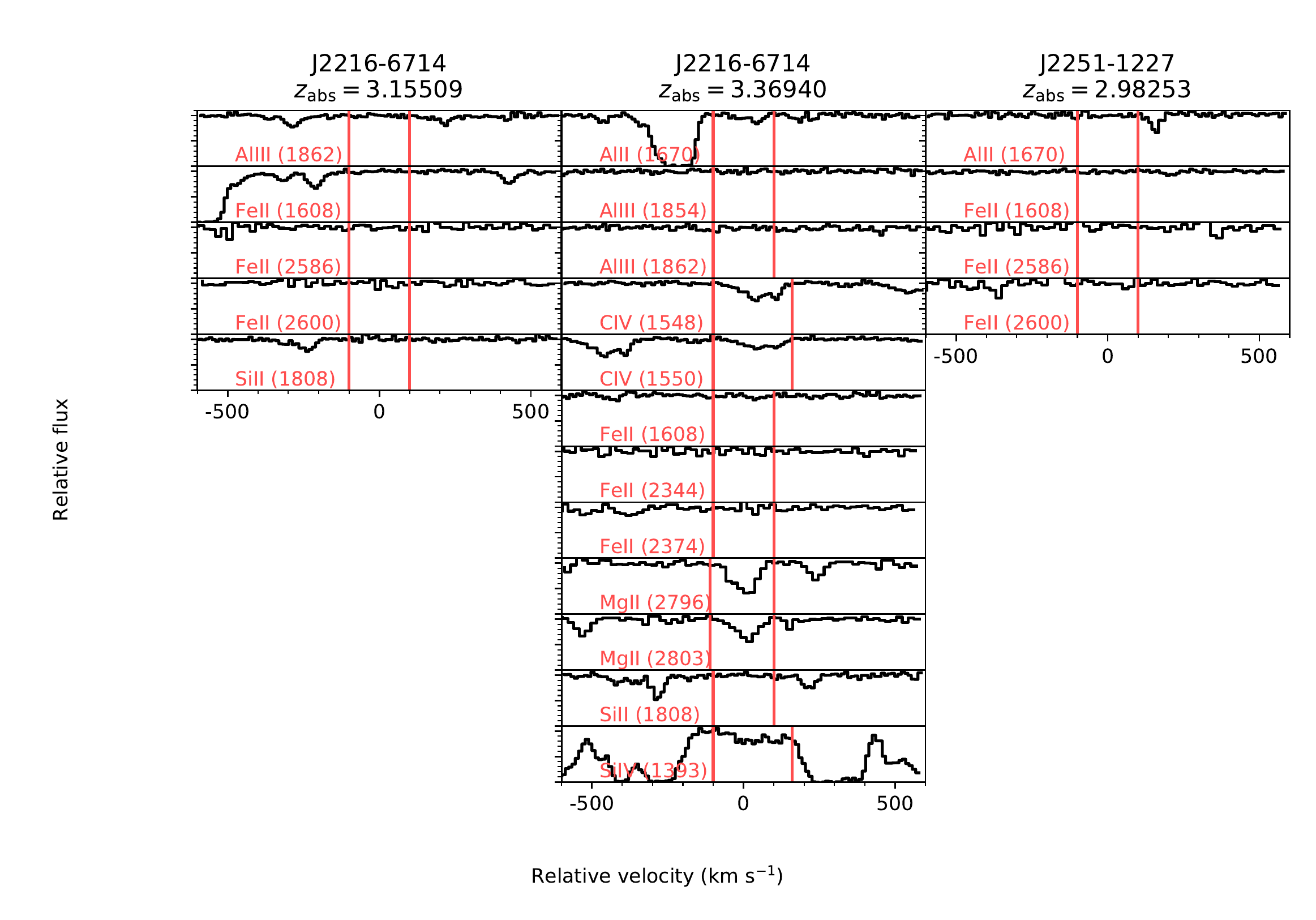}
\end{subfigure}
\caption{(cont'd)}
\end{center}
\end{figure*}

\begin{figure*}
\ContinuedFloat
\begin{center}
\begin{subfigure}{\textwidth}
\includegraphics[width=0.95\textwidth]{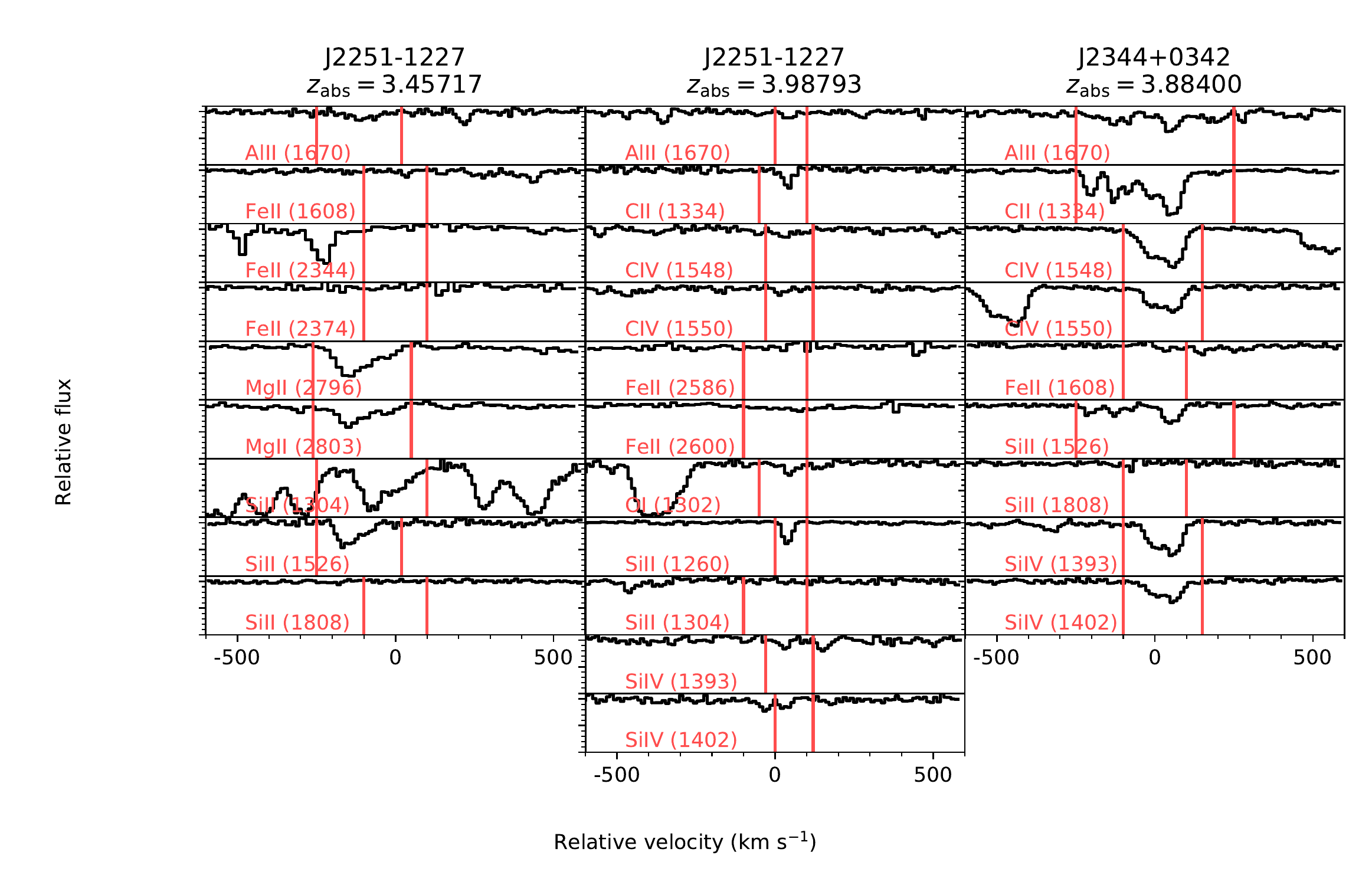}
\end{subfigure}
\caption{(cont'd)}
\end{center}
\end{figure*}

\begin{figure*}
\ContinuedFloat
\begin{center}
\begin{subfigure}{\textwidth}
\includegraphics[width=0.95\textwidth]{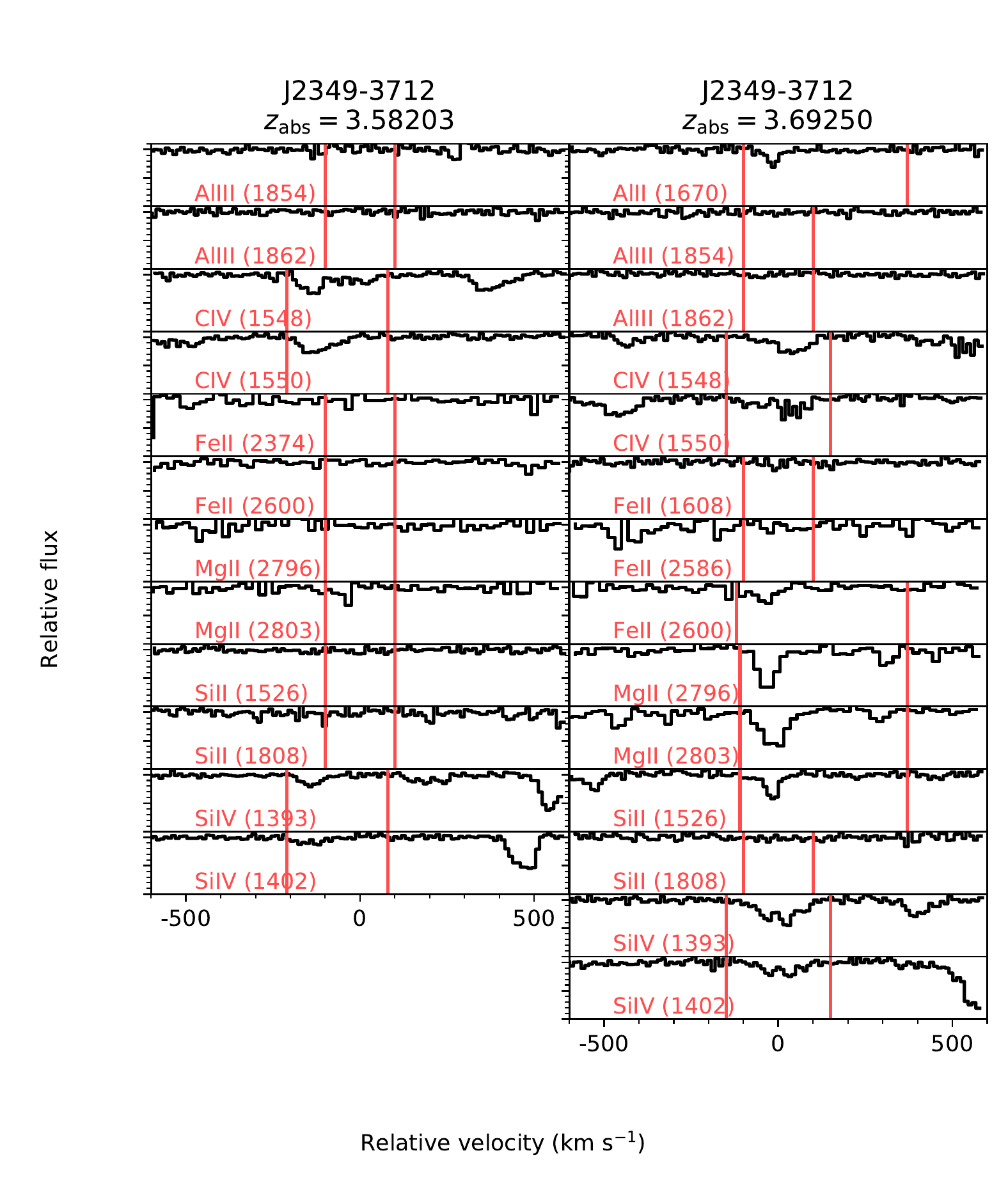}
\end{subfigure}
\caption{(cont'd)}
\end{center}
\end{figure*}

\end{document}